\documentclass[a4paper]{article}
\pdfoutput=1 

\usepackage{jheppub} 
\usepackage[T1]{fontenc} 
\usepackage[italian,english]{babel}
\usepackage{hyperref}
\hypersetup{
	colorlinks=true,
	linkcolor=[rgb]{0.0,0.42,0.89},
	filecolor=magenta,
	urlcolor=[rgb]{0.0,0.42,0.89},
	citecolor=[rgb]{0.64,0.0,0.0},
}

\usepackage{ifpdf}
\usepackage{subfigure}
\usepackage{tikz}
\usepackage[compat=1.1.0]{tikz-feynman}
\usetikzlibrary{arrows,shapes}
\usetikzlibrary{trees}
\usetikzlibrary{matrix,arrows} 				
\usetikzlibrary{positioning}				
\usetikzlibrary{calc,through}				
\usetikzlibrary{decorations.pathreplacing}  
\usepackage{pgffor}							

\usetikzlibrary{decorations.pathmorphing}	
\usetikzlibrary{decorations.markings}
\tikzset{
    vector/.style={decorate, decoration={snake}, draw},
    fermion/.style={draw=black, postaction={decorate}},
    scalar/.style={dashed,draw=black, postaction={decorate}}}
\tikzstyle{block} = [draw, rectangle,
    minimum height=3em, minimum width=6em]

\usepackage{amssymb}
\usepackage{amsfonts}
\usepackage{epsf}
\usepackage{rotating}
\usepackage{graphicx}
\usepackage{amsmath}
\usepackage{fancyhdr}
\usepackage{lineno}
\usepackage{babel}
\usepackage{graphics}
\usepackage{pstricks}
\usepackage{color}
\usepackage{multirow}
\usepackage{array}
\usepackage{diagbox}
\usepackage{cancel}

\usepackage{listings}
\newcolumntype{P}[1]{>{\centering\arraybackslash}p{#1}}
\newcolumntype{M}[1]{>{\centering\arraybackslash}m{#1}}

\newcommand{\lsim}{\mathrel{\mathop{\kern 0pt \rlap
  {\raise.2ex\hbox{$<$}}}
  \lower.9ex\hbox{\kern-.190em $\sim$}}}
\newcommand{\gsim}{\mathrel{\mathop{\kern 0pt \rlap
  {\raise.2ex\hbox{$>$}}}
  \lower.9ex\hbox{\kern-.190em $\sim$}}}

\newcommand{\be}{\begin{equation}}
\newcommand{\ee}{\end{equation}}
\newcommand{\bea}{\begin{eqnarray}}
\newcommand{\eea}{\end{eqnarray}}

\def\gev{\ensuremath{\mathrm{\,Ge\kern -0.1em V\,}}}
\def\tev{\ensuremath{\mathrm{\,Te\kern -0.1em V\,}}}

\title{\boldmath Boosted  displaced decay of right-handed neutrinos at CMS, ATLAS and MATHUSLA}

\author[a]{Priyotosh Bandyopadhyay,}
\author[b]{Eung Jin Chun}
\author[a]{and Chandrima Sen}

\affiliation[a]{Indian Institute of Technology Hyderabad, Kandi,  Sangareddy-502285, Telengana, India}
\affiliation[b]{Korea Institute for Advanced Study, Seoul 02455, Korea}

\emailAdd{bpriyo@phy.iith.ac.in}
\emailAdd{ejchun@kias.re.kr}
\emailAdd{PH19RESCH11014@iith.ac.in}

\preprint{ IITH-PH-0002/22,\\ \hspace*{12.18 cm} KIAS-P22017}

\begin{document}

\abstract{
We investigate  boosted displaced signatures in the Type-I seesaw mechanism associated with the $B-L$ gauge symmetry. Such events arise from decays of right-handed neutrinos depending on their Yukawa couplings and masses. Considering two scenarios: (a) three degenerate right-handed neutrinos whose Yukawa couplings are reconstructed from the observed neutrino masses and mixing; (b) only one right-handed neutrino which decouples from the observed neutrino mass generation and thus its coupling can be arbitrarily small, a detailed PYTHIA based simulation is performed to determine the parameter regions of the $B-L$ gauge boson mass, the neutrino Yukawa couplings, and the right-handed neutrino mass sensitive to CMS, ATLAS, proposed FCC-hh detector and MATHUSLA at the centre of mass energies of 14, 27 and 100 TeV  via displaced signatures.  We also show  in detail how the boost effect enhances the displaced decay lengths, especially for the longitudinal ones, and hinders the probe of Majorana nature of neutrinos.}

\maketitle
\flushbottom

\section{Introduction}\label{introduction}

The discovery of Higgs boson has proved the correctness of the Standard Model (SM) \cite{CMS, ATLAS}.
Its standard decay modes have been discovered mostly at $5\sigma$-level at both CMS and ATLAS detectors reconstructing the events of di-photon \cite{htoggCMS,htoggATLAS}, $W^\pm W^\mp$ \cite{HtoWWCMS,HtoWWATLAS}, $ZZ$ \cite{HtoZZCMS,HtoZZATLAS}, $b\bar{b}$ \cite{HtobbCMS,HtobbATLAS} and $\tau \bar{\tau}$ \cite{HtotautauCMS,HtotautauATLAS}. Although no phenomenon beyond SM has been observed so far,
concerns of various issues viz. light neutrino mass lead us to the extension of SM which still allows
the presence of some non-standard exotic decay modes and extra new particles in a vast mass range.

In this article, we consider a simple extension of SM with the gauge group of $U(1)_{B-L}$.
This requires an additional $B-L$ scalar \cite{BLmodels, BLmodels1,Mohapatra:1980qe}, and
three right-handed neutrinos (RHNs) which are charged under $U(1)_{B-L}$ and generate small neutrino masses  via Type-I seesaw mechanism \cite{ty1d, suchitraBmL,ADas,Liu:2022kid,Das:2022rbl,Padhan:2022fak,FileviezPerez:2009hdc,Kang:2015uoc,Das:2017nvm,Han:2006ip,Dev:2013wba,Accomando:2017qcs}. The $B-L$ scalar is responsible for spontaneous breaking of $U(1)_{B-L}$ and generating Majorana masses of the right-handed neutrinos.  It is also popular to introduce additional scalar as a dark matter candidate which is charged under $U(1)_{B-L}$ \cite{Bandyopadhyay:2020ufc,BmLDM2,Bandyopadhyay:2022xlp}.

Focusing on relatively light RHNs or very small neutrino Yukawa couplings, we investigate displaced leptonic jet signatures. The RHNs, pair-produced from a heavy $B-L$ gauge boson decay, will have a large boost and rather slow decays into lepton plus jets  via (off-shell) $W$, $Z$ and Higgs bosons. As a consequence, there appear displaced vertices producing boosted collimated lepton plus jets \cite{Izaguirre:2015pga}, which  are almost background free and would be a sensitive probe of the scenario.  Such a Fatjet like signature can be easily reconstructed to decrypt the RHN mass.
These signatures may come as well with lepton flavor violation and same-sign di-leptons manifesting the Majorana nature of neutrinos. Combining all of these features we will show the future sensitivity of the model parameter space at HL-LHC \cite{ZurbanoFernandez:2020cco}, HE-LHC \cite{FCC:2018bvk} and  FCC-hh \cite{FCC:2018vvp} with the centre of mass energies of 14, 27 and 100 TeV, respectively.
We will also discuss how the boost effect makes it difficult to probe the Majorana nature of neutrinos. In particular the investigations are done with three generations of RHN considering $U_{\rm PMNS}$ and one generation RHNs with the possibility of the displaced leptons that are detectable within MATHUSLA along with CMS, ATLAS and the proposed FCC-hh detector\,(for 100 TeV centre of mass energy). Though same sign and opposite sign di-leptons coming from RHNs carry the Majorana signature, it can only be probed at the CMS, ATLAS and FCC-hh reference detector. MATHUSLA fails to measure such ratio owing to be situated in one side of the hemispheres of the beam pipe line.

The paper is organized as follows. First we briefly describe main features of the model in \autoref{model}.
We present the results of our study considering two simple scenarios of Type-I seesaw mechanism: three degenerate right-handed neutrinos in \autoref{SC1}; and only one RHN decoupled from the observed neutrino mass generation in \autoref{SC2}. For each scenario, a {\tt PYTHIA8}  based simulation is done with the kinematical distributions to explore displaced vertex signatures at CMS, ATLAS, proposed FCC-hh reference detector and MATHUSLA. We show how the boost effect can skew the Majorana nature of RHNs and how to reinstate it in \autoref{boostfs} and \autoref{ssd_osd}, respectively. Finally we conclude in \autoref{discussion} with discussion.

\section{Model }\label{model}
We extend the Standard Model (SM) with an extra gauge group of $U(1)$ in the minimal $B-L$
framework where the mixing between $U(1)_Y$ and $U(1)_{B-L}$ groups are neglected. Thus the Lagrangian obeys the gauge group $SU(3)_c \times SU(2)_L \times U(1)_Y \times U(1)_{B-L}$.
The three generations of right-handed neutrinos are introduced to generate the light neutrino mass via Type-I seesaw mechanism \cite{TypeI,PBUBmL}. Here the right-handed neutrinos are though SM gauge singlets but charged under $U(1)_{B-L}$. However, the $B-L$ symmetry is broken at relatively higher scale when the $B-L$ scalar $\chi$  gets vacuum expectation value (vev) and spontaneously generates the Majorana mass ($M_N$)term for the right-handed neutrinos ($N$) as well as the $B-L$ gauge boson  $Z_{B-L}$. The charge assignments for the fields are given in \autoref{Table1} such that $B-L$ anomaly cancels and the theory becomes anomaly free.

\begin{table}[h]
	\renewcommand{\arraystretch}{1.2}
	\centering
	\begin{tabular}{|c|c|c|}
		\hline
		& $SU(3)_c \times SU(2)_L \times U(1)_Y$ & $Y_{B-L}$ \\ \hline \hline
		$\Phi$	& $(1,\,2,\,1)$ & $0$  \\	\hline
		$N$	& $(1,\,1,\,0)$ & $-1$  \\\hline
		$L$	& $(1,\,2,\,-1)$ & $-1$  \\ \hline
		$Q$	& $(3,\,2,\,1/3)$ & $1/3$  \\ \hline
		$u_R$	& $(3,\,1,\,4/3)$ & $1/3$ \\ \hline
		$d_R$ & $(3,\,1,\,-2/3)$ & $1/3$  \\ \hline
		$e_R$ & $(1,\,1,\,-2)$ & $1$  \\ \hline
		$\chi$	& $(1,\,1,\,0)$ & $2$  \\ \hline
	\end{tabular}
	\caption{Particle content and there corresponding charges.}  \label{Table1}
\end{table}	

The scalar part of the Lagrangian is given by
\begin{eqnarray}\label{eq1}
\mathcal{L}_S= \left(D^\mu \Phi\right)^\dagger \left(D_\mu \Phi\right)+ (D^\mu \chi)^\dagger (D_\mu \chi)-V(\Phi , \chi),
\end{eqnarray}
with the covariant derivative $D_\mu$ can be written as
\begin{eqnarray}
D_\mu=\partial _\mu + ig_2\,T^a\, W_\mu ^a +ig_1YB_\mu + ig_{BL}\,Y_{B-L}\,B'_\mu.
\end{eqnarray}	
Here  we use  $B'_\mu$  for the $U(1)_{B-L}$  gauge field with strength $g_{BL}$ and hypercharge $Y_{B-L}$.  The total scalar potential in this case can be written as in \autoref{eq1},
\begin{eqnarray}
V(\Phi , \chi)= m_\Phi ^2 \left(\Phi ^\dagger \Phi\right)+ m_\chi ^2 |\chi|^2 + \lambda_1 \left(\Phi ^\dagger \Phi\right)^2 + \lambda_2 |\chi|^4 + \lambda_3 \Phi^\dagger \Phi |\chi|^2,
\end{eqnarray}
where $\Phi$ is the complex scalar  SM Higgs doublet, $\Phi=
\begin{pmatrix}
\phi ^\dagger & \phi ^0
\end{pmatrix}^T$
and $\chi$ is the complex $B-L$ scalar which is singlet under SM gauge group.  The $B-L$ gauge group is broken spontaneously when $\chi$ attains a vev, $v_{BL}$ i.e. at the $B-L$ breaking we can write $\chi = (v_{BL}+ \chi_0 +i\chi ')/\sqrt{2}$. Similarly, the SM gauge group is also broken when $\phi^0$ gets a vev, i.e. at $\phi ^0= (v+h)/\sqrt{2}$ and we are left with only electromagnetic and $SU(3)$ as symmetric gauge group. After symmetry  breaking we have two mass eigenstates $h_{1,2}$, where $h_1\sim h_{\rm SM}$ satisfying the Higgs data \cite{CMS, ATLAS}.

As can be read from \autoref{eq12}, the RHNs couple to SM Higgs doublet  via $Y_{N_{ij}}$  and with $B-L$ scalar via $\lambda_{N_{ij}}$ Yukawa terms respectively, where $i,j$ are generation indices.
\begin{eqnarray}
\mathcal{L}_Y= &-&Y_{ij}^u\, \overline{Q}_i \,\tilde{\Phi}(u_R)_j -Y_{ij}^d\, \overline{Q}_i \,\Phi(d_R)_j -Y_{ij}^e\, \overline{L}_i \,\Phi(e_R)_j \nonumber \\ [3pt]
&-&\underbrace{(Y_N)_{ij}\, \overline{L}_i \,\tilde{\Phi}N_j}_{\text{Dirac mass term}} -\underbrace{(\lambda_N)_{ij}\, \chi \overline{N}_i^C \,N_j}_{\text{Majorana mass term}}, \label{eq12}
\end{eqnarray}
where, $\tilde{\Phi}=i\sigma^2 \Phi ^*$ and $i, \, j$ denote three fermion generations. Here the $B-L$ breaking vev i.e. $v_{BL}$ generates the Majorana neutrino mass term for the RHNs  as $M_{N} = 2\lambda_N \langle \chi \rangle$ and the Dirac mass term is generated via the vev of EWSB i.e. $v$ as $m_D = \frac{Y_N v}{\sqrt{2}}$. The $B-L$ gauge boson also becomes massive by absorbing the Goldstone boson corresponding to the $B-L$ symmetry, which is given as  $M_{Z_{B-L}} = 2g_{BL} v_{BL} $. Thereafter, the three light neutrinos can be generated  through the Type-I seesaw mechanism \cite{TypeI} by diagonalizing \autoref{numatrix}, which gives rise to the mass eigenvalues as shown in \autoref{eigen}

\begin{eqnarray}\label{numatrix}
\mathcal{M}_\nu = \begin{pmatrix}
0 & m_D \\
m_D^T & M_{N}
\end{pmatrix},
\end{eqnarray}
where the obtained  masses of light and heavy neutrinos are,
\begin{eqnarray}\label{eigen}
M_{\nu _L} \simeq \frac{Y_N ^2 \, v^2}{2 M_{N}} ~~~~\text{and},~~~~ M_{\nu _R} \simeq M_{N}.
\end{eqnarray}
Here for simplicity we have dropped the generation indices.

\section{Scenario-1}\label{SC1}

Here  we consider the three generations  of  RHNs considering the light neutrino masses and mixing data  as described below. The case of only one of the  RHNs with light Yukawa is discussed in \autoref{SC2}.
\subsection{Parameter space and benchmark points }\label{BPs_SC1}
We consider first the Type-I seesaw mechanism with three degenerate RHNs where $N_{1,2,3}$ have same mass $M_N$. Adopting the Casas-Ibarra  parameterization \cite{Casas:2001sr}, one can reconstruct their Yukawa matrix given the active neutrino mass matrix compatible with the observations.  Taking the central values of the neutrino oscillation parameters with the normal ordering \cite{Zyla:2020zbs}, we get the following neutrino mass matrix: 
$$M_\nu \approx
\left(
\begin{array}{ccc}
	0.0030 -0.0008 i & 0.0001 +0.0052 i & -0.0043+0.0047 i \\
	0.0001 +0.0052 i & 0.030 +0.00045 i & 0.021  \\
	-0.0043+0.0047 i & 0.021& 0.025
\end{array}
\right)  \mbox{eV}
$$
This can be reproduced by the following Yukawa matrix.
\be\label{Yuk}
 Y_N=
\left(
\begin{array}{ccc}
	0.62 -0.33 i & -0.54+0.30 i & -0.77-0.30 i \\
	-2.01+0.046 i & 0.83 -0.039 i & -2.23-0.09 i \\
	-1.41+0.042 i & 2.34 -0.036 i & -0.93-0.083 i
\end{array}
\right) \times 10^{-7} \sqrt{ M_N\over 100\mbox{GeV}}
\ee
One can see that $\sum_{\alpha=e,\mu,\tau} |Y_{N, \alpha k}|^2=6.54 \times 10^{-14}\, M_N/100\mbox{GeV}$ for each $N_{k}$ ($k=1, \, 2,\, 3$, the indices for three generations of RHN).
Thus, generation wise the branching ratio matrix is given by
\be\label{brf}
 \mbox{BR}_{\alpha k} =
\left(
	\renewcommand{\arraystretch}{1.4}
\begin{array}{ccc}
	0.0758772 & 0.0577112 & 0.104687 \\
	0.618174 & 0.105466 & 0.762472 \\
	0.305949 & 0.836823 & 0.132842
\end{array}
\right) \times B(N \to X Y)_{\rm{diag}},
\ee
where $B(N \to X Y)_{\rm{diag}}$ is the decay branching fraction of the RHN in a particular mode for the  choice of  diagonal Yukawa couplings. For example for  diagonal choice  of the Yukawa $Y_{N}$, and $M_{N}=500$ GeV, the $N_k$ decays to $W^\pm \ell^\mp_k, \, Z \nu_{\ell_k}, \, h \nu_{\ell_k}$ in 2:1:1 ratio.  For a given choice  of  generic Yukawa coupling including off-diagonal entries like \autoref{Yuk}, the generation splitting of a particular decay modes can  be found  from \autoref{brf}.  One can estimate the total contribution for a given decay modes from the three generations of RHN by summing over the RHN generations, for each generation of SM leptons as $\sum_k \mbox{BR}_{\alpha k} = ({0.238, 1.486, 1.276})\times B(N \to X Y)_{\rm{diag}}$.  The average decay branching fraction for a given mode can be obtained as  $\frac{1}{3}\sum_k \mbox{BR}_{\alpha k} $.


For the study of this scenario, we set up the three benchmark points in \autoref{bps}.  The heavy Higgs boson which is also charged under $U(1)_{B-L}$ can in principle give rise to RHN pair  or $N\nu$ via the two-body decays \cite{PBUBmL,Deppisch:2018eth,Liu:2022ugx}, which are taken to be negligible in our study.
Essential bounds on this model come from the searches of  $Z_{B-L}$  via di-leptonic resonance at LEP and LHC that put strong bound on $M_{Z_{B-L}}, \, g_{BL}$.  $Z_{B-L}$ is searched via $e^\pm e^\mp$ and $\mu^\pm \mu^\mp$ modes at the LEP  \cite{Abdallah:2011ew,Carena:2004xs} and a strong bound was put as $M_{Z_{B-L}}/g_{BL}=Y^\chi_{B-L} v_{BL}>6 $ TeV \cite{Abdallah:2011ew,Carena:2004xs}. Similarly, the most recent data of  ATLAS \cite{ATLAS:2019erb} and CMS  \cite{CMS:2021ctt} LHC at 13 TeV  have put new constraints on cross-section$\times$ branching fractions  in the di-lepton  final states till $M_{Z_{B-L}}=6$ TeV. 
However, such bound can be relaxed in our case due to the reduced  decay  branching fraction of $Z_{B-L}$ to di-lepton, owing to the additional decay modes to heavy neutrinos.

\begin{table}[h]
	\renewcommand{\arraystretch}{1.2}
	\centering
	\begin{tabular}{|c|c|c|c|}
		\hline
		Benchmark points& BP1 & BP2 & BP3 \\
		\hline 
		$M_N$ in GeV & 10 & 60 & 100 \\
		\hline
		$M_{Z_{BL}}$ in TeV & 5.0 & 5.0 & 5.0 \\
		 \hline
	\end{tabular}
	\caption{Choice of benchmark points in terms of RHN and $Z_{B-L}$ mass for the collider study. }  \label{bps}
\end{table}

\begin{figure}[bht]
	\centering
	\mbox{\subfigure[]{\includegraphics[width=0.3\linewidth,angle=-0]{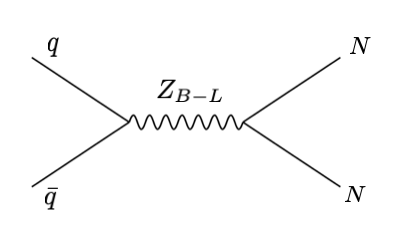}\label{f1}}\quad
		\subfigure[]{\includegraphics[width=0.3\linewidth,angle=-0]{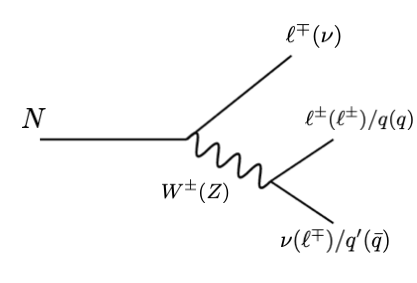}\label{f2}}\quad
	\subfigure[]{\includegraphics[width=0.3\linewidth,angle=-0]{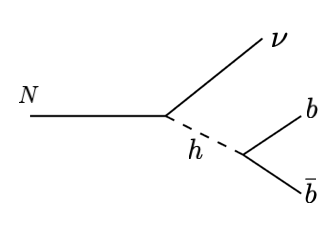}\label{f3}}}
	\caption{Feynman diagrams for generating RHN pairs via $Z_{B-L}$ at the LHC (a) and their further decays (b, c).}\label{Feyn}
\end{figure}
 In our case the right-handed neutrino is charged under $B-L$, which opens up the possibility of  $Z_{B-L}$ decays to $N$ pairs. We see from \autoref{Feyn}(a) that the $Z_{B-L}$ can produce RHN pair, similar to the di-leptonic pair, which further can decay into  $W^{\pm}l^{\mp}$ and $Z\nu$ as shown  in \autoref{Feyn}(b) if RHN is heavier than the $W^\pm, Z$ mass viz., BP3, where RHN mass is 100 GeV.  
 For RHN mass greater than SM Higgs mass the $N \to h \nu$ mode is open as shown in \autoref{Feyn}(c).  The partial decay width for the RHN into on-shell $W^\pm, \, Z, h$ are given by \autoref{pardcy1} - \autoref{pardcy3}.
\begin{eqnarray}\label{pardcy1}
\Gamma(N \to  W^{+}l^{-})= \Gamma(N \to  W^{-}l^{+})&=& \frac{Y_N^2 M_N}{32 \pi} \left( 1- \frac{M_W^2}{M_N^2}\right)^2 \left(1+\frac{2 M_W^2}{M_N^2}\right),\\[5pt] \label{pardcy2}
\Gamma(N \to Z\, \nu)=\Gamma(N \to Z\, \bar{\nu})&=& \frac{Y_N^2 M_N}{64 \pi} \left( 1- \frac{M_Z^2}{M_N^2}\right)^2 \left(1+\frac{2 M_Z^2}{M_N^2}\right),\\[5pt] \label{pardcy3}
\Gamma(N \to \nu \,h)=\Gamma(N \to  \bar{\nu}\,h)&=& \frac{Y_N^2 M_N}{64 \pi} \left( 1- \frac{M_h^2}{M_N^2}\right)^2.
\end{eqnarray}
However, for BP1 and BP2 the RHN masses are 10 GeV and 60 GeV respectively, for which  $W^\pm, \,Z$ both are off-shell, leading to three-body decays of  RHN to light quarks and leptons. The partial three-body decay rates are in this case proportional to the mixing i.e. $Y_N^2$ \cite{N3dcy} as given in  \autoref{pardcy4} - \autoref{pardcy8}. The the corresponding decay branching fractions are incorporated during the  simulation appropriately. 
 \begin{eqnarray}\label{pardcy4}
\Gamma(N\to \ell_1^- \ell_2^+ \nu)=\Gamma(N\to \ell_1^- \ell_2^+ \nu)&=& |Y_N|^2 \frac{G_F^2 M_N^5}{192\pi ^3}, \\[5pt] \label{pardcy5}
\Gamma(N\to \ell^- q_1 \bar{q_2})=\Gamma(N\to \ell^+ \bar{q_1} q_2)&=& |Y_N|^2 \frac{G_F^2 M_N^5}{192\pi ^3}\, N_C\, |K_{q_1q_2}|^2,\\[5pt] \label{pardcy6}
\Gamma(N\to \nu \bar{\ell'} \ell')=\Gamma(N\to \bar{\nu}\ell' \bar{\ell'})&=& |Y_N|^2 \frac{G_F^2 M_N^5}{192\pi ^3}(C_L^2 +C_R ^2),\\ [5pt] \label{pardcy7}
\Gamma(N\to \nu q \bar{q})=\Gamma(N\to \bar{\nu} \bar{q}q)&=& |Y_N|^2 \frac{G_F^2 M_N^5}{192\pi ^3}\,N_C\left[(C_L^q)^2 +(C_R ^q)^2\right], \\[10pt] \label{pardcy8}
\Gamma(N\to \nu_{\ell} \nu_{\ell'} \bar{\nu_{\ell'}}=\Gamma(N\to \bar{\nu_{\ell}} \nu_{l'}\bar{\nu_{\ell'}})&=& |Y_N|^2 \frac{G_F^2 M_N^5}{192\pi ^3}\,C_{\nu}^2,
\end{eqnarray}
where $N_C=3$, number of color degrees of freedom of quark and $K_{q_1, q_2}$ is the CKM matrix element. The various couplings in terms of $C$ are given by
\begin{eqnarray}
C_L&=& -\frac{1}{2}+\sin^2\theta_W,~~~~ C_R= \sin^2 \theta_W,~~~ C_{\nu}=\frac{1}{2},\\
C_L^u&=& \frac{1}{2}-\frac{2}{3}\sin^2 \theta_W, ~~~~C_R^u= -\frac{2}{3} \sin^2 \theta_W,\\
C_L^d&=&-\frac{1}{2} + \frac{1}{3} \sin ^2 \theta_W, ~~C_R^d = \frac{1}{3}\sin^2 \theta_W.
\end{eqnarray}

\begin{figure}[h]
	\centering
	\includegraphics[width=0.6\linewidth,angle=0]{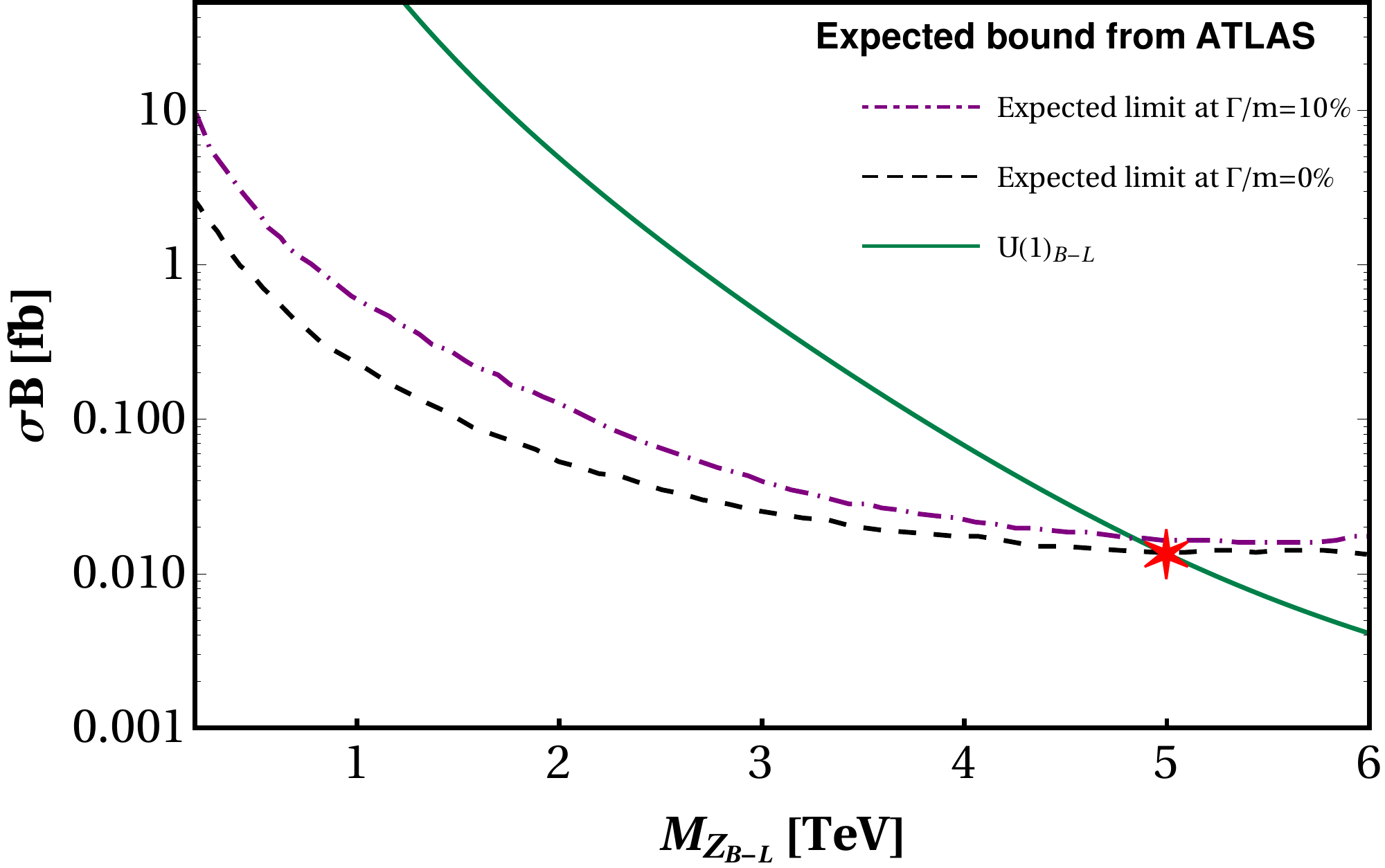}
	\caption{Constraints from $\sigma B$ coming from di-leptonic final state at the LHC with the centre of mass energy of 13 TeV and luminosity of 139 fb$^{-1}$ \cite{ATLAS:2019erb}. The black dashed and purple dot-dashed lines are for expected limits on the cross-section, at the relative width-signals of zero and $10\%$, respectively. The green line represents the theoretical prediction of $U(1)_{B-L}$ model. The red star is the chosen benchmark point for the collider study.}\label{BmLbound}
\end{figure}

We calculate the $\sigma \times \mathcal{B}(Z_{B-L}\to \ell^+ \ell^-)$ for the benchmark points mentioned in \autoref{bps} for 13\,TeV centre of mass energy and presented them against the 13 TeV bounds from ATLAS\cite{ATLAS:2019erb} in \autoref{BmLbound} and it can be seen that our chosen point is allowed by the ATLAS bound.  In this figure the black dashed and purple dot-dashed lines are for expected limits on the cross-section, at the relative width-signals of zero and $10\%$, respectively \cite{ATLAS:2019erb}.
The green line is for the $U(1)_{B-L}$ Model, where the gauge couplings are taken as $g_{BL}=0.3$. It can be seen that the chosen benchmark point is allowed by both the  LEP data \cite{Lepbound} and collider limits from LHC \cite{ATLAS:2019erb, CMS:2021ctt}.

\begin{table}[h]
	\begin{center}	
		\renewcommand{\arraystretch}{1.2}
		\begin{tabular}{ |c|c|c|c|  }
			\hline			
			{\multirow{2}{*}{Benchmark points}}& \multicolumn{3}{c|}{Centre of mass energy}\\
			\cline{2-4}
			& 14\,TeV    & 27\,TeV & 100\,TeV   \\ \hline 
			BP1 & $0.083 $ & $ 1.63 $ &  $ 45.75 $  \\ \hline
			BP2& $ 0.077 $ & $ 1.61 $   & $ 45.66 $  \\ \hline
			BP3 & $ 0.073$ & $ 1.60 $ & $ 45.51 $  \\ 		
			\hline
		\end{tabular}
		\caption{Drell-Yan cross-section in fb for $p\,p \to N\, N$ for the three generations at LHC. We calculate it with three different center of mass energies, $14\,\rm{TeV}$, $27\,\rm{TeV}$ and $100\,\rm{TeV}$ for chosen benchmark points.}  \label{PrCross}
	\end{center}		
\end{table}	


The chosen benchmark points are further looked into for different final state topologies at the LHC with centre of mass energies of 14 TeV, 27 TeV and 100 TeV, respectively. Once the RHNs are produced via $Z_{B-L}$ they would decay to $\ell^\mp W^{(*)\pm}$, $\nu Z^{(*)} $, or $\nu h^{(*)}$ as shown in \autoref{Feyn} (b) and (c). The on-shell gauge bosons can further decay hadronically or they can produce leptons and missing energy. Since we choose BP1 and BP2 for comparatively low mass of RHN, the intermediate gauge bosons are off-shell. We can see the evidences of off-shell decay in the subsequent  sections  as the RHN mass  is  very small around 10 GeV, it decays to $W^{* \pm} \ell^\mp$ and can produce displaced di-leptons and a neutrino. In this case due to extra boost coming from RHNs, the leptons will be very co-linear and form almost a {\tt di-leptonic jet} structure.

Contrary to the only  Type-I seesaw mechanism, where the  pair production of RHN  solely depends on  the mixing angle $\theta^2 \sim (\frac{Y_N v}{\sqrt{2}M_N})^2$, here it can be mediated via the $Z_{B-L}$ gauge boson enhancing the production cross-section.  The production cross-section for three benchmark points with three different centre of mass energies i.e., $14\,\rm{TeV}$, $27\,\rm{TeV}$ and $100\,\rm{TeV}$,  respectively at the LHC /FCC are shown in  \autoref{PrCross}. We generate the model files in { \tt SARAH} \cite{sarah} and they are fed to { \tt CalcHEP} \cite{CalcHEP}, which is used to calculate the cross-section with the {\tt NNPDF23\_LO} as parton distribution function \cite{pdf} using the renormalization/factorization scale of $\sqrt{\hat{s}}$, where $s$ is the partonic level centre of mass energy of the interaction. We can see that the production cross-section is a bit low at the LHC for $14\,\rm{TeV}$ centre of mass energy because of the high mass of $Z_{B-L}$. As we  increase the RHN mass from BP1 to BP3 keeping the $Z_{B-L}$ mass fixed, the cross-section changes infinitesimally, which is not very significant from the collider perspective. It can be noticed that for 100 TeV centre of mass energy the cross-section is enhanced to 45\,fb.

\subsection{Collider simulation and kinematical distributions}\label{cuts}

In this section, we perform the collider simulation at the HL-LHC/HE-LHC/FCC for three different centre of mass energies i.e. 14, 27 and 100 TeV, respectively for  the chosen benchmark points and also discuss their displaced vertex signatures at the CMS, ATLAS, proposed FCC-hh reference detector \cite{FCC:2018vvp, Aleksa:2019pvl,FCChh:talk} and MATHUSLA \cite{Curtin:2018mvb}  in the next sections. For BP1 and BP2  the RHNs decay via three-body decays. On the contrary for BP3, the RHNs decay via two-body decays. We will explore how the boost effect influences the final state topologies of these two distinct types of scenarios. For our study the RHN mass is less than the SM Higgs boson mass  for all BPs and thus it cannot be produced on-shell. However, during the sensitivity reach plots in \autoref{region_SC1} and \autoref{region_SC2}, the appropriate decay branching fractions  are taken into account. 

For the collider simulation we generate the events (.lhe files) via {\tt CalcHEP} and simulate those via {\tt PYTHIA8} \cite{Pythia8.2} with initial state and final state radiations. Hadronization and its decays are also done by {\tt PYTHIA8}, whereas  for the jet formation we use { \tt Fastjet-3.0.3} \cite{FastJET}  with the {\tt ANTI-KT} algorithm with the jet cone size $R=0.5$.  Initial state radiation (ISR) and final state radiation (FSR) are switched on during the collider simulation. The other basic cuts are as follows:

\begin{itemize}
	\item The pseudorapidity for the calorimeter coverage is $|\eta|< 4.5\,(6)$ for 14, 27\,(100) TeV colliders.
	
	\item The minimum transverse momentum for jets is $p_{T,\text{min}}^{\text{jet}}= 20\,(25)\, \text{GeV}$ for 14, 27\,(100) TeV colliders.
	
	\item We choose the leptons with $p_{T}>10\,(3)\,\text{GeV}$ and $|\eta_{\text{max}}|=3.8\,(4.8)$ for 14, 27\,(100) TeV colliders.
	
	\item For a selection of clean leptons, we put a cut on the total transverse momentum of the hadrons within the cone $\Delta R =0.3$, demanding it to be $\leq 0.15\; p_{T}^{l}$. Here $p_{T}^{l}$ is the transverse momentum for the leptons within that specified cone.
	
	\item Additionally, the leptons are isolated from the clustered jets with $\Delta R_{lj}\geq 0.1 $.
\end{itemize}

\begin{figure}[hbt]
	\begin{center}
		\mbox{\subfigure[]{\includegraphics[width=0.45\linewidth,angle=0]{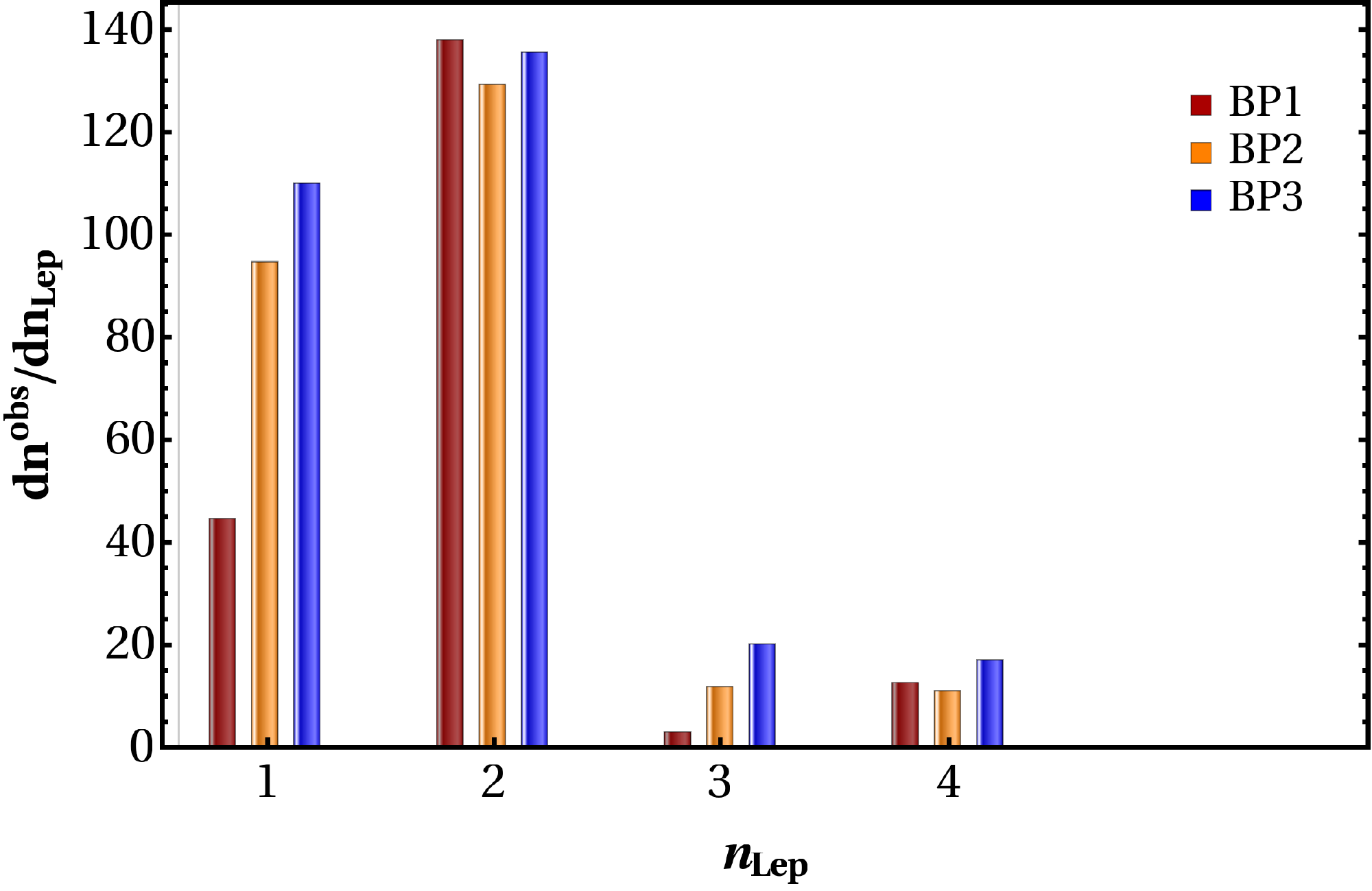}\label{14dcyl}}\quad \quad
			\subfigure[]{\includegraphics[width=0.45\linewidth,angle=0]{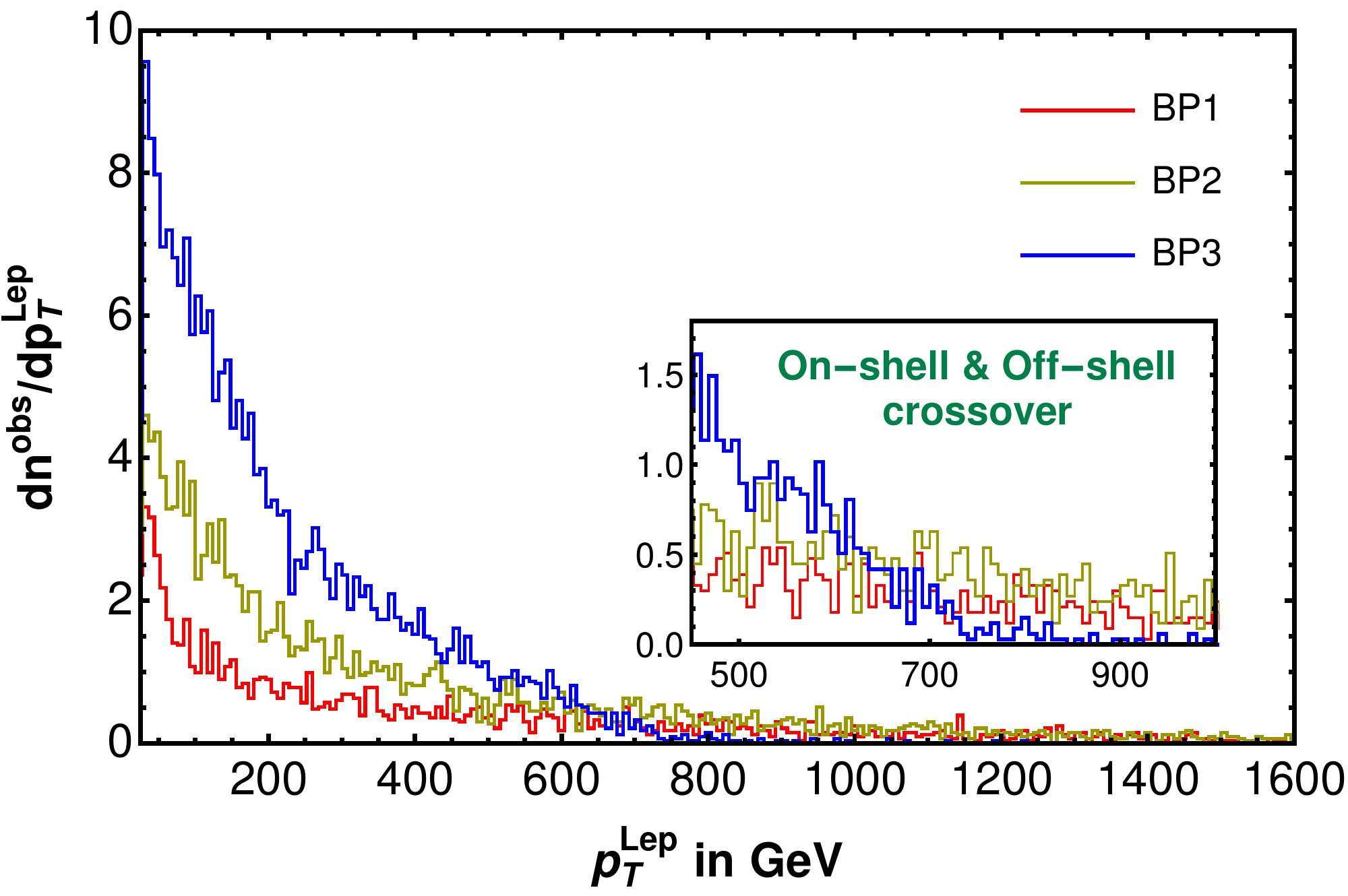}\label{30dcyl}}}			        		
		\caption{The lepton $(e^{\pm}, \mu ^{\pm})$ multiplicity distributions, $n_{\rm Lep}$ (a), and the lepton $p_{T}$ distributions, $p_T^{\rm Lep}$ for the first lepton only (b), for the benchmark points at the centre of mass energy of 14 TeV with the integrated luminosity of 3\,ab$^{-1}$.}\label{Lepdist}
	\end{center}
\end{figure}

The isolated charged leptons $(e^{\pm} ~ \text{and}~\mu^{\pm})$ are tagged as the final state leptons. Those can come directly from the right-handed neutrinos or from the decay of intermediate gauge bosons $W^{\pm},\, Z$. Below we describe  the kinematical distributions at the LHC only with centre of mass energy of 14 TeV for the benchmark points. Here the kinematical distributions show the normalised event numbers for a given integrated luminosity as $ n^{\rm obs}= \frac{1}{N}\times n^{\rm sim} \times  \sigma \times \int \mathcal{L} dt. $

\autoref{Lepdist}(a)  describes the lepton multiplicity distribution for the benchmark points. For the benchmark points the lepton at higher multiplicity are significantly low in numbers due to the low effective branching and also due to the isolation cuts as mentioned above. The multiplicity distributions peak around  $n_{Lep}=2$ for all three cases, as the $W^\pm$ bosons mostly  decay hadronically.  In \autoref{Lepdist}(b) we present the transverse momentum distributions of the first lepton i.e., $p^\ell_T$ for the benchmark points. In case of BP3, the RHN ($N$) has a on-shell decay to either $\ell^\pm W^\mp$ or $Z\nu$. Thus, the leptons come from the on-shell decays of these gauge bosons have shorter tail in $p^\ell_T$ compared to the ones coming directly from RHNs in case of BP1 and BP2 due to the direct boost effect from the RHN.  For BP1 and BP2, the RHNs go through off-shell three-body decays, where we cannot distinguish the two leptons coming from a single RHN leg and thus share the momentum equally  leading to a larger tail for the higher momentum as can be seen from the inset of \autoref{Lepdist}(b). Thus a cross-over around $p^\ell_T \sim 700$ GeV among BP3 and BP1 or BP2 can be noticed.

\begin{figure}[hbt]
	\begin{center}
		\mbox{\subfigure[]{\includegraphics[width=0.465\linewidth,angle=0]{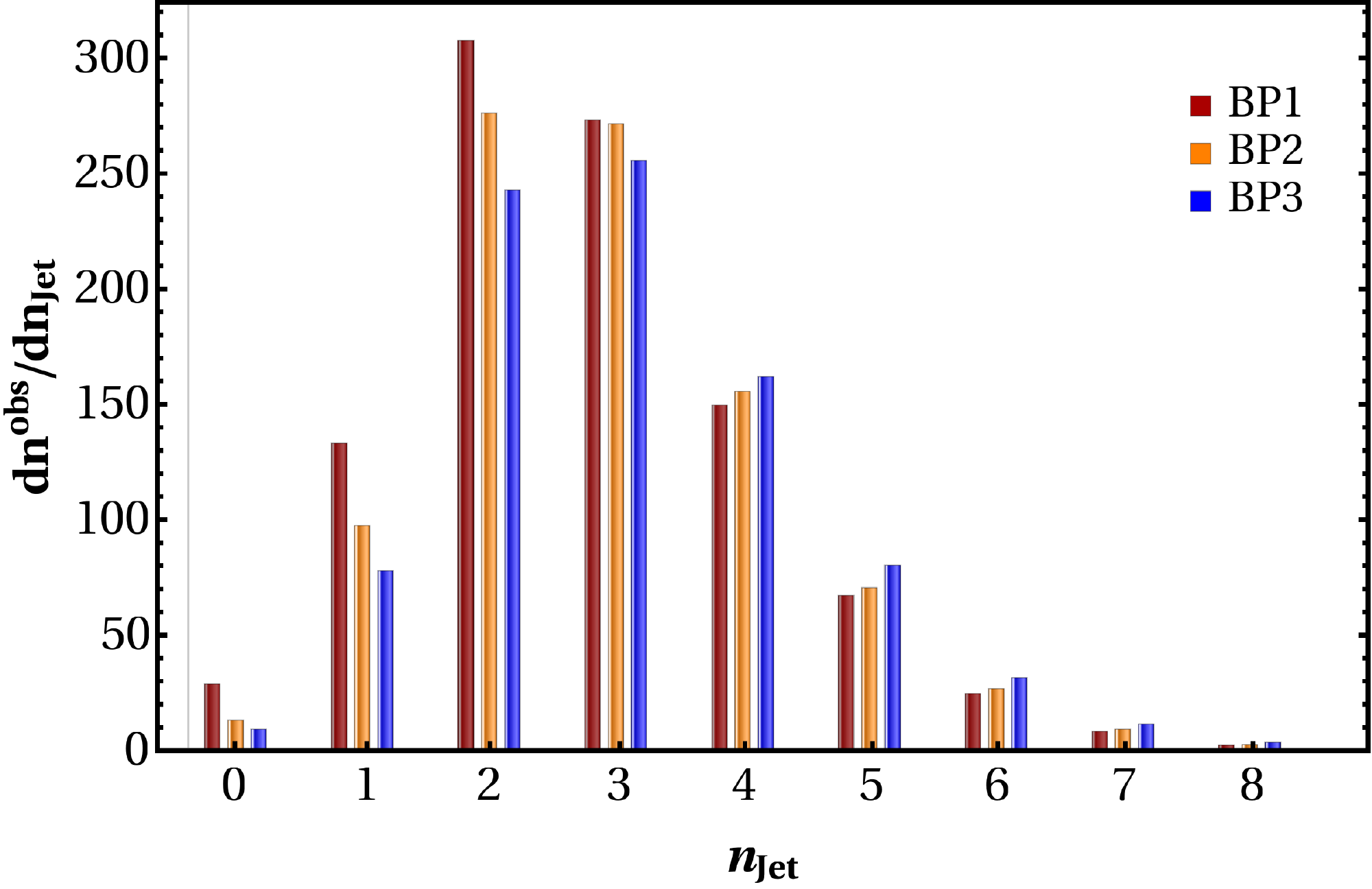}\label{14dcyl}}\quad \quad
			\subfigure[]{\includegraphics[width=0.45\linewidth,angle=0]{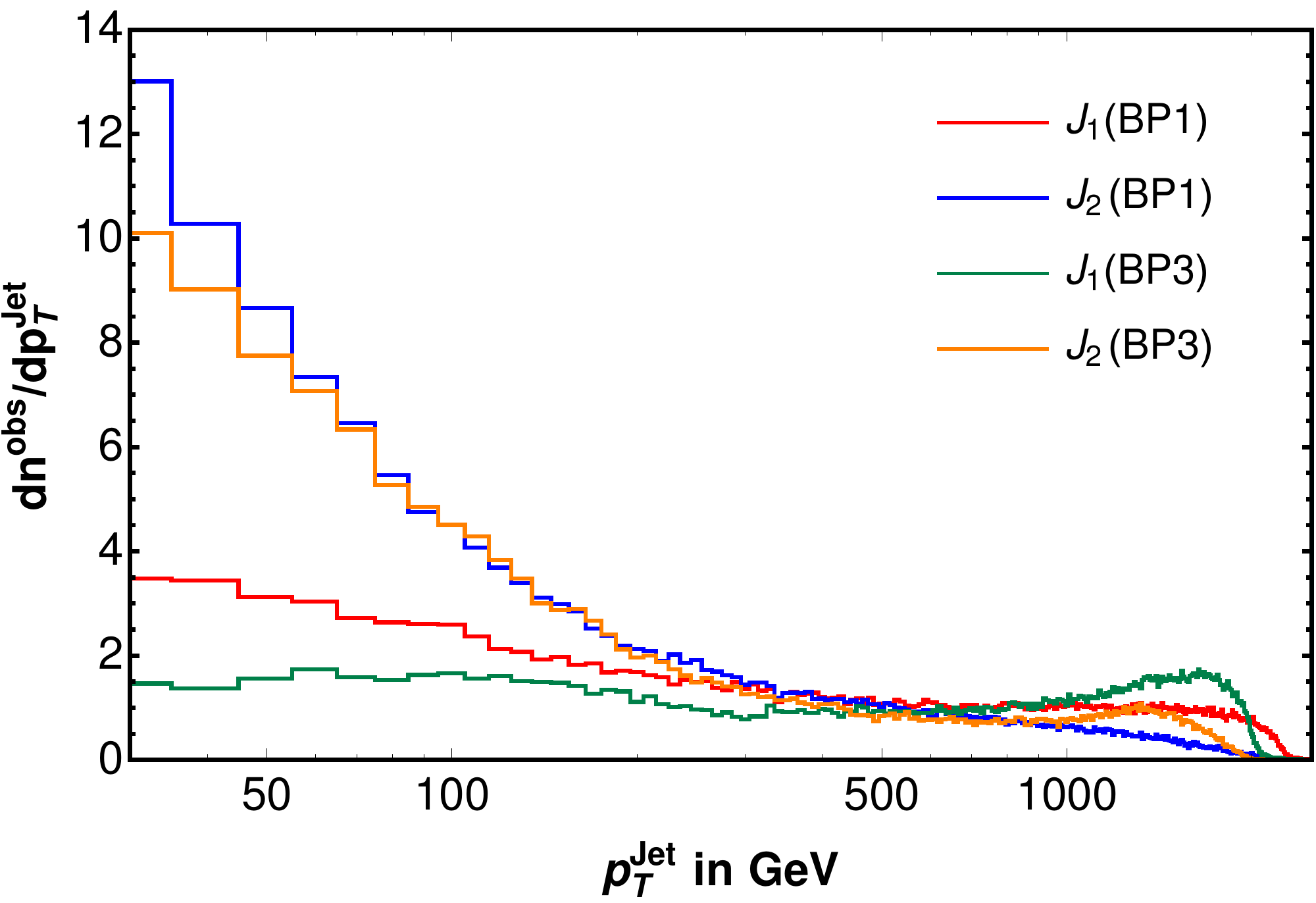}\label{30dcyl}}}			        		
		\caption{The Jet multiplicity distribution ($n_{\rm jet}$) for three benchmark points (a), and the $p_{T}$ distribution ($p_T^{\rm jet}$) of the $1^{st}$ and $2^{nd}$ Jets with BP1 and BP3 (b) are shown at the centre of mass energy of 14 TeV with the integrated luminosity of 3\,ab$^{-1}$.}\label{Jetdist}
	\end{center}
\end{figure}

We depict the jet multiplicity distribution for the benchmark points in \autoref{Jetdist}(a). We see that for all three BPs the multiplicity distributions peaks around two or three, which is expected when only one of the gauge boson (either $W^\pm$ or $Z$) decays hadronically along with a ISR/FSR jet or both the gauge bosons decay hadronically but one jet is missed or two jets are merged as one. The maximum partonic jet multiplicity should be around four; however, due to ISR/FSR the jet multiplicity can go up to eight.  In \autoref{Jetdist}(b) we describe the first and second $p_T$ ordered jets for BP1 and BP3, respectively. A cross over of $p^{\rm jet}_T$ like lepton $p_T$ around $700$ GeV is also noticed.

\begin{figure}[hbt]
	\begin{center}
		\hspace*{-1.0cm}
		\mbox{\subfigure[]{\includegraphics[width=0.35\linewidth,angle=0]{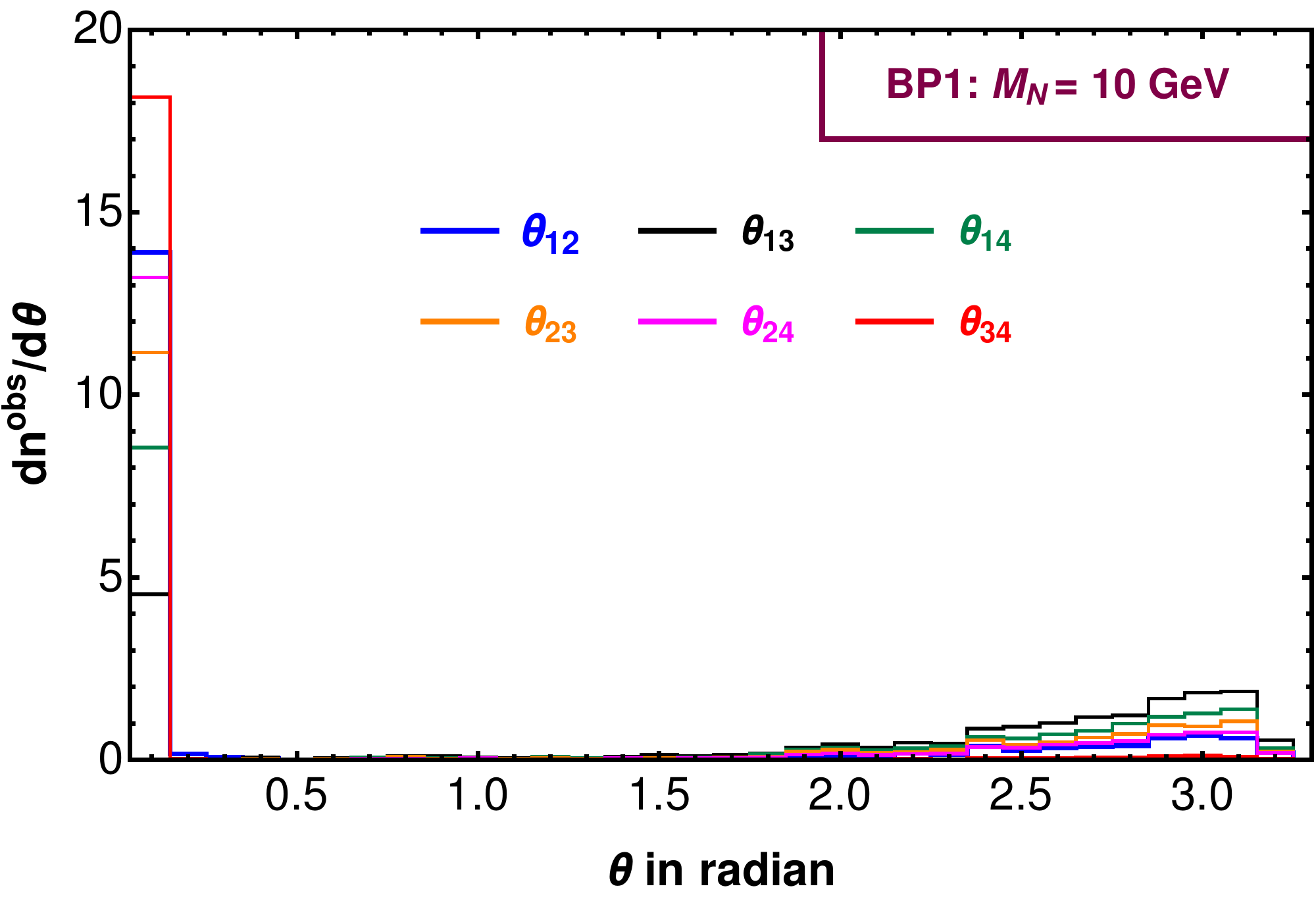}\label{14dcyl}}\quad
			\subfigure[]{\includegraphics[width=0.35\linewidth,angle=0]{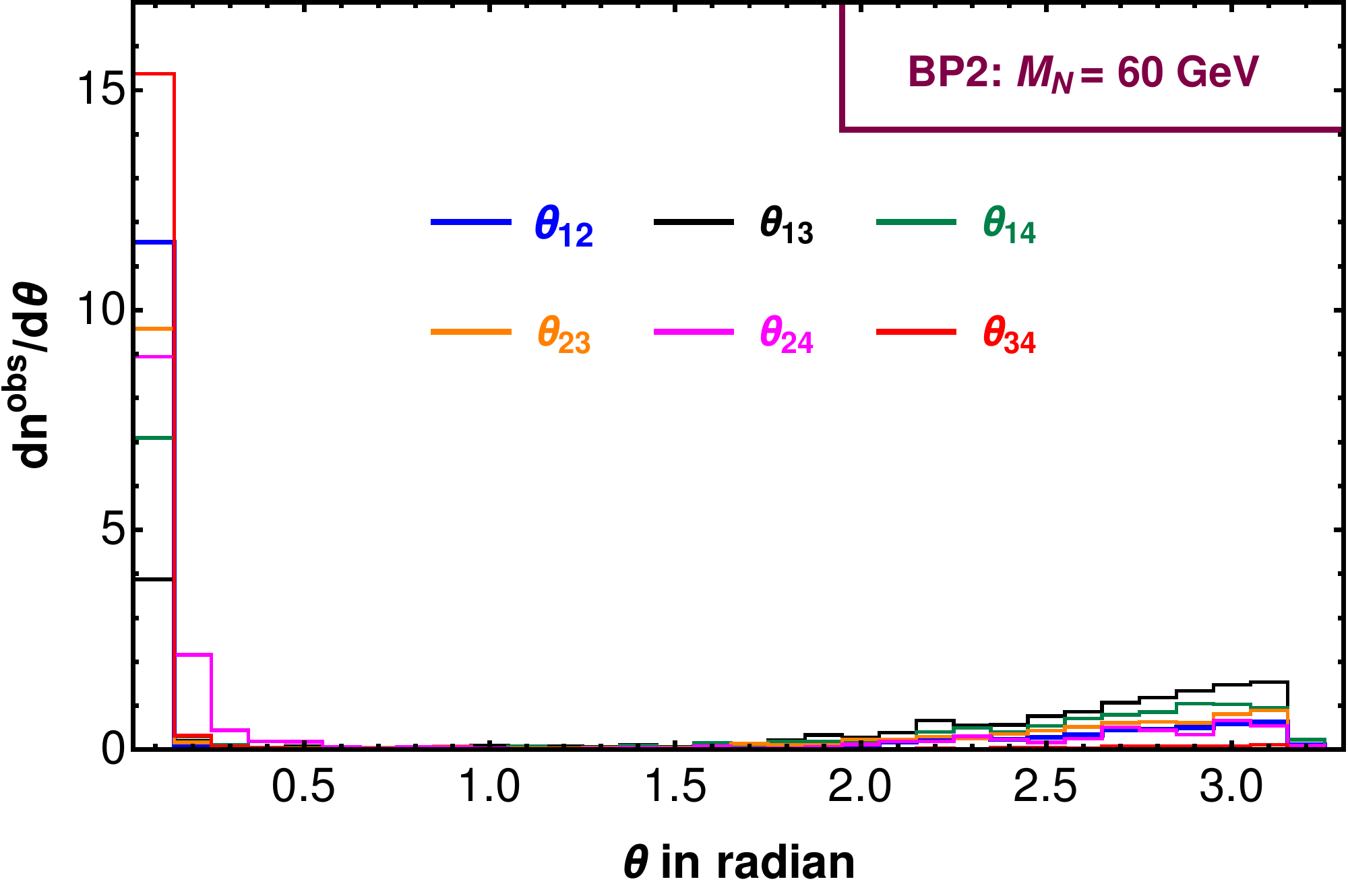}\label{30dcyl}}\quad
			\subfigure[]{\includegraphics[width=0.35\linewidth,angle=0]{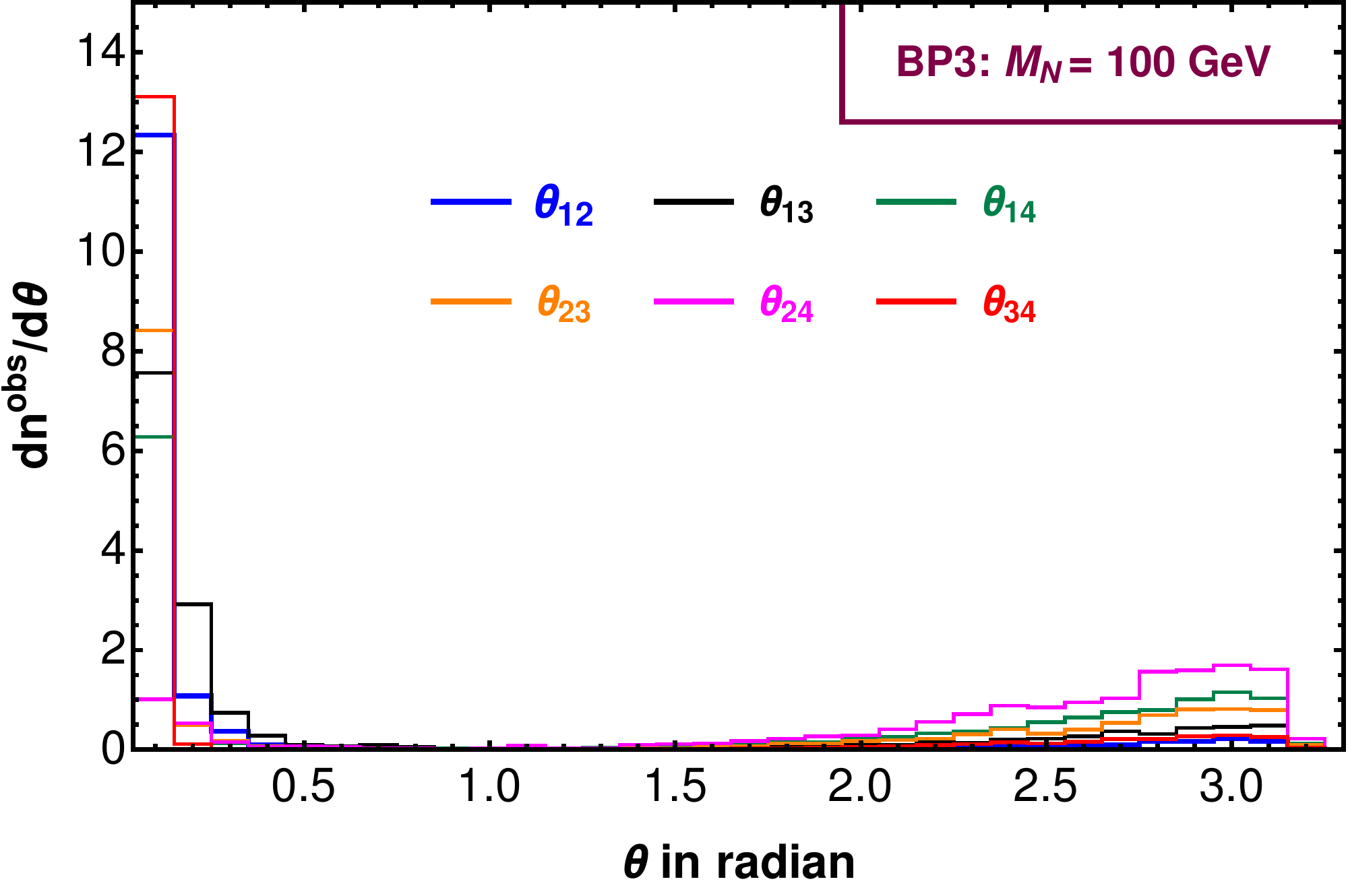}\label{30dcyl}}}        		
		\caption{The angular distribution ($\theta$) of the angles between two leptons at the LHC with the centre of mass energy of 14 TeV and the integrated luminosity of 3\,ab$^{-1}$ for the benchmark points.}\label{opangll}
	\end{center}
\end{figure}
After dealing with the leptons and jet kinematic distributions, we focus on the angles between the leptons ($\theta_{ij}$) which are already isolated from the jets. The leptons coming from the decays of RHNs can be very co-linear with the leptons that is coming  from on- or off-shell $W^\pm$ of the same RHN leg, due to the boost effect. This can form a scenario where two extremely co-liner charged leptons can come as {\tt leptonic jet}. In \autoref{opangll} we show how the opening angles among the leptons are distributed for three different benchmark points. It is evident to see that for $\theta_{12}$ (in blue) and $\theta_{34}$ (in red) peaks around zero, which points out the fact that $\ell_1, \ell_2$ are coming from same RHN leg and $\ell_3, \ell_4$ are coming from different RHN leg. Thus, the distribution helps us to figure out the correct leg which forms the {\tt leptonic jet}. However, from  \autoref{opangll}(a) one can see that such effects are more prominent for the case $M_N=10$ GeV (BP1) and the effect gets reduced as we increase the RHN mass as can be seen from \autoref{opangll}(c) for $M_N=100$ GeV (BP3).

\subsection{Reconstruction of RHN mass}\label{RHN_mass}
\begin{figure}[hbt]
	\begin{center}
		\hspace*{-0.5cm}
		\mbox{\subfigure[]{\includegraphics[width=0.45\linewidth,angle=0]{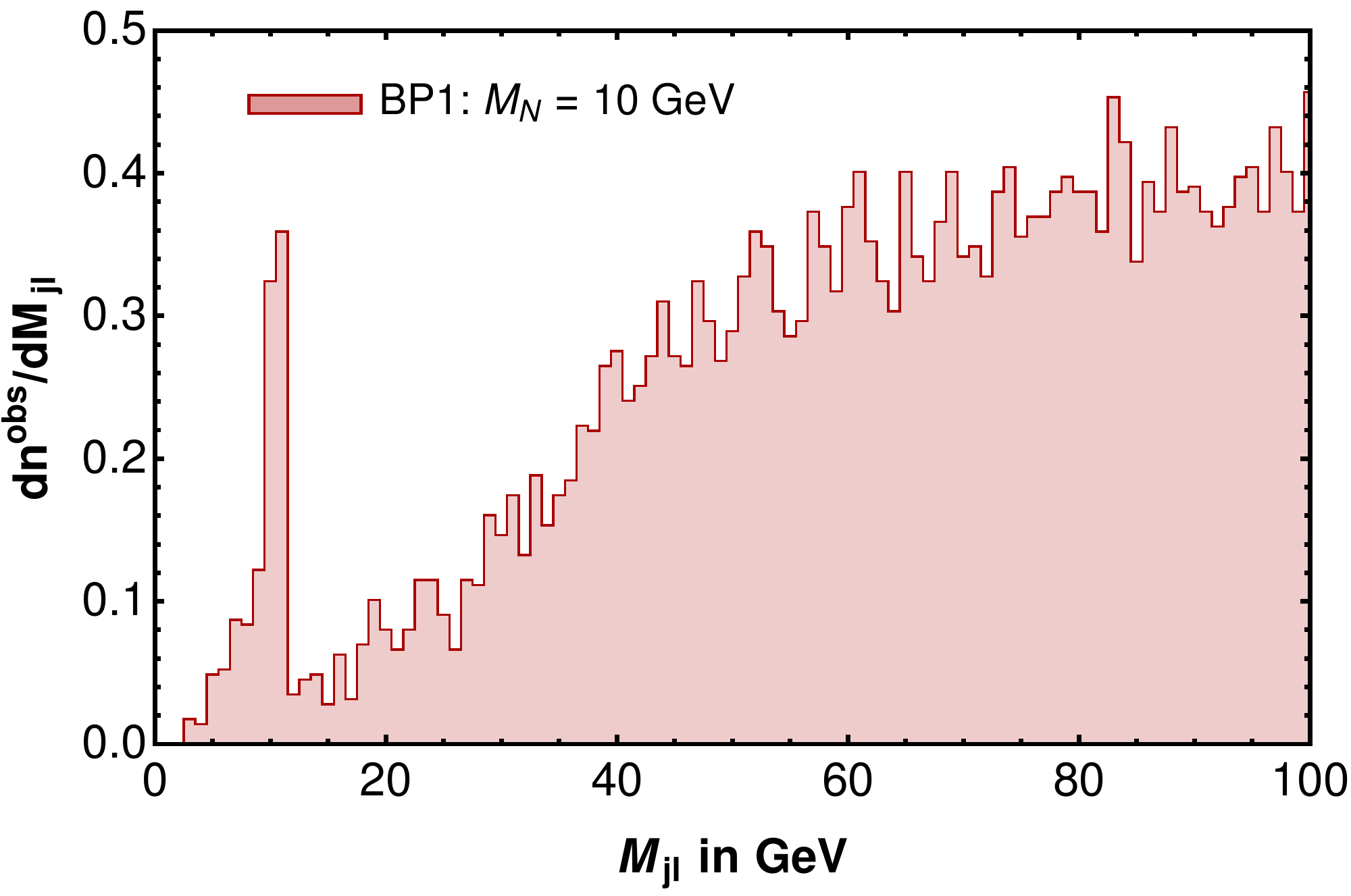}\label{14dcyl}}\quad \quad
		\subfigure[]{\includegraphics[width=0.45\linewidth,angle=0]{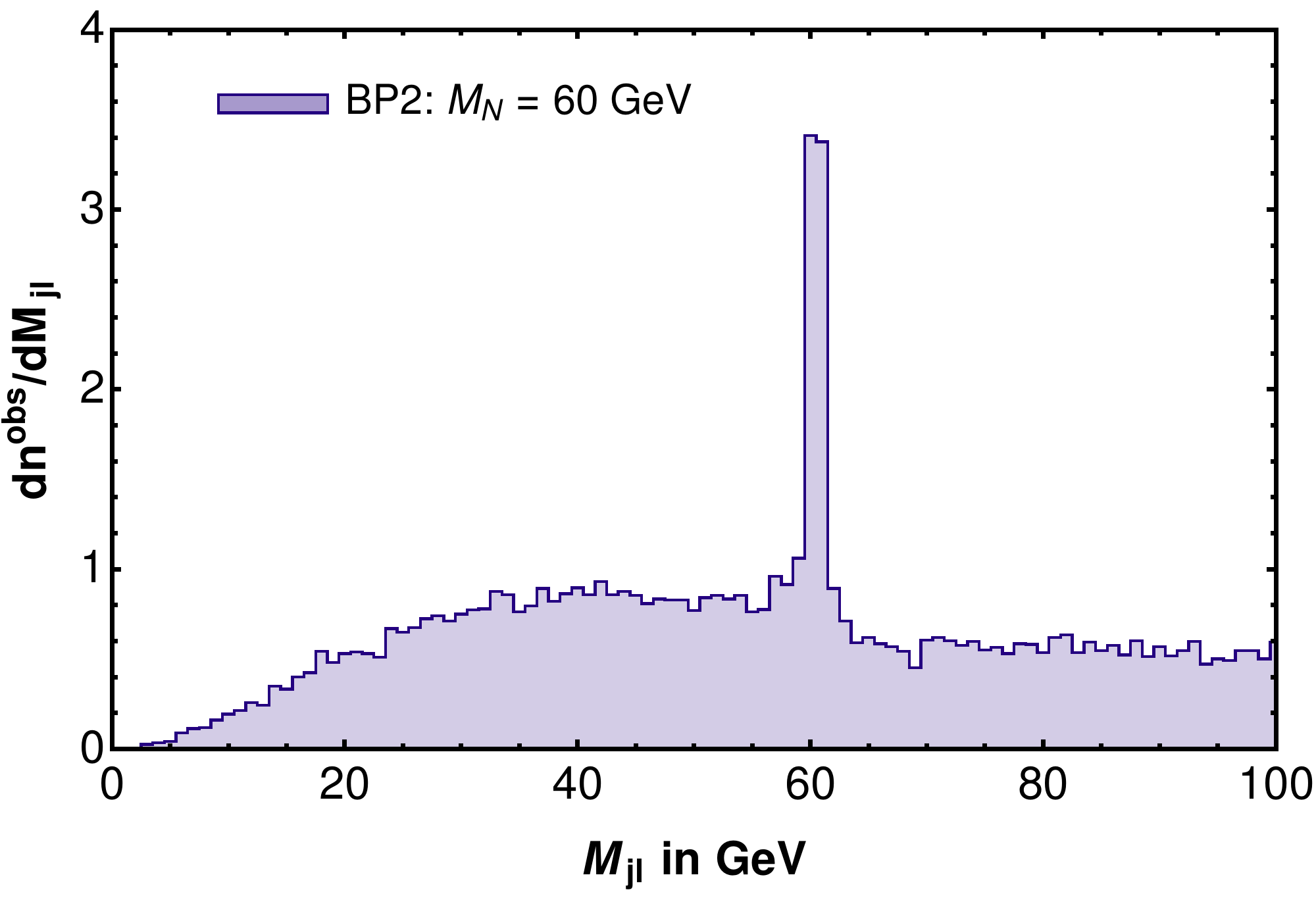}\label{30dcyl}}}
		\hspace*{-0.5cm}
		\mbox{\subfigure[]{\includegraphics[width=0.45\linewidth,angle=0]{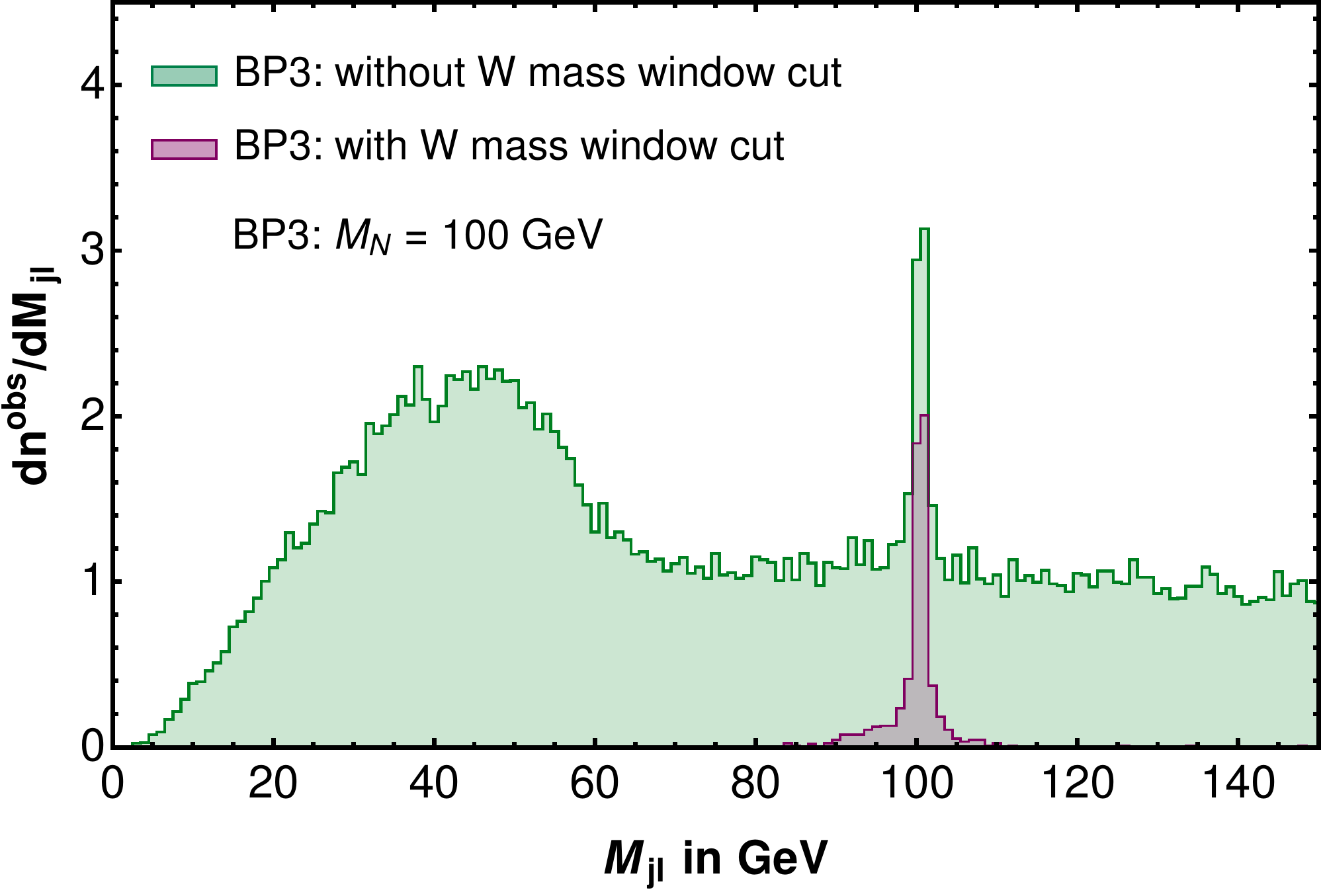}\label{100dcyl}}\quad \quad
		\subfigure[]{\includegraphics[width=0.45\linewidth,angle=0]{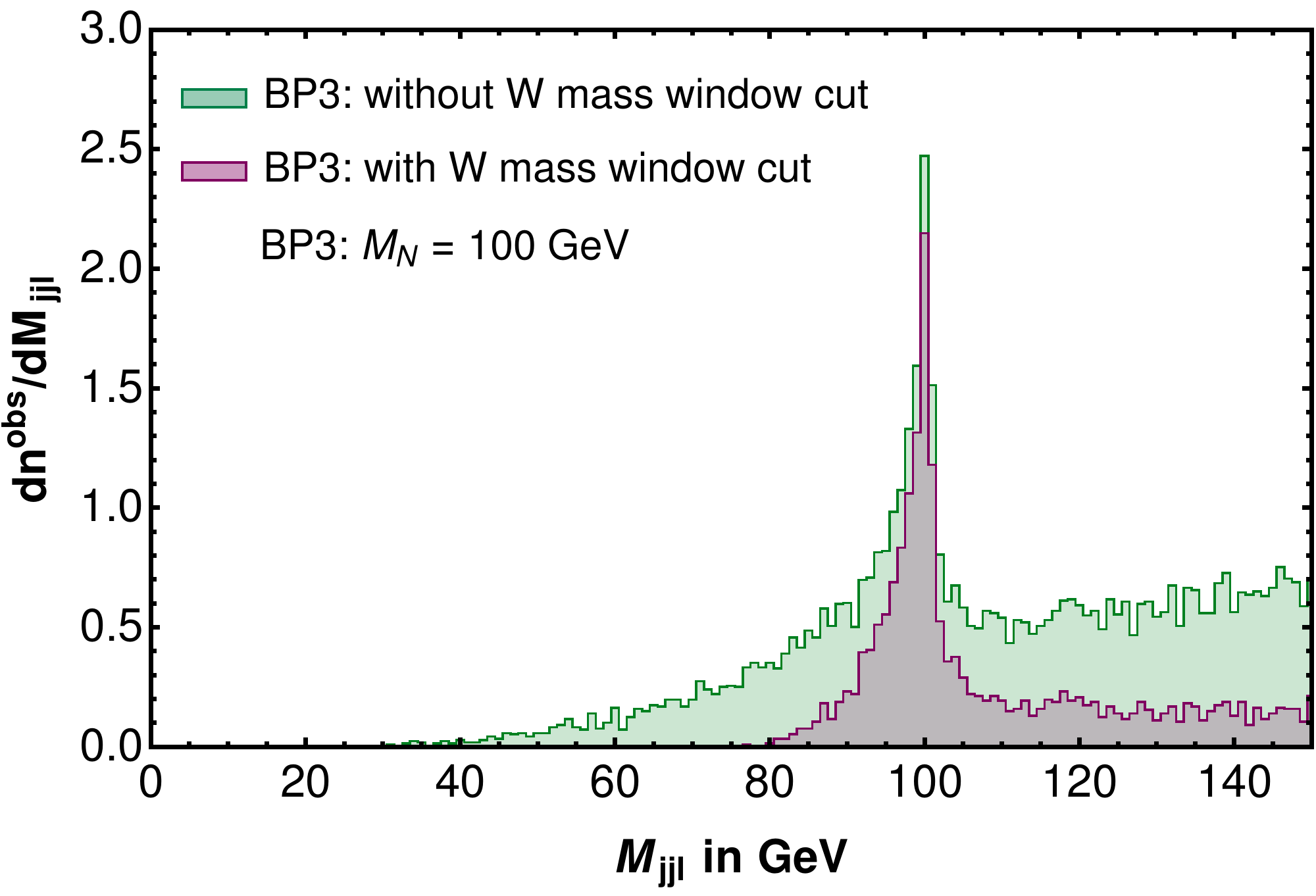}\label{30dcyl}}}
		\caption{Reconstruction of RHN mass from heavily boosted Fatjet and mono-lepton ($M_{jl}$) for BP1, BP2, BP3 (a, b, c) and mono-lepton-di-jet invariant mass for BP3 (d) at the LHC with the centre of mass energy of 14\,TeV and the integrated luminosity of 3\,ab$^{-1}$. The purple colour represents the distribution after applying a window cut of 10\,GeV around $W^{\pm}$ mass, while the green represents the distribution of $M_{jl}$ and $M_{jjl}$ for BP3 in (c) and (d), respectively.    }\label{InvM}
	\end{center}
\end{figure}

One of the parameters which can lead to the discovery of RHN is the invariant mass distribution. In \autoref{InvM}, we summarize the invariant mass distributions towards the reconstructions of the $W^\pm$ boson for the benchmark points. For BP1 and BP2 the RHN mass is 10, 60 GeV, respectively, which are less than the $W^\pm$, making it off-shell. \autoref{InvM}(a) depicts the scenario of BP1 ($M_N=10$ GeV), where the two-jets coming from an off-shell $W^\pm$ boson are collimated forming a single jet, known as Fatjet \cite{Bandyopadhyay:2010ms,Bhardwaj:2018lma,Chakraborty:2018khw,Ashanujjaman:2022cso,Ashanujjaman:2021zrh}. Thus, an invariant mass distribution of $M_{j\ell}$ shows the peak around 10 GeV in \autoref{InvM}(a), which designates the RHN mass for the BP1.
Similar situation can also be realized for BP2, where $M_{j\ell}$ peaks around 60 GeV as can be seen in \autoref{InvM}(b). In \autoref{RHNTab} the reconstructions are demonstrated via the number of events in $M_{j\ell}$ distributions after the window cut around the mass peak for BP1, BP2 with the centre of mass energies of 14,  27, 100\,TeV at the integrated luminosities of ($\mathcal{L}_{\rm int}=$) 3, 10 and 30 ab$^{-1}$, respectively considering the goal of the different colliders, HL-LHC, HE-LHC and FCC-hh \cite{FCChh:talk}. For BP2 and with centre of mass energies of 27, 100  TeV look promising.

\begin{table*}[h]	
	\begin{center}
		\hspace*{-1cm}
		\renewcommand{\arraystretch}{1.2}
		\begin{tabular}{|c|c|c|c|c|}
			\hline
			Benchmark&\multirow{2}{*}{$M_{j\ell}$} & \multicolumn{3}{c|}{Centre of mass energy}\\
			\cline{3-5}
			Points & &
			14\,TeV  & 27\,TeV  & 100\,TeV \\
			\hline
			\multirow{2}{*}{BP1} &\multirow{2}{*}{$|M_{j\ell}-10.0|\leq 5$\,GeV  }& \multirow{2}{*}{0.8} & \multirow{2}{*}{13.1} & \multirow{2}{*}{611.3} \\	
			& & & & \\
			\hline
			\multirow{2}{*}{BP2} & \multirow{2}{*}{$|M_{j\ell}-60.0|\leq 5$\,GeV } & \multirow{2}{*}{8.7} & \multirow{2}{*}{388.4} & \multirow{2}{*}{36911.7} \\
			& & & & \\
			\hline
		\end{tabular}
		\caption{The number of events in $M_{j\ell}$ distributions after the window cut around the mass peak for BP1, BP2 with the centre of mass energies of 14\,TeV, 27\,TeV, 100\,TeV at the integrated luminosities of ($\mathcal{L}_{\rm int}=$) 3, 10 and 30 ab$^{-1}$, respectively. } \label{RHNTab}
	\end{center}	
\end{table*}


\begin{table*}[h]	
	\begin{center}
		\hspace*{-1cm}
		\renewcommand{\arraystretch}{1.2}
		\begin{tabular}{|c|c|c|c|c|}
			\hline
			Benchmark&\multirow{2}{*}{$M_{j\ell}/ M_{jj\ell}$} & \multicolumn{3}{c|}{Centre of mass energy}\\
			\cline{3-5}
			Point & &
			14\,TeV  & 27\,TeV  & 100\,TeV \\
			\hline
			\multirow{4}{*}{BP3} &$|M_{j}-80.0|\leq 10$\,GeV\, \&  & \multirow{2}{*}{5.0} & \multirow{2}{*}{175.2} & \multirow{2}{*}{16401.7} \\
			&  $|M_{j\ell}-100.0|\leq 10$\,GeV &  &  &  \\	
			\cline{2-5}
			& $|M_{jj}-80.0|\leq 10$\,GeV\, \&& \multirow{2}{*}{4.5} & \multirow{2}{*}{156.5} & \multirow{2}{*}{10674.3} \\
			&  $|M_{jj\ell}-100.0|\leq 10$\,GeV&  &  &  \\
			\hline
		\end{tabular}
		\caption{The number of events in $M_{j\ell}$ and $M_{jj\ell}$ distributions after the window cut around the mass peak for BP3 with the centre of mass energies of 14\,TeV, 27\,TeV, 100\,TeV at the integrated luminosities of ($\mathcal{L}_{\rm int}=$) 3, 10 and 30 ab$^{-1}$, respectively. An additional window cut of 10\,GeV around the $W^{\pm}$ mass peak is also implemented while reconstruction of RHN mass. }  \label{RHNTabBP3}
	\end{center}	
\end{table*}


However, the situation changes a lot, when $W^\pm$ is produced on-shell from the RHN decay in the case of BP3. Two quarks coming from the $W^\pm$, can be collimated and form a single jet or they can also form two different jets. As the previous case the analysis would be similar and they are presented in \autoref{InvM}(c). Contrary to the previous two cases (for BP1 and BP2), here a single jet mass can be around the on-shell $W^\pm$ mass, and a reconstruction of RHN mass with the jet coming from the $\pm 10$ GeV of the $W^\pm$ boson mass, is presented by the purple graph. Whereas, the green curve shows such reconstruction without any demand of $W^\pm$ mass reconstruction. Though the number of events in the previous case is little lower than the later one, the peak is sharper in the previous case and we consider that for final state number presented latter in the text. The corresponding events number for $M_{j\ell}$ for $2\ell +2j$ final state are given in \autoref{RHNTabBP3} after the window cut around the mass peak for BP3 with the centre of mass energies of 14\,TeV, 27\,TeV, 100\,TeV at the integrated luminosities of ($\mathcal{L}_{\rm int}=$) 3, 10 and 30 ab$^{-1}$, respectively. 
	
The other possibility, where the two quarks are not boosted enough and form two separate jets, can establish $W^\pm$ mass peak via $M_{jj}$, which can be seen in \autoref{InvM}(d). Following similar approaches of with and without the $W^\pm$ mass window cuts, one can reconstruct the RHN mass around 100 GeV, for the BP3. The events corresponding to  $M_{jj\ell}$ distributions  for $2\ell +4j$ are given in \autoref{RHNTabBP3} after the window cut around the mass peak for BP3 with the centre of mass energies of 14\,TeV, 27\,TeV, 100\,TeV at the integrated luminosities of ($\mathcal{L}_{\rm int}=$) 3, 10 and 30 ab$^{-1}$, respectively.


\subsection{Displaced vertex signatures}\label{DisVartex_SC1}

A particle, in this case the RHN, can have displaced decay with rest mass decay length  $L_0=c \tau_0=\frac{c}{\Gamma_0}$, where $\Gamma_0$ is the rest decay width. Generic decays follow the exponential distribution as given below
\be
N(\tau) \propto \exp(-\tau/\tau_0),
\ee
where $N$ is number corresponding to the actual decay life time $\tau$, and $\tau_0$ is the rest mass decay time. On top of that, the boost effect can enhance such decay lengths further, i.e. the resultant decay length is given by \cite{Bierlich:2022pfr}
\be
L_\tau=c\tau \beta \gamma= \tau \frac{p}{m},
\ee
where $p$ is three momentum of the particle and $m$ is the rest mass. The momentum measured in the transverse and longitudinal directions can lead to displaced decay lengths of $L_\perp, \, L_{||}$ as given by 
\be
L_\perp=\tau \frac{p_\perp}{m}, \quad L_{||}=\tau \frac{p_{||}}{m}.
\ee

\begin{figure*}[hbt]
	\begin{center}
		\hspace*{-1.0cm}
		\mbox{\subfigure[]{\includegraphics[width=0.35\linewidth,angle=-0]{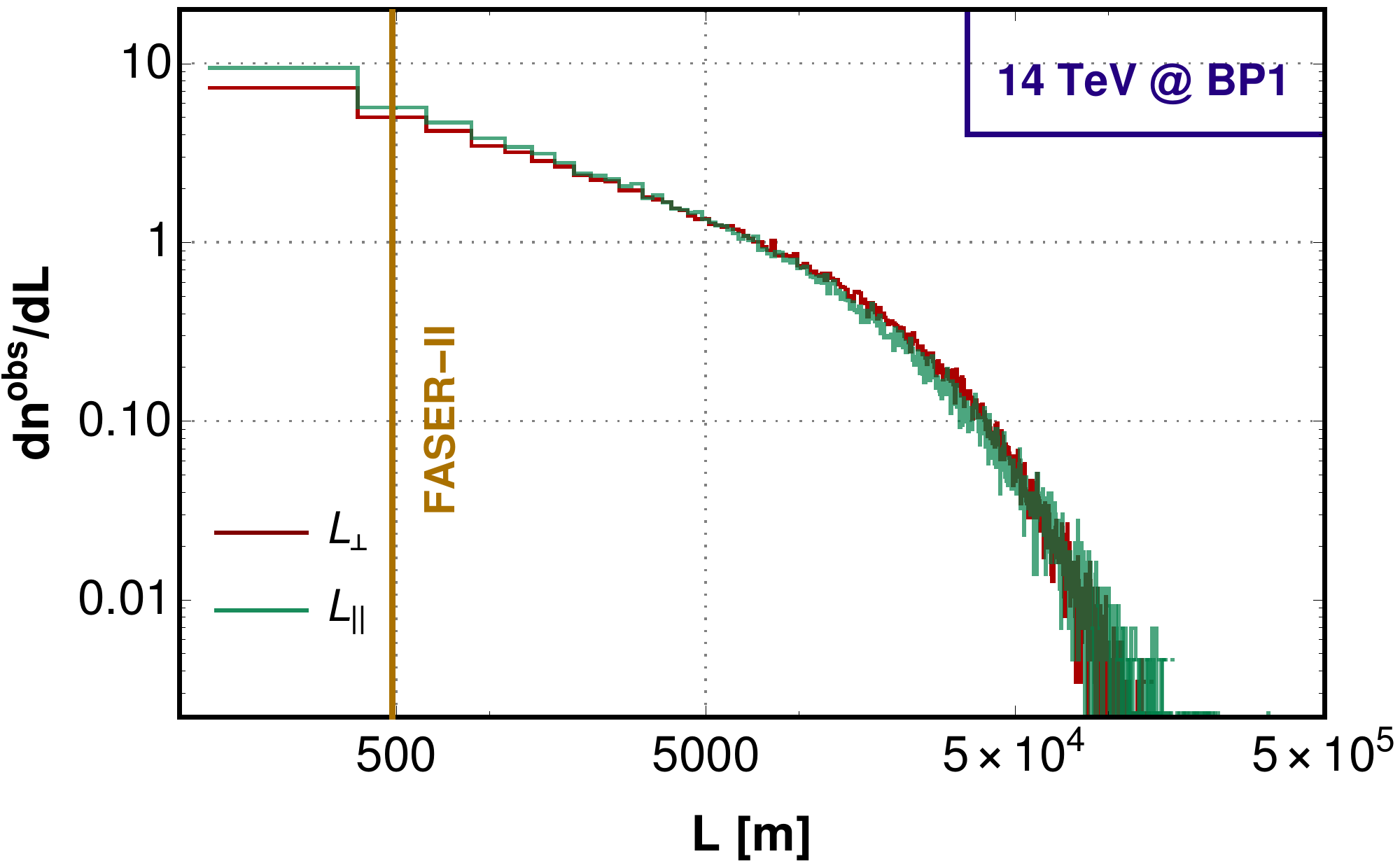}\label{}}\quad
			\subfigure[]{\includegraphics[width=0.35\linewidth,angle=-0]{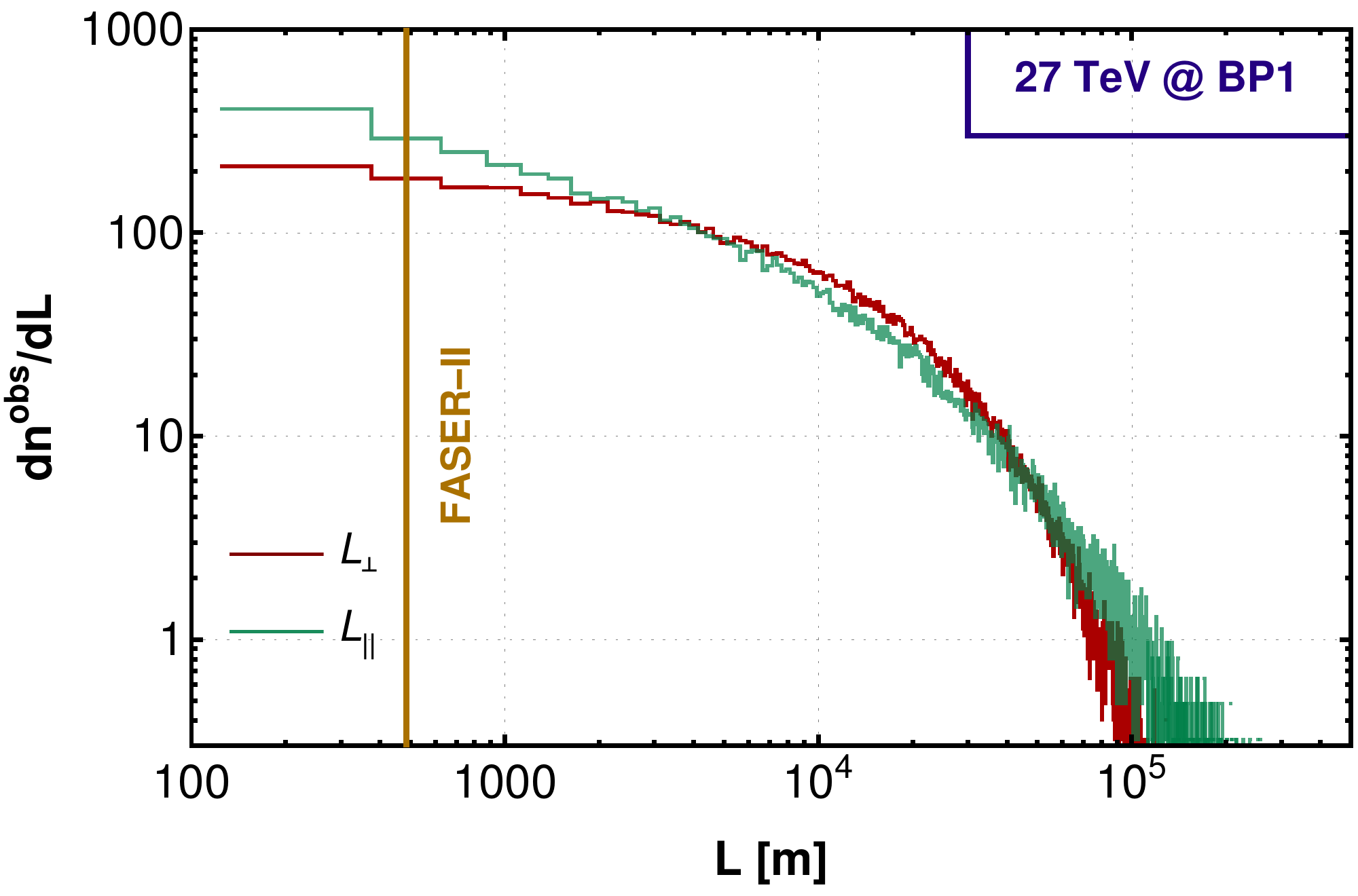}\label{}}\quad
			\subfigure[]{\includegraphics[width=0.35\linewidth,angle=-0]{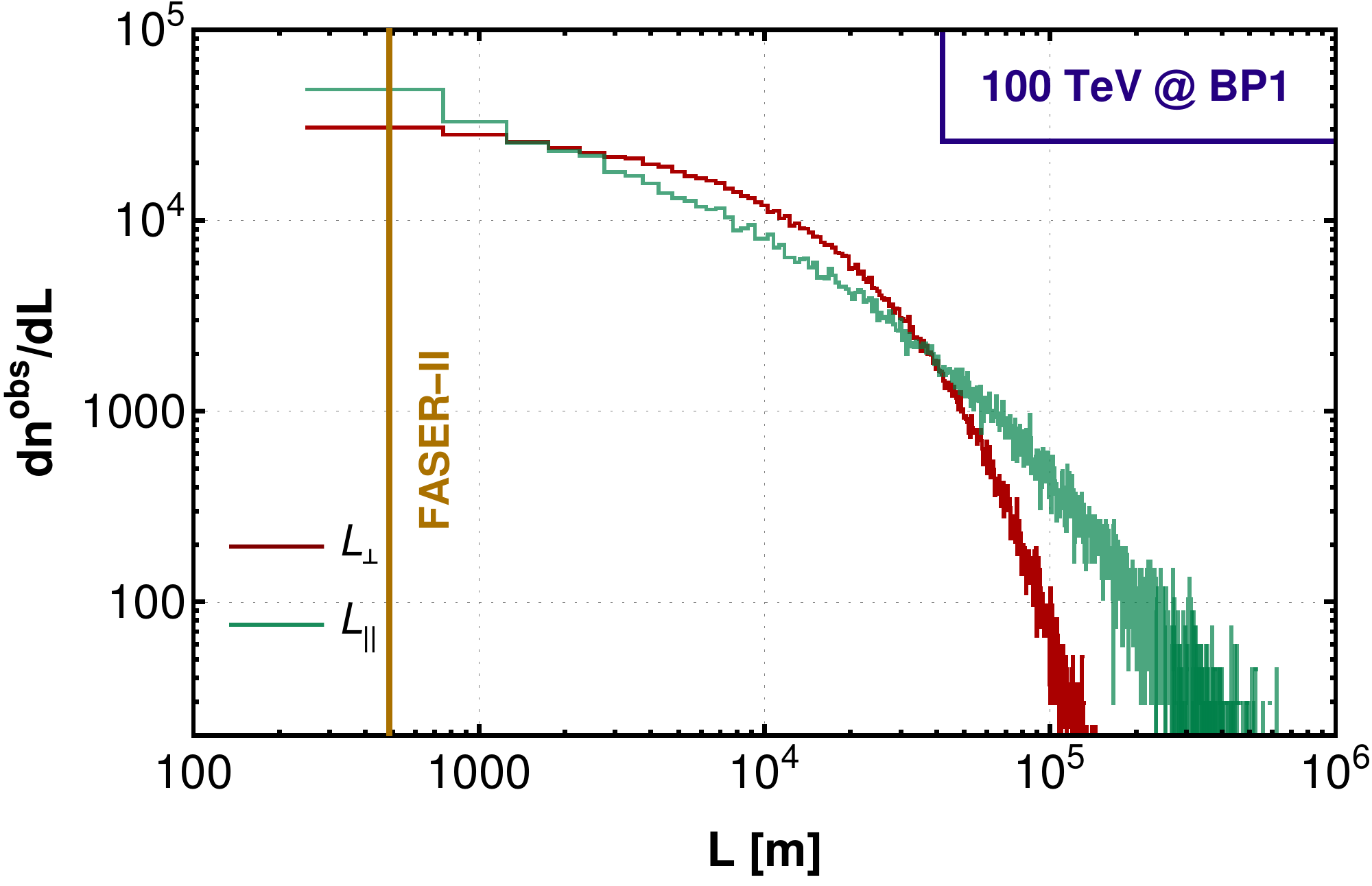}\label{}}}
		\hspace*{-1.0cm}
		\mbox{\subfigure[]{\includegraphics[width=0.35\linewidth,angle=-0]{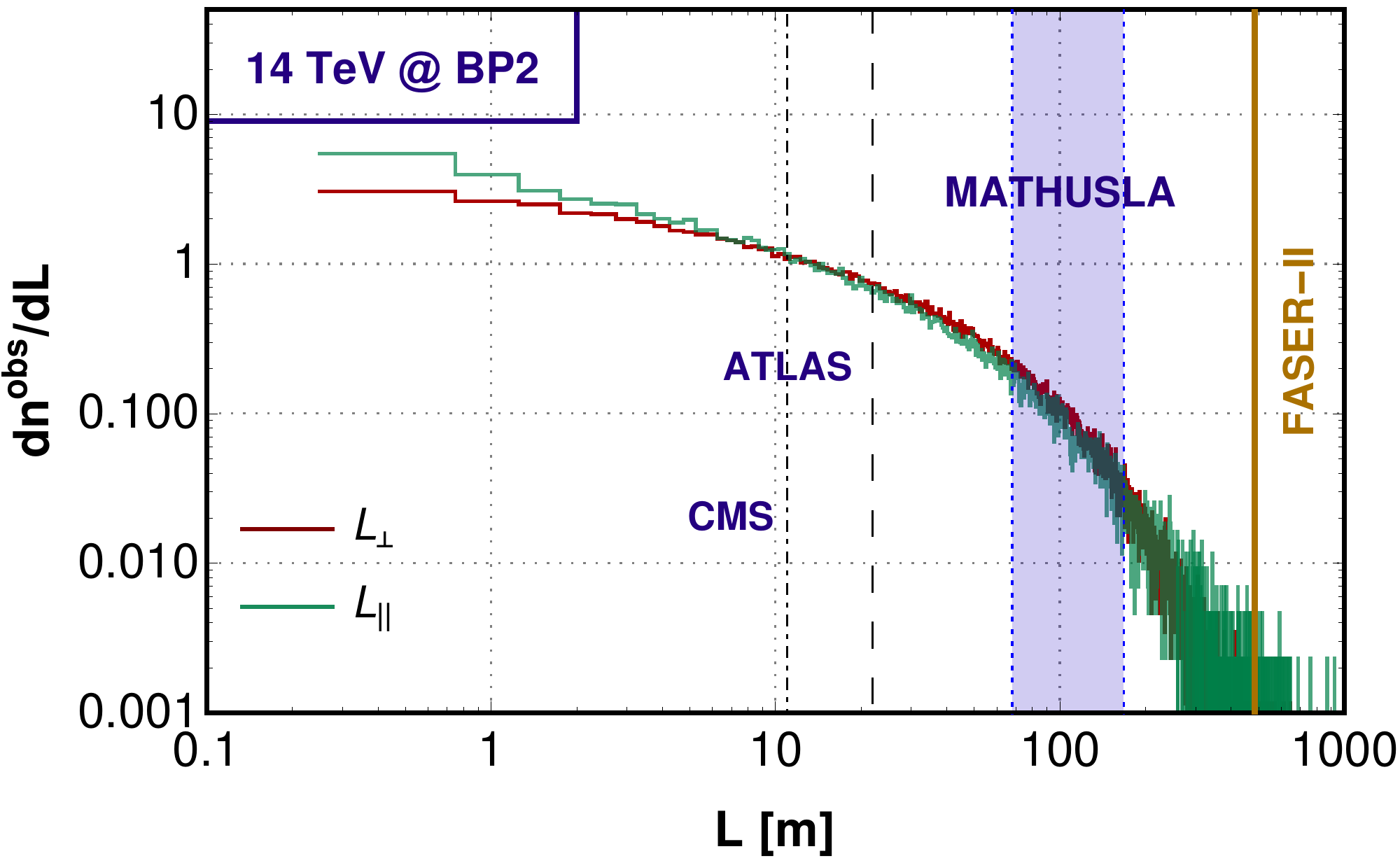}\label{}}\quad
			\subfigure[]{\includegraphics[width=0.35\linewidth,angle=-0]{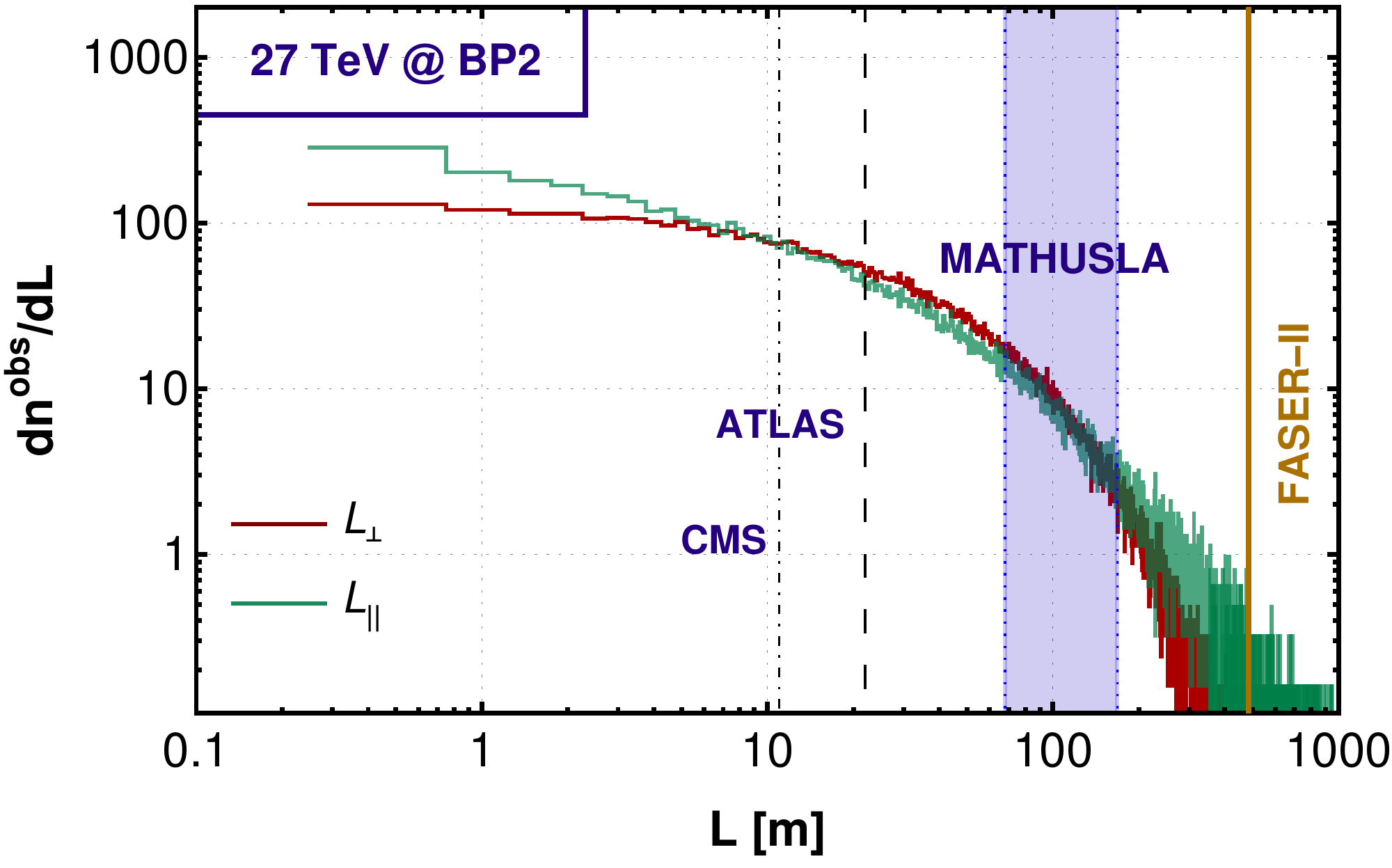}\label{}}\quad
			\subfigure[]{\includegraphics[width=0.35\linewidth,angle=-0]{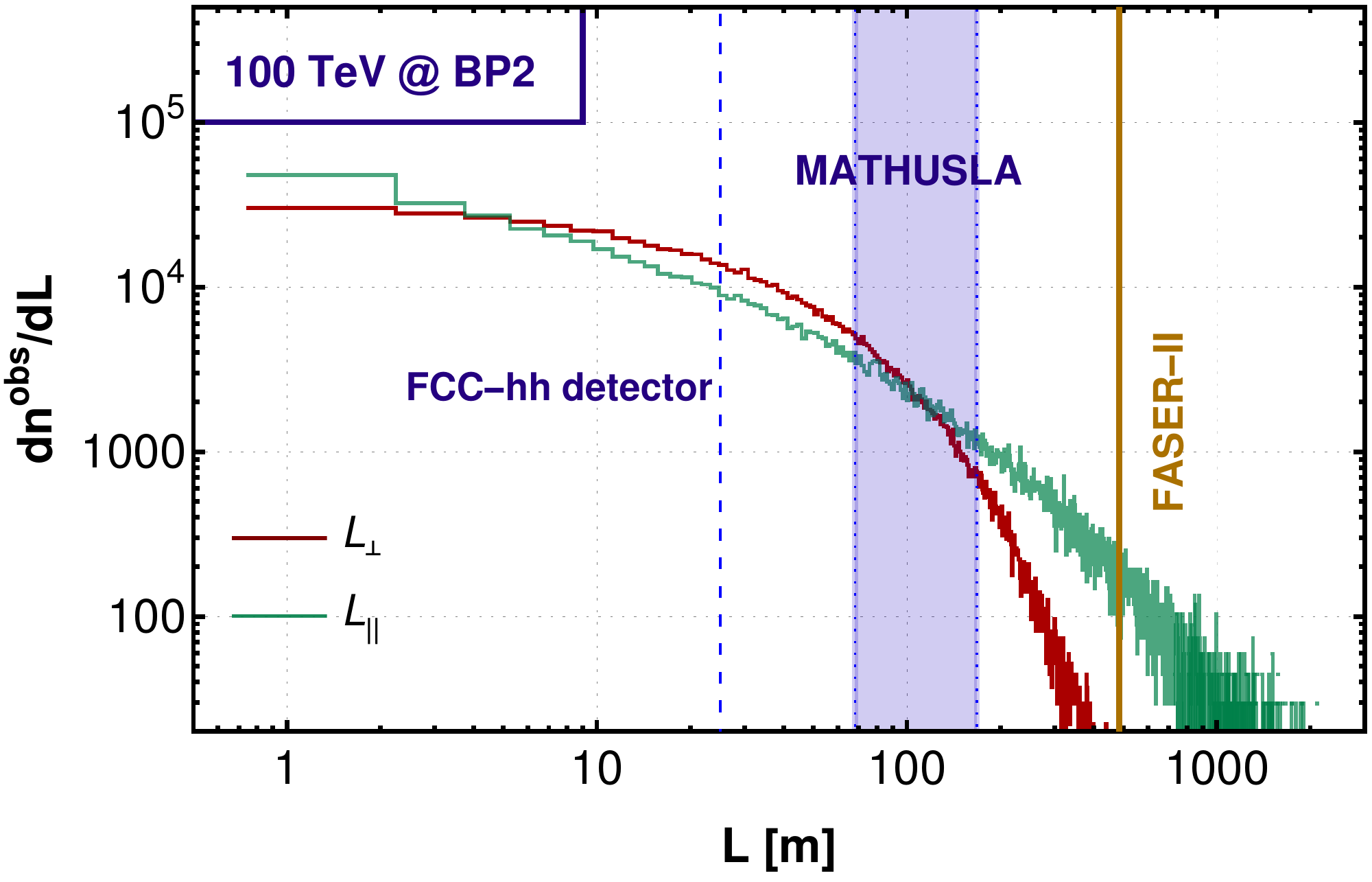}\label{}}}
		\hspace*{-1.0cm}
		\mbox{\subfigure[]{\includegraphics[width=0.35\linewidth,angle=-0]{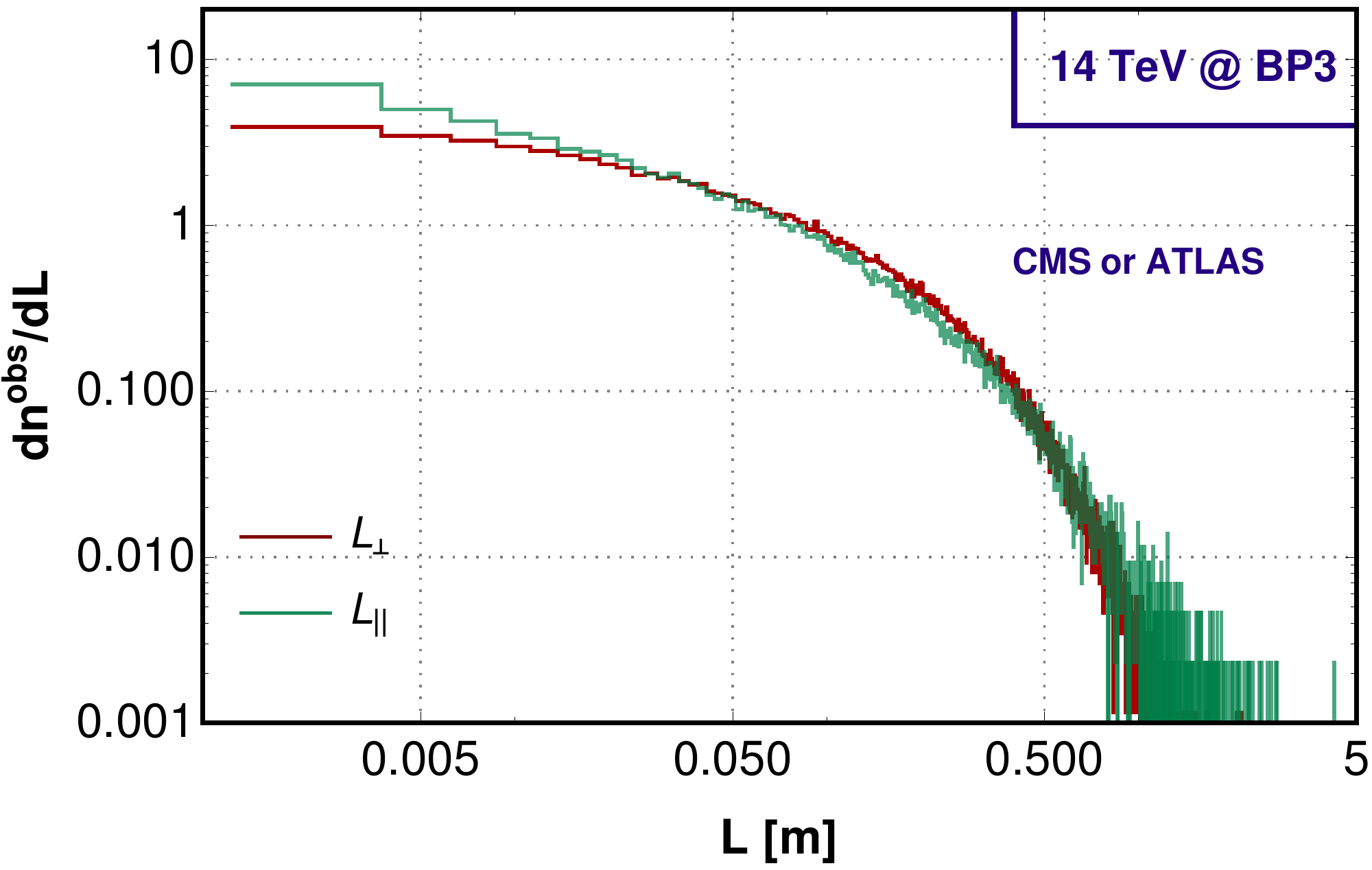}\label{}}\quad
			\subfigure[]{\includegraphics[width=0.35\linewidth,angle=-0]{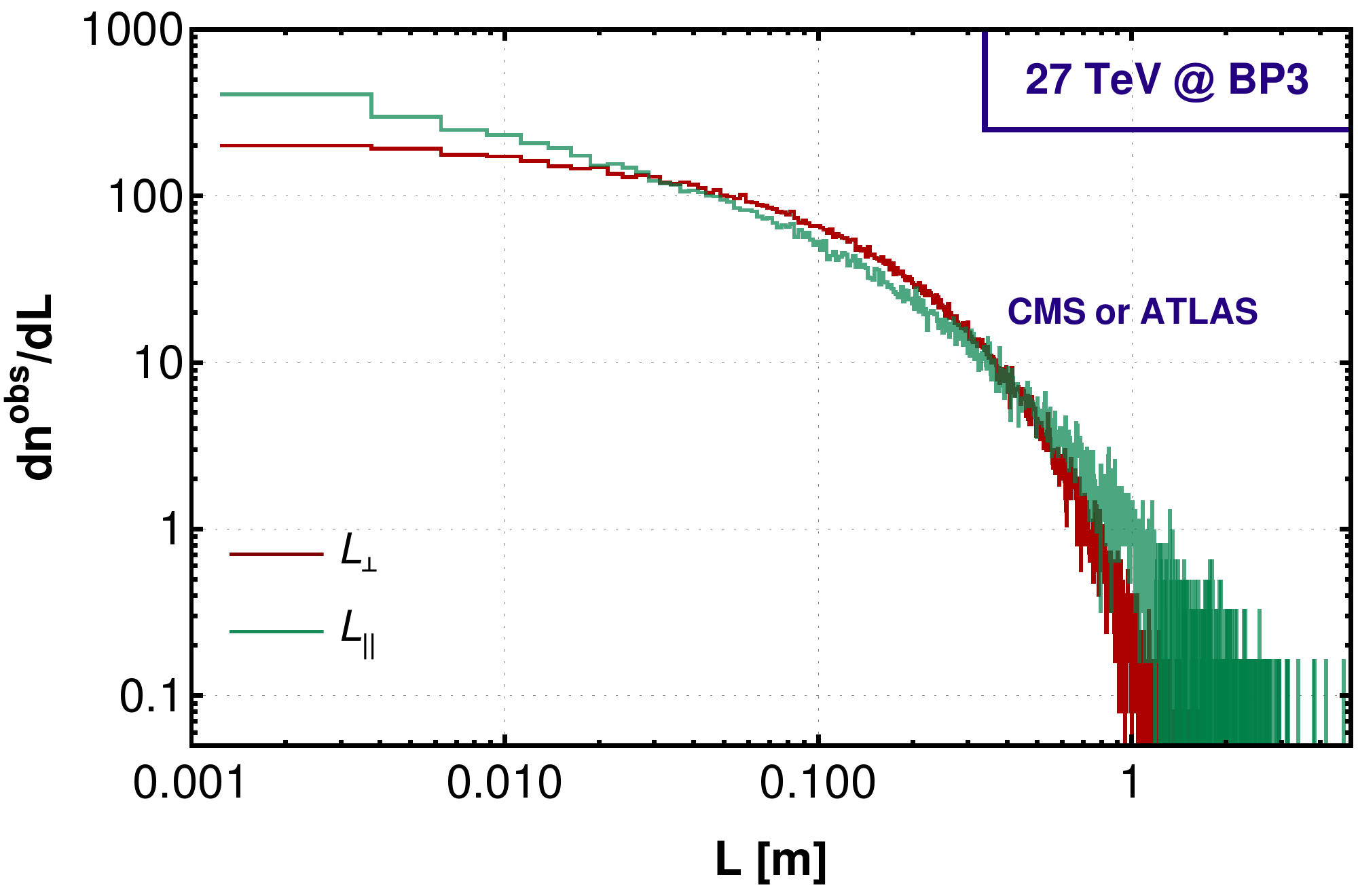}\label{}}\quad
			\subfigure[]{\includegraphics[width=0.35\linewidth,angle=-0]{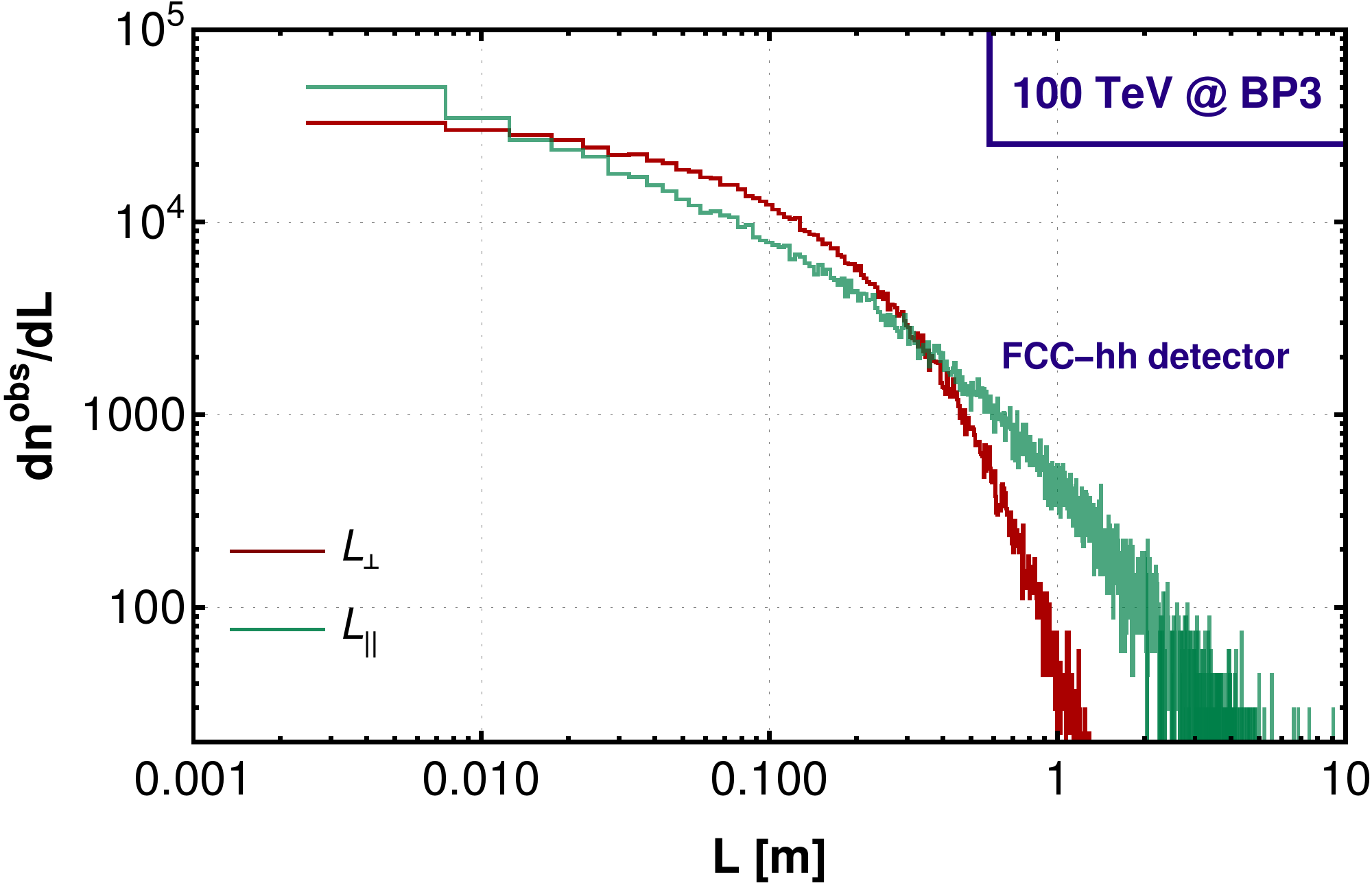}\label{}}}
		\caption{Displaced transverse ($L_\perp$ in red) and longitudinal ($L_{||}$ in green) decay length distributions for $N$, coming from the pair productions at the LHC with the centre of mass energies of 14\,TeV (a, d, g), 27\,TeV (b, e, h) and 100\,TeV (c, f, i) for the integrated luminosities of 3, 10 and 30 ab$^{-1}$, respectively. The first, second and third rows delineate the distributions for BP1, BP2 and BP3, respectively. The black dotted-dashed and dashed lines indicate the upper limit of CMS and ATLAS, respectively, whereas, the blue dashed line specify the upper limit of the proposed FCC-hh detector. The light blue band (68\,m$-$168\,m) denotes the MATHUSLA region and the golden yellow strip denotes the FASER-II region. All of the regions depicted here are in the longitudinal direction.}\label{DcyLT}
	\end{center}
\end{figure*}

Defining the transverse and longitudinal displaced decay lengths considering the corresponding boosts, we will see that LHC at  higher centre  of mass energies can lead to more longitudinal boost governed by the parton distribution function than the transverse one, which is mainly dependent on the uncertainty principle.
We show  $L_\perp, \, L_{||}$ distributions (in \autoref{DcyLT} and in \autoref{DcyLT1gen}) separately in order to show the different boost effects for the perpendicular and longitudinal decay lengths. However, for the event number inside the detector of length $L$, we consider the total decay length of the RHN  $L_\tau = \tau \frac{p}{m}$, and the number of events inside the detector is given by $N(L; L_\tau)=N_0 (1- \exp(-L/L_\tau))$, where $N_0$ is the initial number of particles. Here we pick a conservative minimum value for the boosted decay length of the RHNs of 1 mm \cite{CMS:2021kdm,ATLAS:2020wjh} is taken into account as displaced.

We consider two scenarios to study the displaced decays of the RHNs. For the three BPs in scenario-1 , the root sum square values of the Yukawa couplings are $Y_N=8.08\times 10^{-8}$, $1.98\times 10^{-7}$, and $2.56\times 10^{-7}$
for BP1, BP2, and BP3, respectively.  The corresponding decay lengths are tabulated in \autoref{TabRestDecay}. In scenario-2, we consider BP2 and BP3 only, with the Yukawa couplings $Y_N=5\times 10^{-8}$, $5\times 10^{-9}$ and the details are discussed in \autoref{SC2}.


\begin{table}[h]
	\begin{center}	
		\renewcommand{\arraystretch}{1.2}
		\begin{tabular}{ |c|c|c|  }
			\hline			
			Benchmark  & Decay width & {Rest mass decay } \\
			points & (in GeV) & length (in m) \\
			\hline
			BP1 & $1.02 \times 10^{-18}$ & 193.00 \\ \hline
			BP2& $ 4.97 \times 10^{-17} $ & 3.96   \\ \hline
			BP3 & $ 9.25 \times 10^{-15} $ & 0.02 \\ 		
			\hline
		\end{tabular}
		\caption{Decay widths and rest mass decay lengths for the benchmark points.}  \label{TabRestDecay}
	\end{center}		
\end{table}	

\begin{figure}[hbt]
	\begin{center}
		\hspace*{-1.0cm}
		\mbox{\subfigure[]{\includegraphics[width=0.55\linewidth,angle=-0]{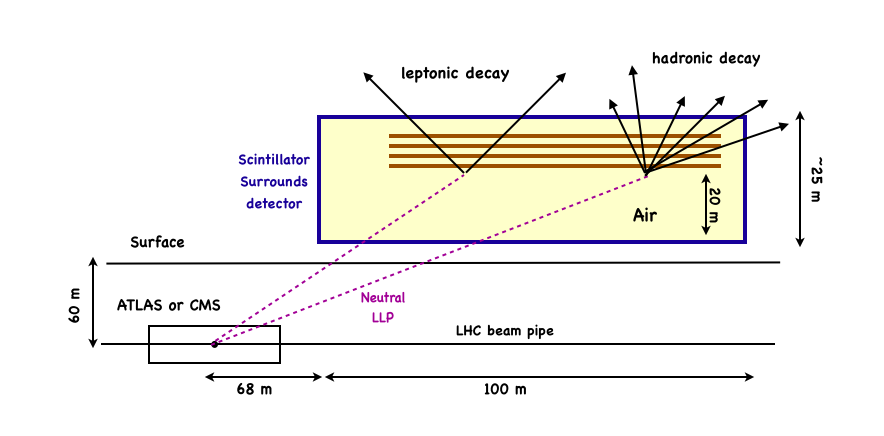}\label{f5}}\quad
			\subfigure[]{\includegraphics[width=0.55\linewidth,angle=-0]{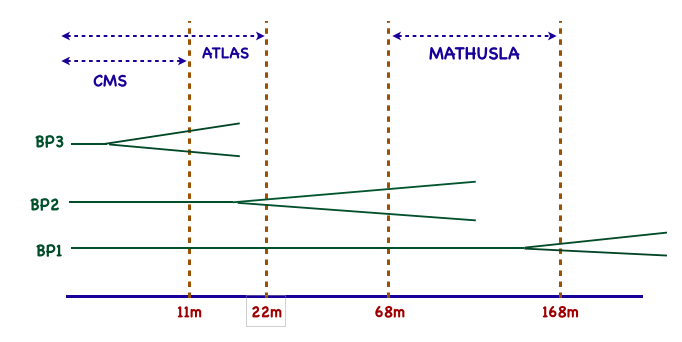}\label{f6}}}
		\caption{A schematic diagram for MATHUSLA detector \cite{Alpigiani:2020iam} and the rest decay points of chosen BPs.}\label{mathusla}
	\end{center}
\end{figure}

The CMS \cite{CMS:2007sch} and ATLAS \cite{ATLAS:design} detectors at the LHC can detect displaced decay signatures for the long lived particles  up to 7.5 ($L_{\perp}$), 12.5 ($L_{\perp}$) meters, and 11 ($L_{||}$), 22 ($L_{||}$) meters, respectively from the interaction point (IP) in the transverse ($L_{\perp}$) and  the longitudinal  ($L_{||}$) directions. The corresponding lengths for the proposed detector of FCC-hh \cite{FCC:2018vvp, Aleksa:2019pvl, FCChh:talk} is 10 ($L_{\perp}$) and 25 ($L_{||}$) meters, respectively from the IP. The limitations of detecting displaced vertex signatures with larger decay lengths $\mathcal{O}(100)$ m at the CMS, ATLAS and FCC-hh detector led to the proposal of   MATHUSLA \cite{Curtin:2018mvb,Lubatti:2019vkf,Chou:2016lxi,Coccaro:2016lnz,Alpigiani:2020iam} detector, which will be placed on the LHC beam pipe, 68 m away from the centre of the LHC detectors as depicted in \autoref{mathusla}(a). It  should be notated that MATHUSLA detector is planned to be placed in one side of the hemispheres, thus it can only tag maximum one RHN at a time. Thus the same and opposite sign leptons coming from both the legs of RHN remain illusive for MATHUSLA.

According to a recent update, the MATHUSLA detector will be 68 meters from the CMS/ATLAS interaction point in the longitudinal direction, 60 meters in the transverse direction and will have a volume of $25\times 100\times 100$\,m$^3$, with a access over the azimuthal angle of 0.27. The corresponding $\eta$ cuts are taken care of by demanding the transverse and longitudinal decay lengths within the detectors simultaneously.   The detecting efficiency for the hadronic as well as leptonic decays are almost $100\%$ \cite{Curtin:2018mvb,cartin}. In \autoref{mathusla}(b)  the rest mass decay lengths are exhibited for the benchmark points as mentioned in \autoref{TabRestDecay}. As evident from \autoref{TabRestDecay}, for BP2 and BP3 the rest mass decay length $\lesssim 10$ m and fall within the range of CMS and ATLAS detector. For the boost effect, we can still get some events in the MATHUSLA region for BP2.  However, for BP1, it is $\sim 193$ m, which falls beyond the MATHUSLA region. It is worth mentioning here that, as the 100 TeV FCC-hh is not built yet, one can take the liberty of suggesting LLP (Long  Lived  particle) detectors with location and geometry suited to one's model and search strategies for example in \cite{Bhattacherjee:2021rml}. However, the MATHUSLA geometry is better  for the LLPs in our consideration even at 100 TeV FCC-hh.

\begin{figure}[hbt]
	\centering
	\includegraphics[width=0.6\linewidth,angle=0]{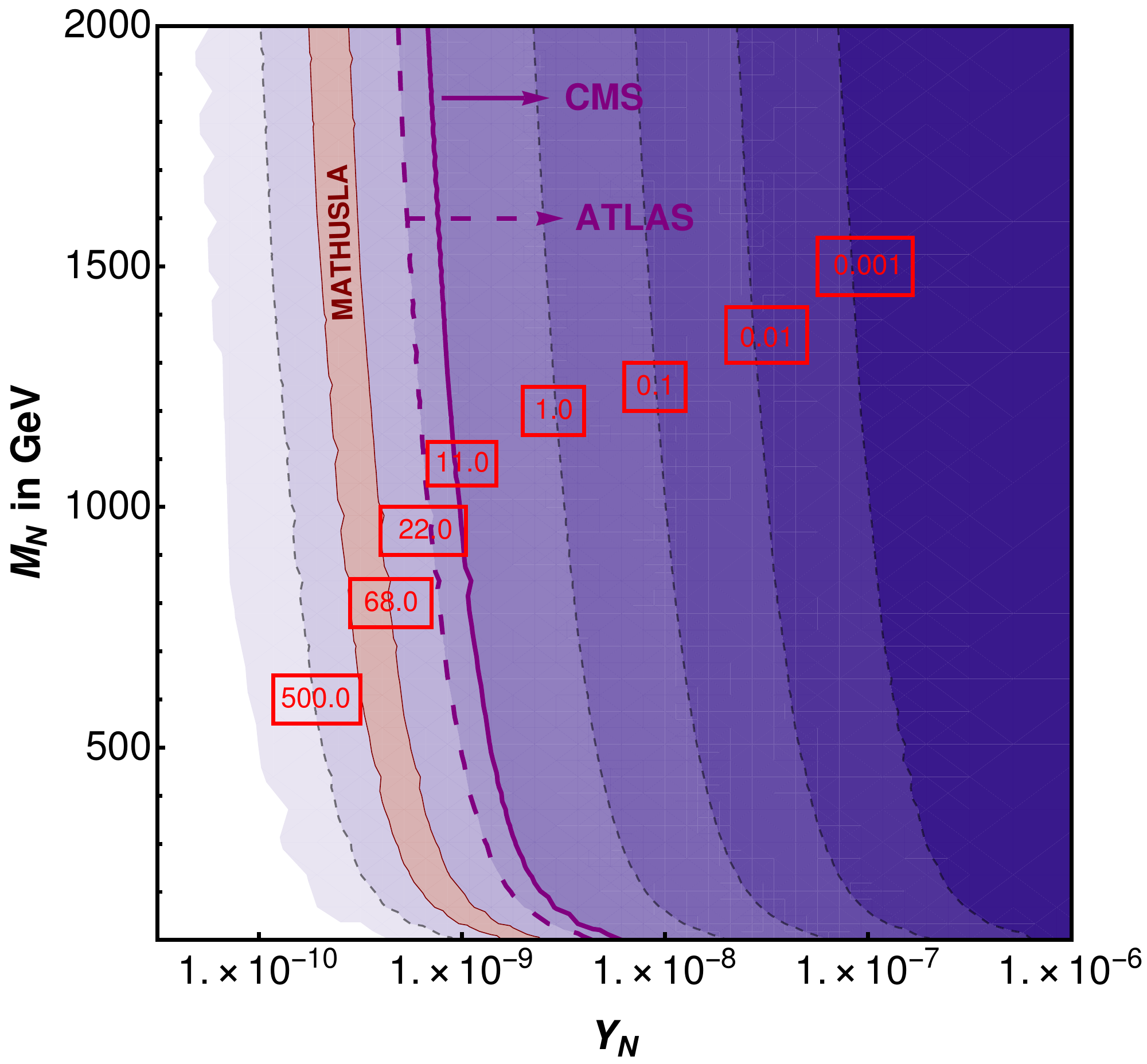}
	\caption{The displaced decay length of RHN is presented in $Y_N - M_N$ plane . Different regions of decay length from 0.001\,m to 500\,m are marked with red box. The CMS, ATLAS and MATHUSLA boundaries are shown in magenta solid, magenta dashed and brown thick lines, respectively.}\label{DecayRegion}
\end{figure}

In \autoref{DecayRegion} we present the displaced decay length contours of RHN from $0.001\,\rm{m}$ to $500.0\,\rm{m}$ in $M_N-Y_N$ plane with different shaded regions. We also mention the boundaries of CMS and ATLAS detectors with magenta solid and dashed lines, respectively with the regions in the right that can be explored. MATHUSLA will be implemented approximately $68\,\rm{m}$ away from LHC and the region that can be traversed is shown by the  light  brown band in \autoref{DecayRegion}. Along with MATHUSLA, FASER-II \cite{FASER:2019aik,FASER:2018eoc} is also proposed to be placed at a distance of 480\,m in the longitudinal direction from ATLAS interaction point. This detector will be only 10\,m long and its radius is proposed to be only 1\,m. It can detect the events with $|\eta | \geq 3.5$.
Thus the main motivation of FASER-II is to detect the soft events coming along with beam pipe line.


Equipped with the collider set up and with the knowledge of the rest mass decay lengths, we plot the transverse ($L_{\perp}$ in red) and longitudinal ($L_{||}$ in green)  decay length distributions for the benchmark points at the LHC/FCC  for the centre of mass energies of 14 (a, d, g), 27 (b, e, h), 100 TeV (c, f, i) with the integrated luminosities of 3, 10 and 30 ab$^{-1}$, respectively in \autoref{DcyLT}. The plots are generated by PYTHIA8 \cite{Pythia8.2}, where the boost effects and the decay distributions are included dynamically event by event. The CMS and ATLAS boundaries in the longitudinal direction are shown with black dotted-dashed and dashed lines, respectively, while the corresponding boundary for the proposed FCC-hh reference detector is specified with blue dashed line. The  regions for MATHUSLA is exhibited in the light-blue band. The golden yellow strip depicts the tiny region that can be explored by the  FASER-II detector. The transverse regions also can be read from the \autoref{DcyLT}, however  the detector ranges are not shown explicitly in the figures. 

In \autoref{DcyLT} (a, b, c), we have presented the transverse and longitudinal decay length distributions of BP1 ($M_N=10$\,GeV). Because of the low Yukawa coupling and low mass, the decay length reaches up to 100\,km in both the directions for 14\,TeV centre of mass energy. Since the boost effect is stronger in the longitudinal direction compared to the transverse one, we can see the enhancement of decay length in the longitudinal direction for 27 and 100\,TeV centre of mass energies(second and third columns). Due to the large decay lengths in BP1, most of the events fall outside the reach of CMS, ATLAS or MATHUSLA detectors. On the contrary, BP2 ($M_N=60$\,GeV) is the most suitable to be detected by all three detectors simultaneously. Here the transverse and longitudinal decay length distributions for BP2 are depicted by \autoref{DcyLT} (d, e, f).  Though the rest mass decay length for this benchmark point is around 4\,meter (\autoref{TabRestDecay}), the displaced longitudinal decay length can reach up to 1\,km for 100\,TeV centre of mass energy due to the larger boost. Finally,  the displaced transverse and longitudinal decay length distribution for BP3 ($M_N=100$\,GeV) are shown in \autoref{DcyLT} (g, h, i). Here the maximum displacement can occur  around 10 meters, resulting from all of the events inside the CMS, ATLAS or the proposed FCC-hh reference detector. We would like to mention that due to very small detectable range for the FASER-II, the event numbers are abysmally low even though they fall under the detectable ranges for BP1, BP2.

\begin{table*}[hbt]	
	\begin{center}
		\renewcommand{\arraystretch}{1.2}
		\begin{tabular}{|c|c|c|c|}
			\cline{2-4}
			\multicolumn{1}{c|}{}  & \multicolumn{3}{c|}{Events inside MATHUSLA}\\
			\cline{2-4}
			\multicolumn{1}{c|}{} &	\multicolumn{3}{c|}{Centre of mass energy} \\
			\cline{2-4}
			\multicolumn{1}{c|}{}& \multicolumn{1}{c|}{14\,TeV}&\multicolumn{1}{c|}{27\,TeV}&\multicolumn{1}{c|}{100\,TeV}\\
			\hline
			BP2  & 4.7 & 385.7  & 38029.2 \\
			\hline
		\end{tabular}
		\caption{The inclusive number of events inside the proposed detector, MATHUSLA, for the second benchmark point (BP2), at the centre of mass energies of 14, 27 and 100 TeV at an integrated luminosity of 3\,ab$^{-1}$, 10\,ab$^{-1}$ and  30\,ab$^{-1}$, respectively.} \label{MATHUSLATab}
	\end{center}	
\end{table*}


\autoref{MATHUSLATab} presents the inclusive number of events that can be obtained inside MATHUSLA for BP2 with 14, 27 and 100 TeV  centre of mass energies, at an integrated luminosity of 3\,ab$^{-1}$, 10\,ab$^{-1}$ and 30\,ab$^{-1}$, respectively. BP1 and BP3 are beyond the reach of MATHUSLA;   former having the decay length beyond the reach and the later having them within the range of 10\,m.

\subsection{Number of displaced leptonic events}\label{result_SC1}

In this section, we present the number of displaced leptonic events collected by the CMS, ATLAS and MATHUSLA detectors for 14 and 27 TeV colliders. As the detector specifications change from 14 TeV to 100 TeV, we provide the number of events with FCC-hh reference detector at 100 TeV (specified earlier) and MATHUSLA, taking into account the larger forward cover of the projected detector shape.

RHNs produced via $Z_{B-L}$ has dominant decays to $\ell^{\pm} W^\mp$ for the chosen benchmark points. Thus multi-leptonic final states are mostly common, when the gauge bosons also decay leptonically. In the following tables we present the final state numbers coming from different multi-leptonic  final state topologies.

\begin{table}[h]	
	\begin{center}
		\renewcommand{\arraystretch}{1.2}
		\begin{tabular}{ |c||c|c|c|c| }
			\hline
			{\multirow{2}{*}{\diagbox[width=5.0cm]{$4\ell$}{Displaced decay}}}&
			Benchmark &\multicolumn{3}{c|}{Centre of mass energy}\\
			\cline{3-5}
			& points  & 14\,TeV & 27\,TeV & 100\,TeV  \\
			\hline
			\multirow{3}{*}{CMS} & BP1 & 0.0 & 0.4 & $-$ \\
			\cline{2-5}
			& BP2 & 0.5  & 33.2 & $-$\\
			\cline{2-5}
			& BP3 & 2.6 & 196.3 &  $-$ \\	
			\hline 
			\multirow{1.5}{*}{ATLAS} & BP1 & 0.0 & 0.6  & 17.7  \\
			\cline{2-5}
			\multirow{1.1}{*}{\&} & BP2 & 0.7  & 49.4 & 4393.44 \\
			\cline{2-5}
			\multirow{0.5}{*}{FCC-hh reference detector} & BP3 & 2.6 & 196.3 & 17032.68  \\	
			\hline 
			\multirow{3}{*}{MATHUSLA} & BP1 & 0.0 & 0.3 & 33.7 \\
			\cline{2-5}
		    & BP2 & 0.0 & 3.2 & 236.52 \\
			\cline{2-5}
			& BP3 & 0.0  & 0.0 & 0.0 \\	
			\hline
			
		\end{tabular}
		\caption{Number of events in  $4\ell $ final state for the benchmark points with the center of mass energies of 14\,TeV, 27\,TeV and 100\,TeV at the integrated luminosities of $3\,\text{ab}^{-1}$, $10\,\text{ab}^{-1}$ and $30\,\text{ab}^{-1}$, respectively. The numbers are given separately for CMS, ATLAS, FCC-hh reference detector (for 100 TeV) and MATHUSLA.}  \label{FS1}
	\end{center}	
\end{table}

\begin{table}[h]	
	\begin{center}
		\renewcommand{\arraystretch}{1.2}
		\begin{tabular}{ |c||c|c|c|c| }
			\hline
			{\multirow{2}{*}{\diagbox[width=5.0cm]{$3\ell + (\geq 1j)$}{Displaced decay}}}&
			Benchmark &\multicolumn{3}{c|}{Centre of mass energy}\\
			\cline{3-5}
			& points  & 14\,TeV & 27\,TeV & 100\,TeV  \\
			\hline
			\multirow{3}{*}{CMS} & BP1 & 0.0 & 0.6 & $-$ \\
			\cline{2-5}
			& BP2 & 1.9 & 88.7 & $-$ \\
			\cline{2-5}
			& BP3 & 5.4 & 383.8 &  $-$ \\	
			\hline 
			\multirow{1.5}{*}{ATLAS} & BP1 & 0.0  & 0.8  & 41.8  \\
			\cline{2-5}
			\multirow{1.1}{*}{\&} & BP2 & 2.3 & 121.3 & 10872.4 \\
			\cline{2-5}
			\multirow{0.5}{*}{FCC-hh reference detector} & BP3 & 5.4 & 383.8 & 35228.0 \\		
			\hline 
			\multirow{3}{*}{MATHUSLA} & BP1 & 0.0  & 0.5  & 60.5 \\
			\cline{2-5}
			& BP2 & 0.1 & 5.3 & 635.0 \\
			\cline{2-5}
			& BP3 & 0.0 & 0.0 & 0.0 \\	
			\hline
			
		\end{tabular}
		\caption{Number of events in  $3\ell + (\geq 1j) $ final state for the benchmark points with the center of mass energies of 14\,TeV, 27\,TeV and 100\,TeV at the integrated luminosities of $3\,\text{ab}^{-1}$, $10\,\text{ab}^{-1}$ and $30\,\text{ab}^{-1}$, respectively. The numbers are given separately for CMS, ATLAS, FCC-hh reference detector (for 100 TeV) and MATHUSLA.}  \label{FS2}
	\end{center}	
\end{table}

Four lepton final states can arise from leptonic decays of both the $W^\pm$s  or one of the $Z$ bosons coming  from the RHN. In \autoref{FS1}, we present the number of inclusive $4\ell$ final state which are displaced for the chosen benchmark points for the centre of mass energies of 14, 27 and 100 TeV, at an integrated luminosities of 3, 10 and 30 ab$^{-1}$, respectively.

The $4\ell$ final states are generally with very little SM backgrounds, especially after the reconstruction  of the RHN invariant mass. In this  case  due to the  displaced vertex signatures  the final state  is background free.  We also segregate the event numbers for the CMS, ATLAS and MATHUSLA detectors. It is interesting to note that for BP1 ($M_N=10$ GeV), the rest mass decay length is 193 m, which is already out of the ranges of CMS, ATLAS and MATHUSLA and thus boost effect further makes it undetectable. As mentioned earlier in the text, for BP2 all three detectors fall in the regions of the displaced decays specially for 27, 100 TeV centre of mass energies, however, for BP3, the displaced decay lengths are restricted to CMS and ATLAS and fails to reach in the MATHUSLA range. 

\begin{table}[h]	
	\begin{center}
		\renewcommand{\arraystretch}{1.2}
		\begin{tabular}{ |c||c|c|c|c| }
			\hline
			{\multirow{2}{*}{\diagbox[width=5.0cm]{$2\ell + (\geq 2j)$}{Displaced decay}}}&
			Benchmark &\multicolumn{3}{c|}{Centre of mass energy}\\
			\cline{3-5}
			& points  & 14\,TeV & 27\,TeV & 100\,TeV  \\
			\hline
			\multirow{3}{*}{CMS} & BP1 & 0.0 & 2.3 & $-$ \\
			\cline{2-5}
			& BP2 & 7.5 & 442.2 & $-$ \\
			\cline{2-5}
			& BP3 & 14.1  & 1066.7 & $-$ \\	
			\hline 
			\multirow{1.5}{*}{ATLAS} & BP1 & 0.1  & 3.4  & 308.7 \\
			\cline{2-5}
			\multirow{1.1}{*}{\&} & BP2 & 9.8 & 623.7 & 57844.8 \\
			\cline{2-5}
			\multirow{0.5}{*}{FCC-hh reference detector} & BP3 & 14.1  & 1066.7 & 102492.2  \\	
			\hline 
			\multirow{3}{*}{MATHUSLA} & BP1 & 0.0 & 5.3  & 443.9  \\
			\cline{2-5}
			& BP2 & 0.4  & 30.3  & 3354.2 \\
			\cline{2-5}
			& BP3 & 0.0 & 0.0  & 0.0 \\	
			\hline
		\end{tabular}
		\caption{Number of events in  $2\ell + (\geq 2j) $ final state for the benchmark points with the center of mass energies of 14\,TeV, 27\,TeV and 100\,TeV at the integrated luminosities of $3\,\text{ab}^{-1}$, $10\,\text{ab}^{-1}$ and $30\,\text{ab}^{-1}$, respectively. The numbers are given separately for CMS, ATLAS, FCC-hh reference detector (for 100 TeV) and MATHUSLA.}  \label{FS3}
	\end{center}	
\end{table}

After $4\ell$ signature we move to $3\ell + (\geq 1j) $ signature, which results in when one of the gauge boson decays hadronically, as we present the numbers in \autoref{FS2}.  The inclusive $3\ell + (\geq 1j) $  numbers are predicted for the LHC/FCC with centre of mass energy of 14, 27 and 100 TeV at an integrated luminosities of $3\,\text{ab}^{-1}$, $10\,\text{ab}^{-1}$ and $30\,\text{ab}^{-1}$, respectively. For CMS and ATLAS only BP2 and BP3 have healthy event numbers. MATHUSLA fails to register any significant event numbers for all three benchmark points, except for BP2 at higher energies.

In \autoref{FS3}, we describe the event numbers for $2\ell + (\geq 2j)$ final state for the benchmark points with the center of mass energies of 14\,TeV, 27\,TeV and 100\,TeV at the integrated luminosities of $3\,\text{ab}^{-1}$, $10\,\text{ab}^{-1}$ and $30\,\text{ab}^{-1}$, respectively. Unlike earlier two final states, here we have significant number of events for BP1 and BP2 for higher energies. For CMS and ATLAS, BP2 and BP3 can be probed very easily. Due to highest branching fraction of $W^\pm$ to hadronic mode and because of the Fatjet signature, the event numbers for $ 2\ell + \geq 2j$ are largest for all the benchmark points.

\subsection{Sensitivity regions at different energies}\label{region_SC1}

The dominant decay mode of a single RHN is $N \to W^{\pm}\ell^{\mp}$, which gives $1\ell+ 2j$ signature, results in  the RHN pair with $2\ell +4j$ final state. However, for the lower RHN mass, due to the large boost,  the  two-jets coming from the on- or off-shell $W^\pm$ boson are collimated to form a single Fatjet  even with the jet radius of 0.5,  catalyses  the final state  of $ 2\ell +2j$ as  we elaborate in the  following paragraphs.

In this section, we draw the sensitivity regions in the $M_{Z_{B-L}}-M_N$ plane at the LHC/FCC with  centre of mass energies of 14, 27, 100 TeV for $ 2\ell +2j$ final state (dark shaded region) and for $ 2\ell +4j$ final state (light shaded region). Here we follow the  prescription  of  Poisson distribution in order to estimate the $95\%$ confidence level sensitivity plots for the non-observation of signals \cite{pdg,pdg1} with zero backgrounds.

\begin{figure}[hbt]
	\begin{center}
		\hspace*{-1.0cm}
		\mbox{\subfigure[]{\includegraphics[width=0.36\linewidth,angle=0]{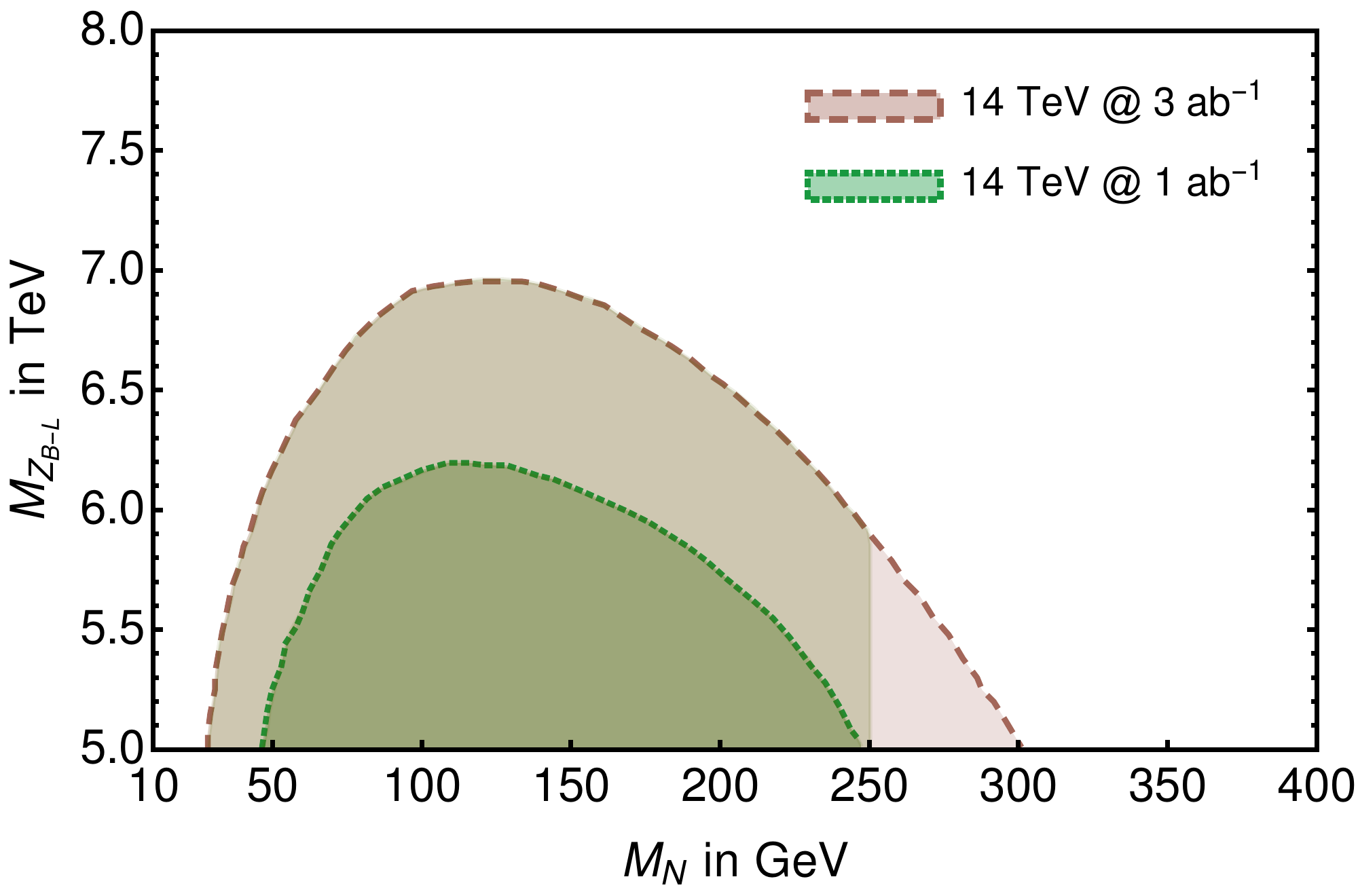}\label{14dcyl}}\quad
		\subfigure[]{\includegraphics[width=0.36\linewidth,angle=0]{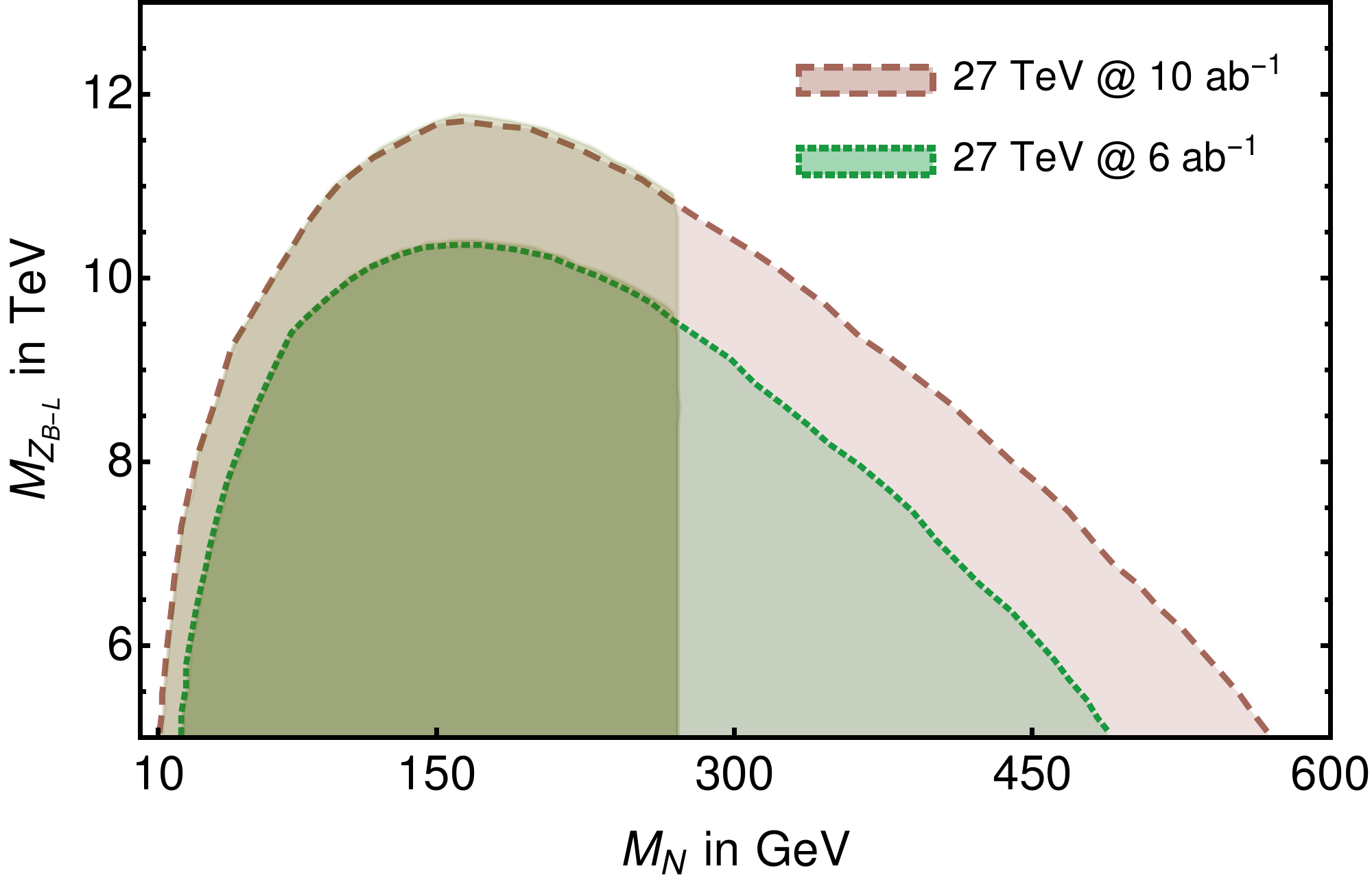}\label{30dcyl}}\quad
		\subfigure[]{\includegraphics[width=0.35\linewidth,angle=0]{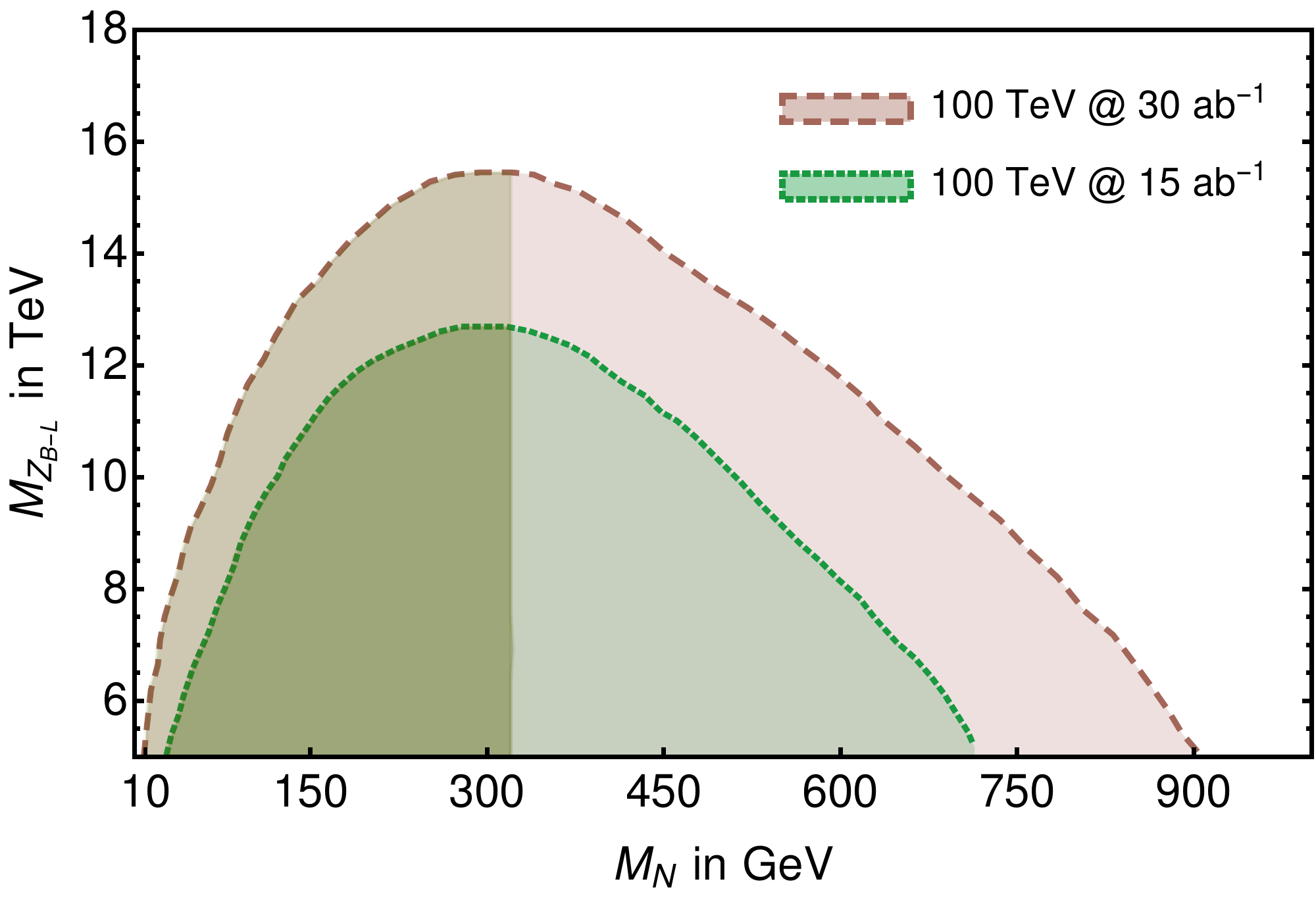}\label{30dcyl}}}
		\caption{Limits obtained via the inclusive measurements from RHNs decay to displaced $2\ell + 2j$ final state (dark shaded region), displaced $2\ell + 4j$ final state (light shaded region) and it is presented in $M_{Z_{B-L}}$ versus $M_N$ plane at $95\%$ CL. The shaded regions can be probed at any of the detectors CMS, ATLAS and MATHUSLA for  14, 27\,TeV centre of mass energies, and at either of the FCC-hh reference detector and MATHUSLA for 100\,TeV centre of mass energy, considering the root sum square values of the Yukawa couplings discussed in \autoref{BPs_SC1}. }\label{ReachUPMNS}
	\end{center}
\end{figure}

In \autoref{ReachUPMNS},  we plot the regions in the $M_{Z_{B-L}}-M_N$ plane which can be probed at  $95\%$ confidence levels with the centre of mass energies of 14 TeV, 27\,TeV and 100 TeV at the integrated luminosities of  $1\,(3),\,\, 6\,(10),$ and $ 15\,(30)\,\rm{ab}^{-1}$,  respectively. These regions include two different final state topologies: the darker green (darker brown) corresponds to $2\ell +2j$ and the lighter green (lighter brown) refers to $2\ell +4j$ final state.  For lower RHN mass ($\lesssim 250\,$GeV for 14\,TeV centre of mass energy), Fatjets forms and thus the final state of $2\ell+2j$ occurs more often than $2\ell+4j$, which is more probable for higher RHN mass as discussed in \autoref{RHN_mass}. Such analysis shows that  a very light RHN i.e. $M_N \sim 30$ GeV can be probed along with $M_{Z_{B-L}} \sim 7$ TeV for 14 TeV centre of mass energy. However, at 100 TeV, a very low RHN  mass of $\sim 5$ GeV  can be probed along with a maximum of $900$ GeV and $M_{Z_{B-L}} \sim 15.5$ TeV.  For very low RHN masses, the decay widths become small, resulting in a very long displaced decay length which is  outside any of the detectors.

\section{Scenario-2}\label{SC2}
In this scenario, we focus  on the case where one of the RHNs  decouples from the observed neutrino mass generations, with the possibility of much smaller Yukawa coupling, while the rest  two can explain light neutrino masses and mixing. Thus collider predictions here is solely  for one generation of RHN. 
\subsection{Parameter space and benchmark points}
Scenario-2 assumes the presence of a RHN which couples dominantly to electron or muon and whose Yukawa coupling is too small to produce a sizeable neutrino mass scale. This implies that the observed neutrino mass matrix is generated by the other two RHNs. We further assume that they do not produce displaced vertices.  For the analysis of this scenario,  we take two benchmark points: BP2 ($M_N=60$ GeV) and BP3 ($M_N=100$ GeV) with two different values of the Yukawa coupling: $Y_N= 5 \times 10^{-8}$ and $5\times 10^{-9}$.

\subsection{Displaced vertex signatures}\label{DisVartex_SC2}


\begin{figure*}[h!]
	\begin{center}
		\hspace*{-1.0cm}
		\mbox{\subfigure[]{\includegraphics[width=0.35\linewidth,angle=-0]{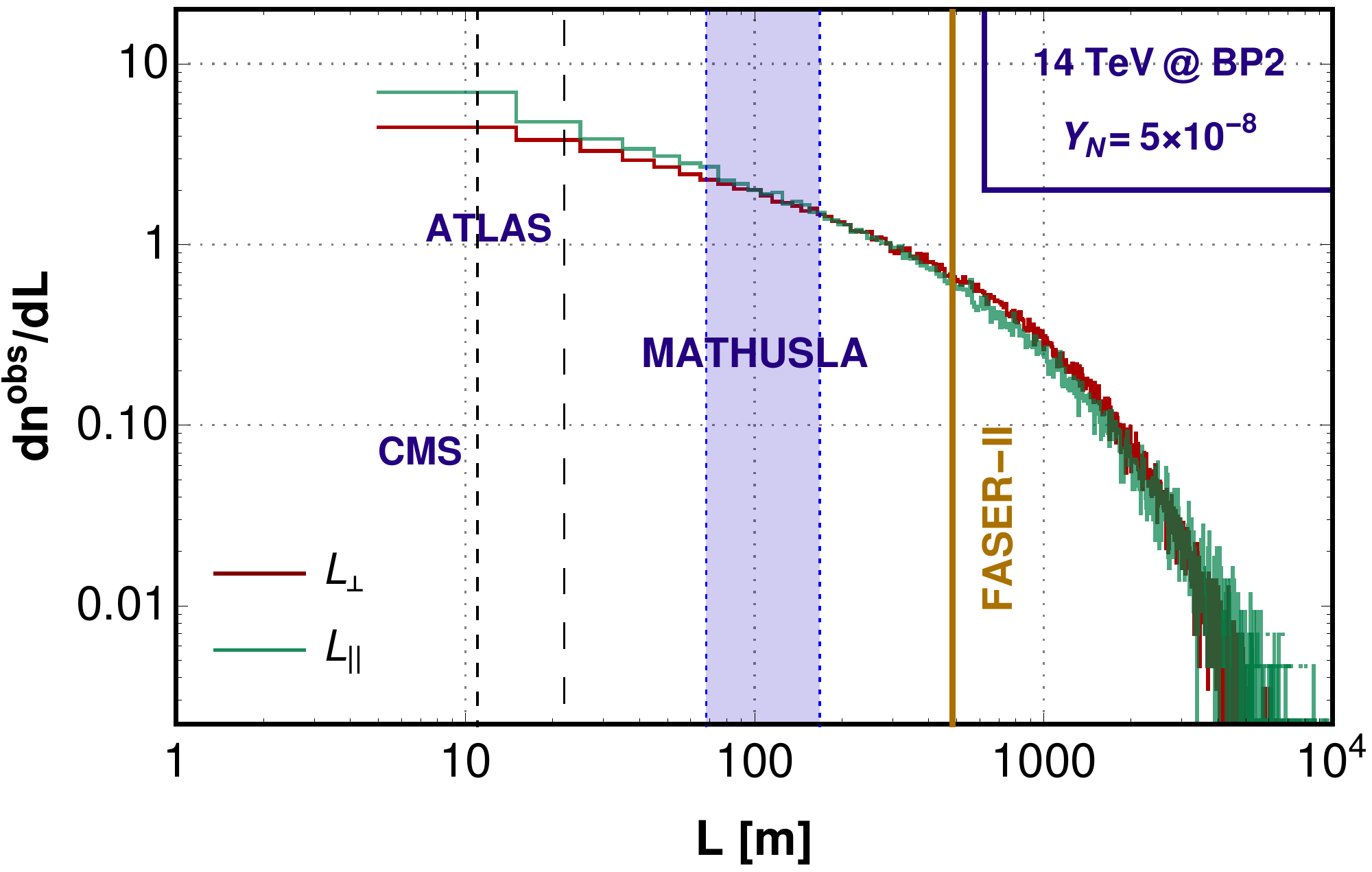}\label{}}\quad
			\subfigure[]{\includegraphics[width=0.35\linewidth,angle=-0]{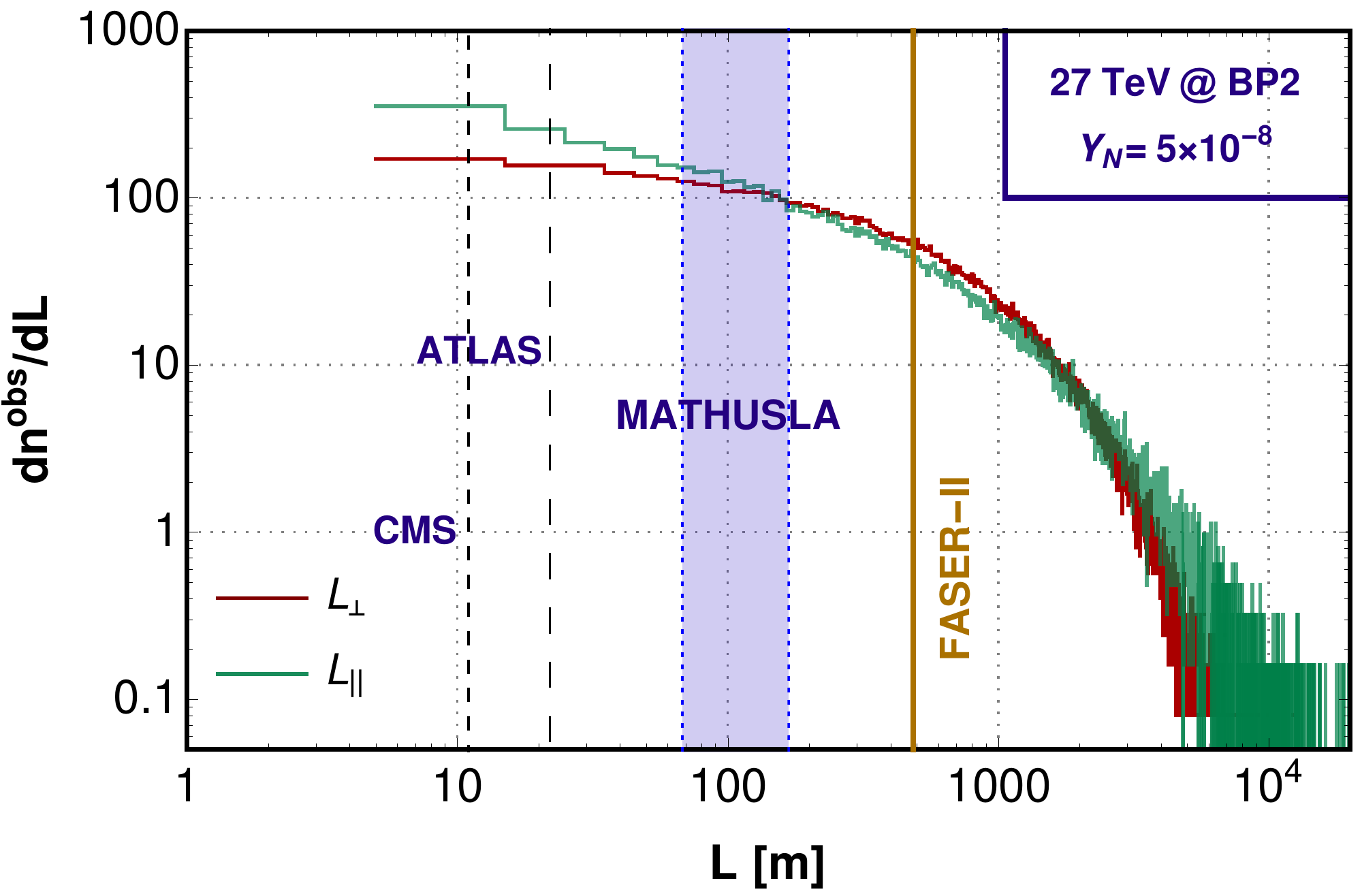}\label{}}\quad
			\subfigure[]{\includegraphics[width=0.35\linewidth,angle=-0]{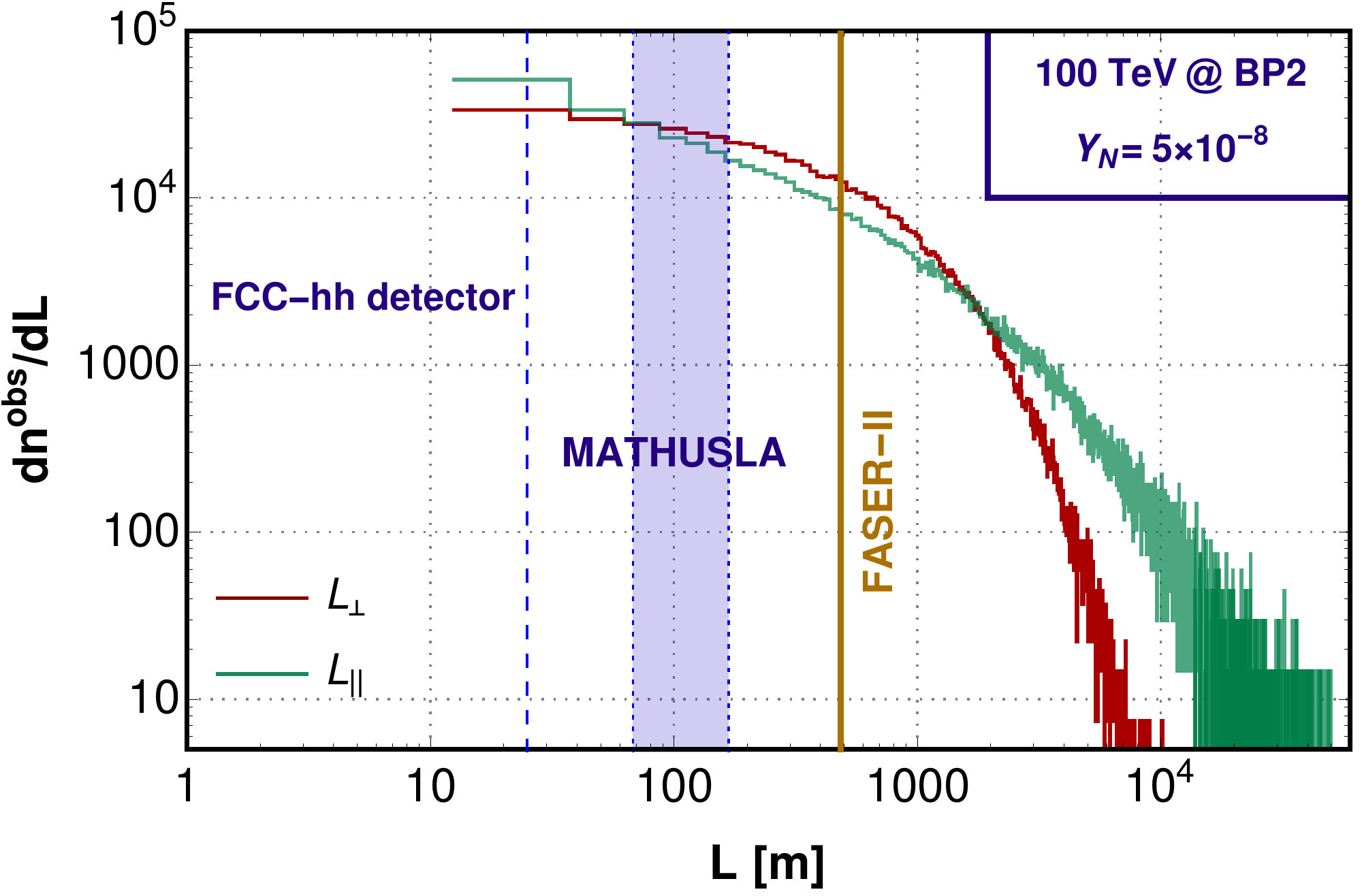}\label{}}}
		\hspace*{-1.0cm}
		\mbox{\subfigure[]{\includegraphics[width=0.35\linewidth,angle=-0]{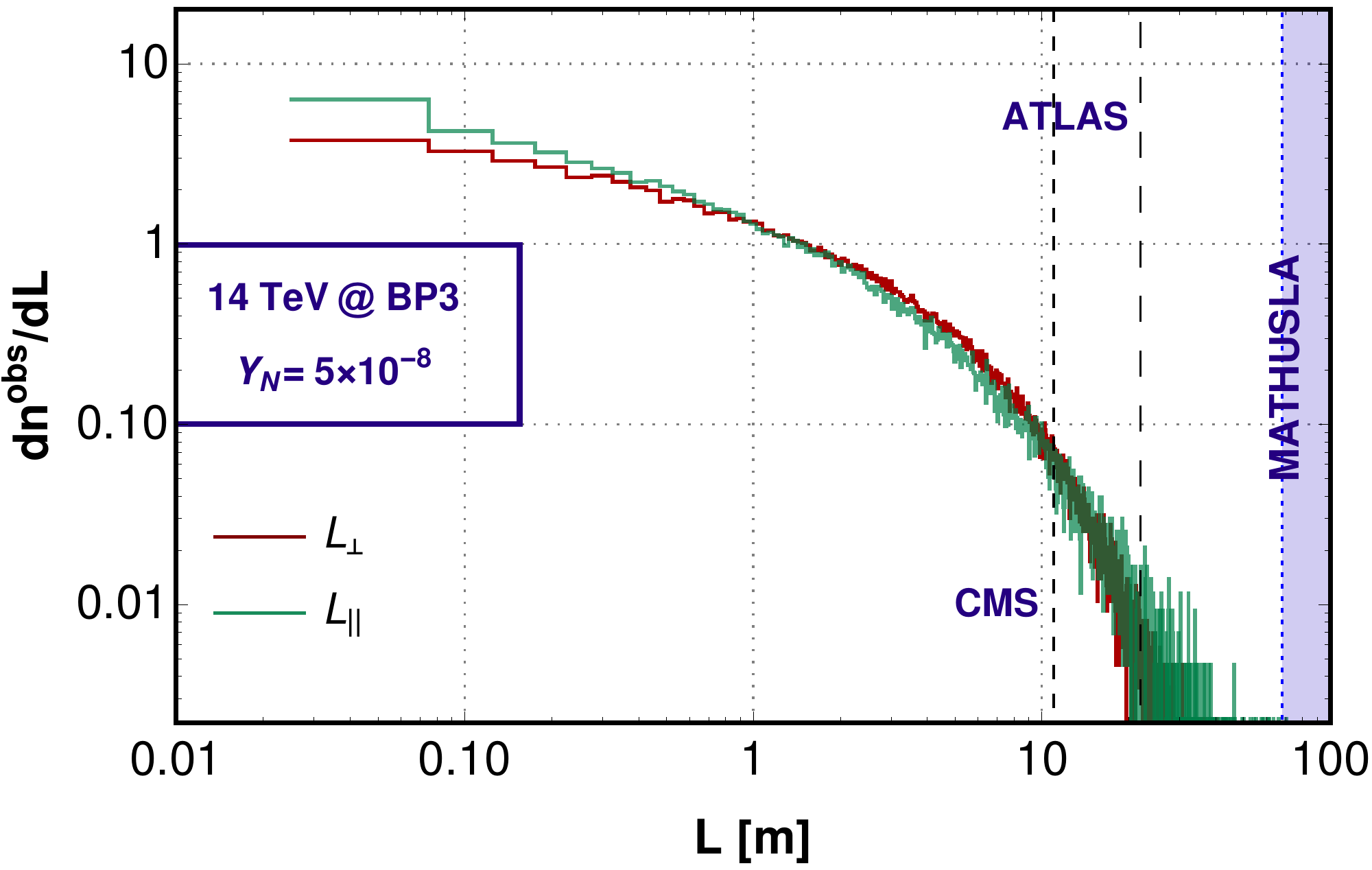}\label{}}\quad
			\subfigure[]{\includegraphics[width=0.35\linewidth,angle=-0]{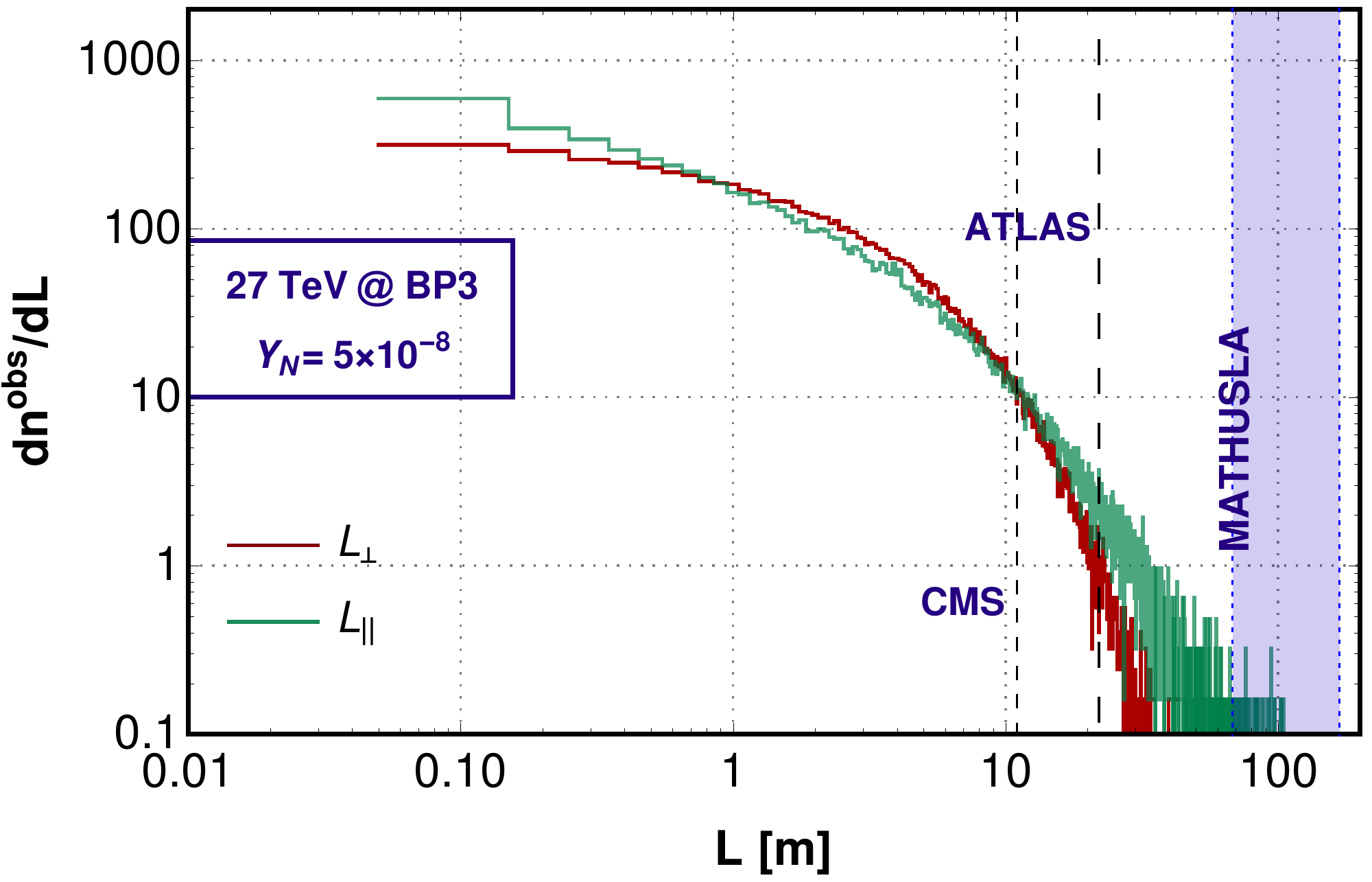}\label{}}\quad
			\subfigure[]{\includegraphics[width=0.35\linewidth,angle=-0]{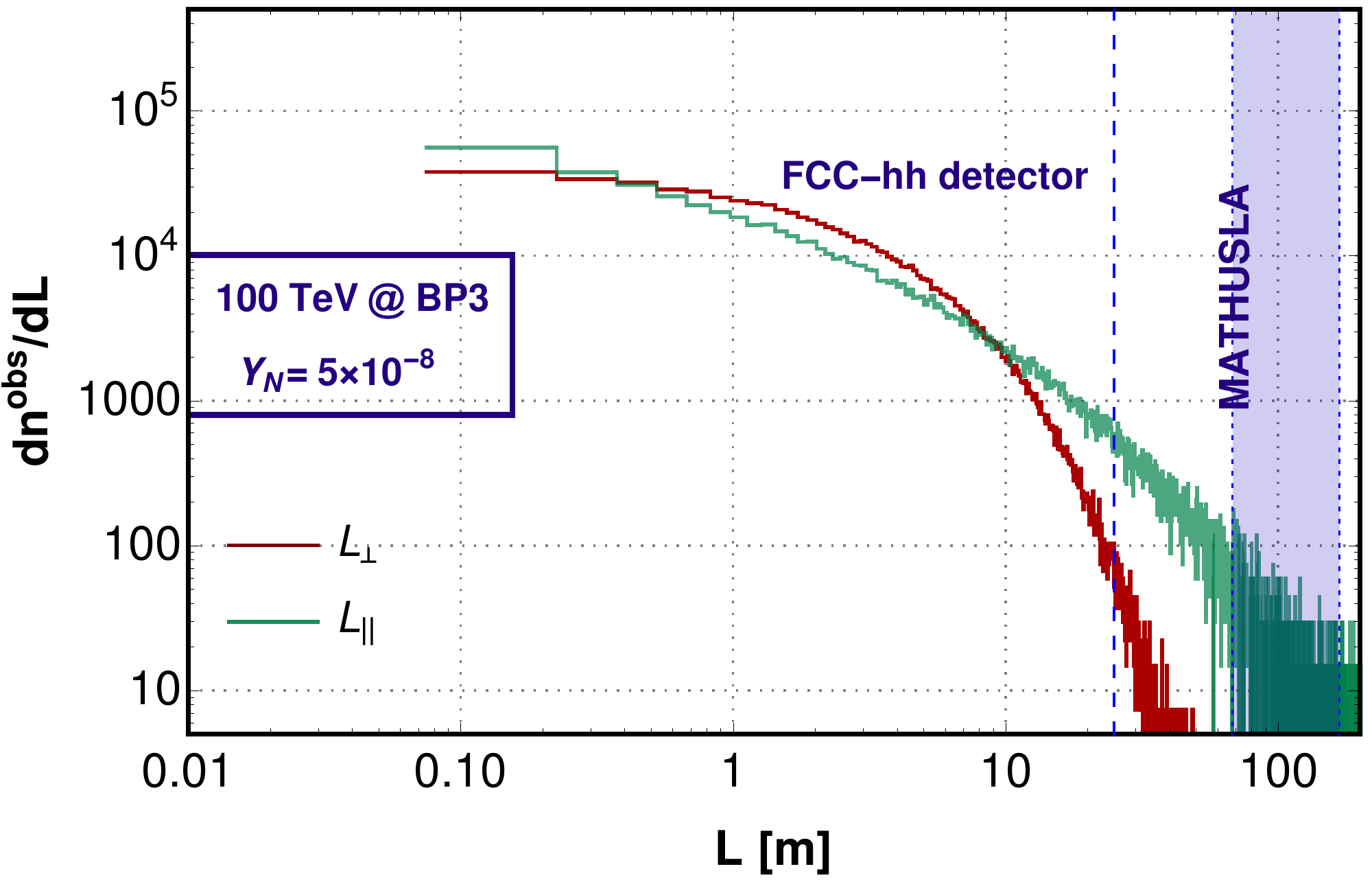}\label{}}}
		\hspace*{-1.0cm}
		\mbox{\subfigure[]{\includegraphics[width=0.35\linewidth,angle=-0]{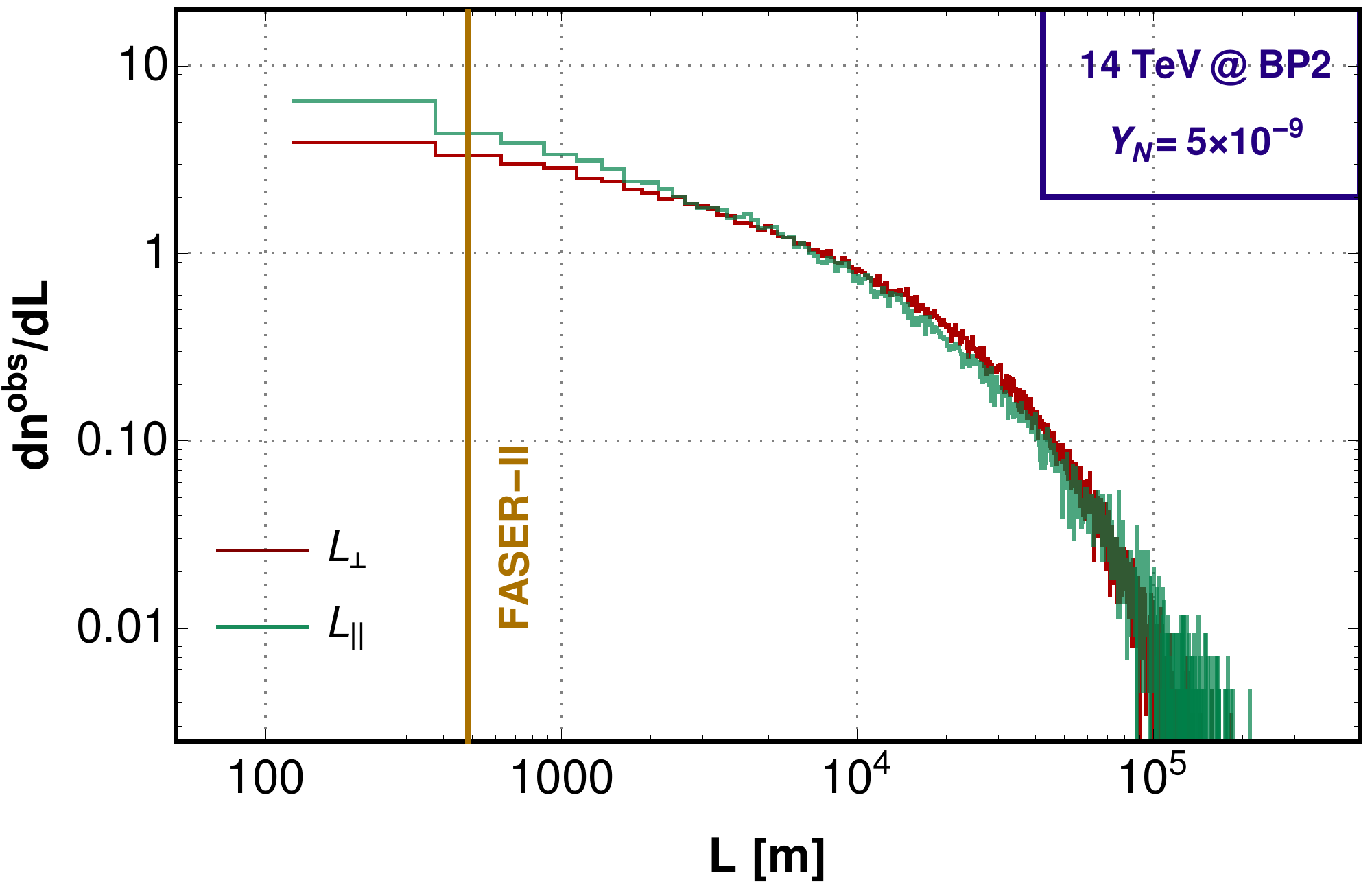}\label{}}\quad
			\subfigure[]{\includegraphics[width=0.35\linewidth,angle=-0]{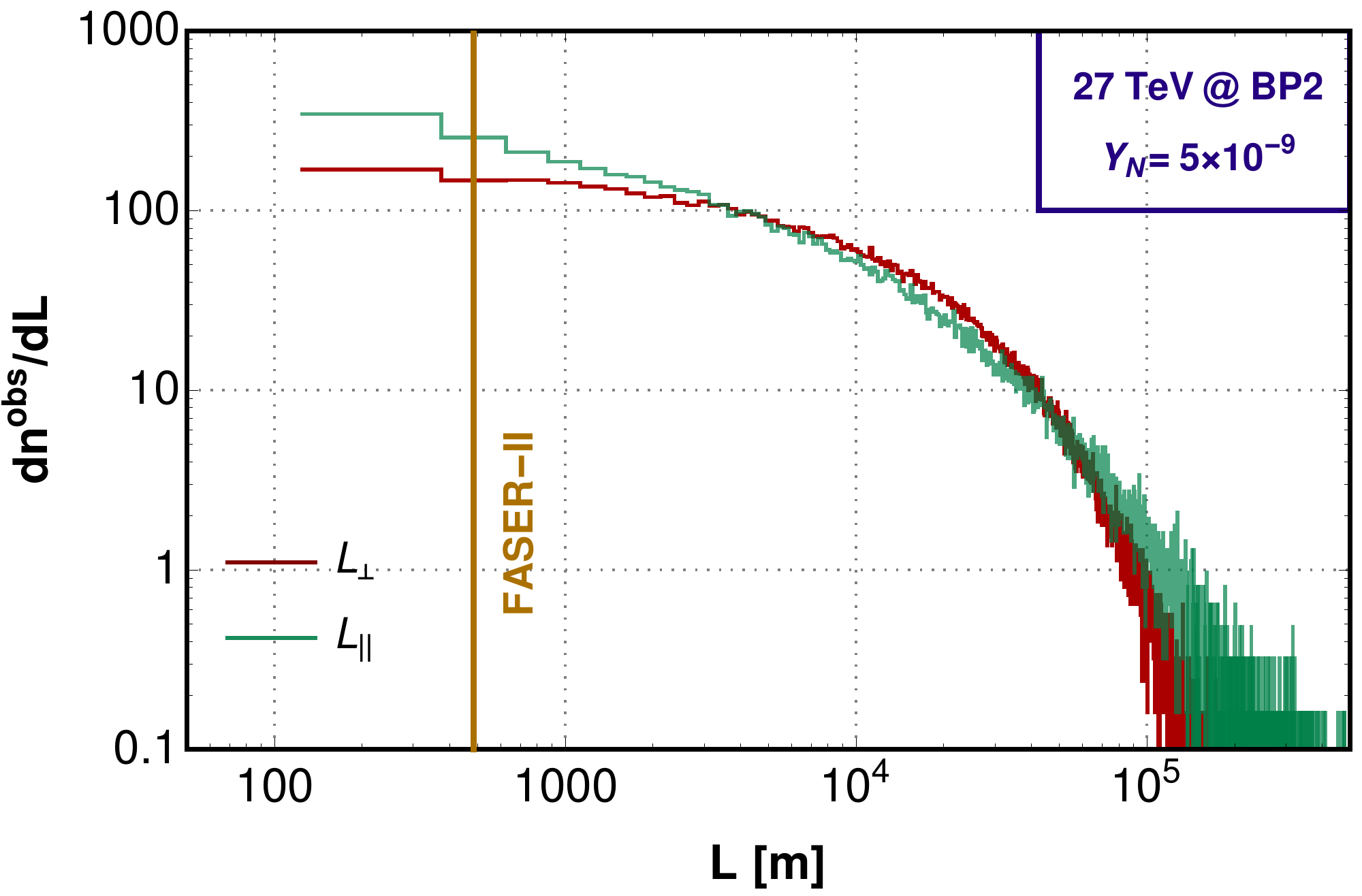}\label{}}\quad
			\subfigure[]{\includegraphics[width=0.35\linewidth,angle=-0]{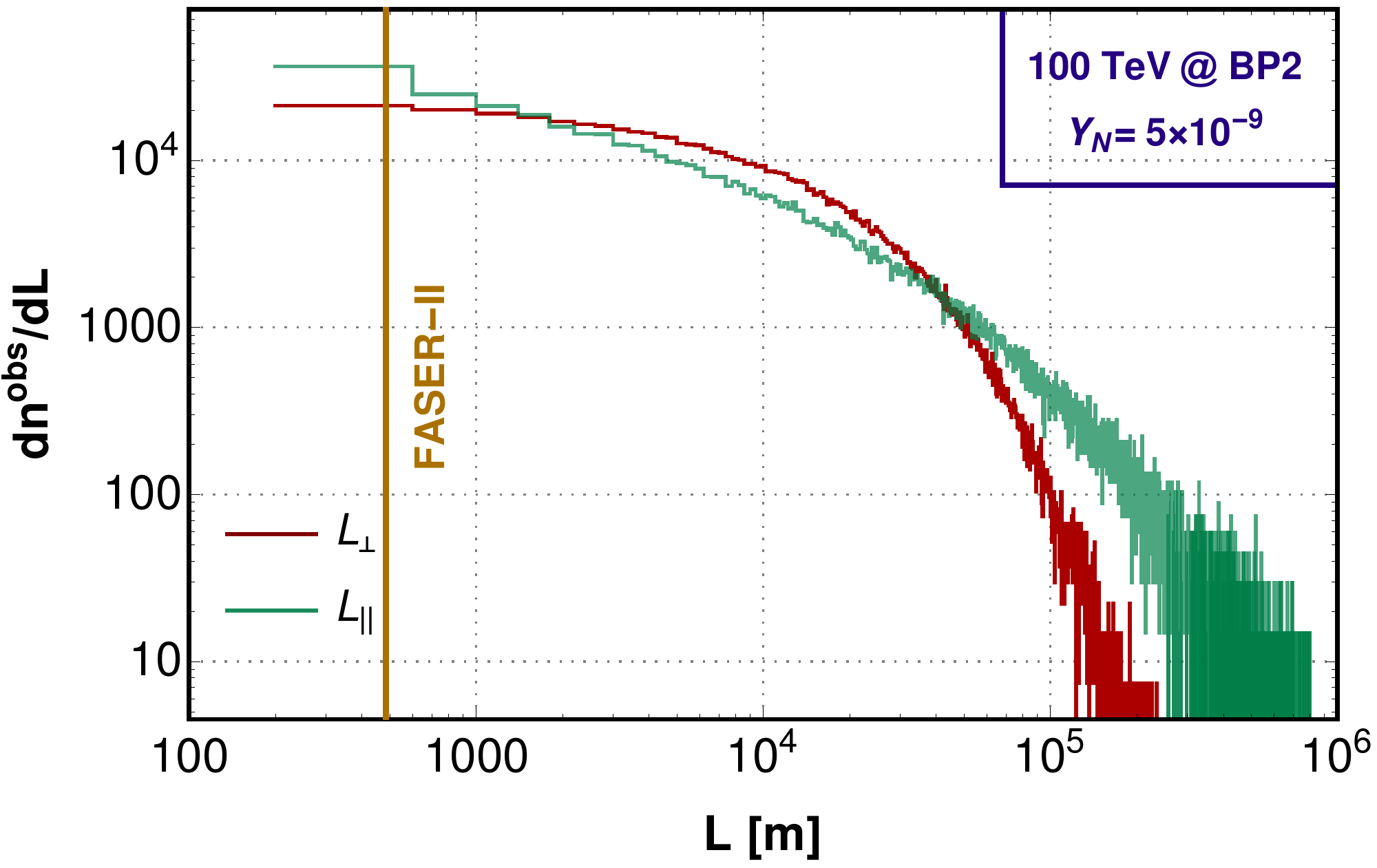}\label{}}}
			\hspace*{-1.0cm}
		\mbox{\subfigure[]{\includegraphics[width=0.35\linewidth,angle=-0]{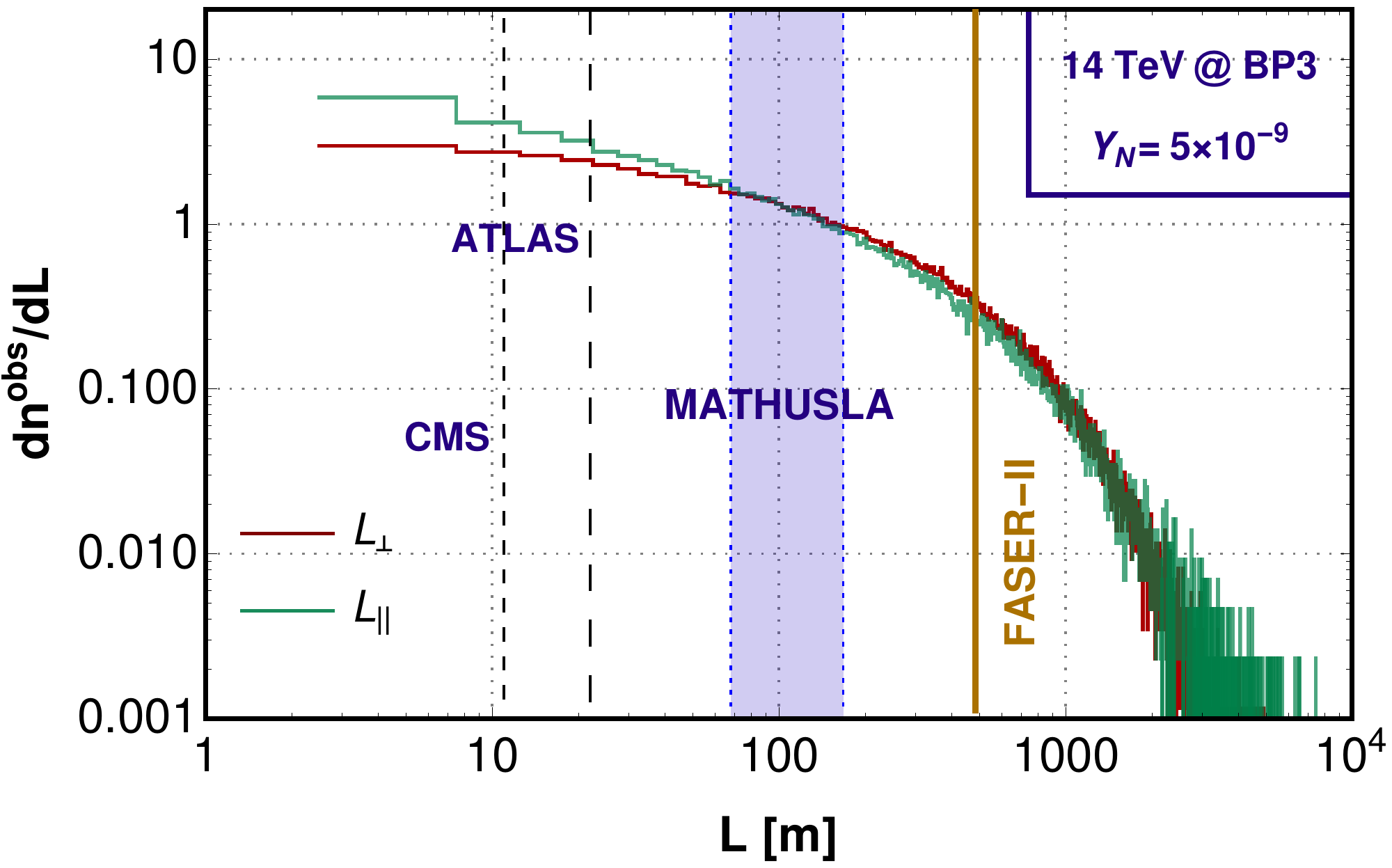}\label{}}\quad
			\subfigure[]{\includegraphics[width=0.35\linewidth,angle=-0]{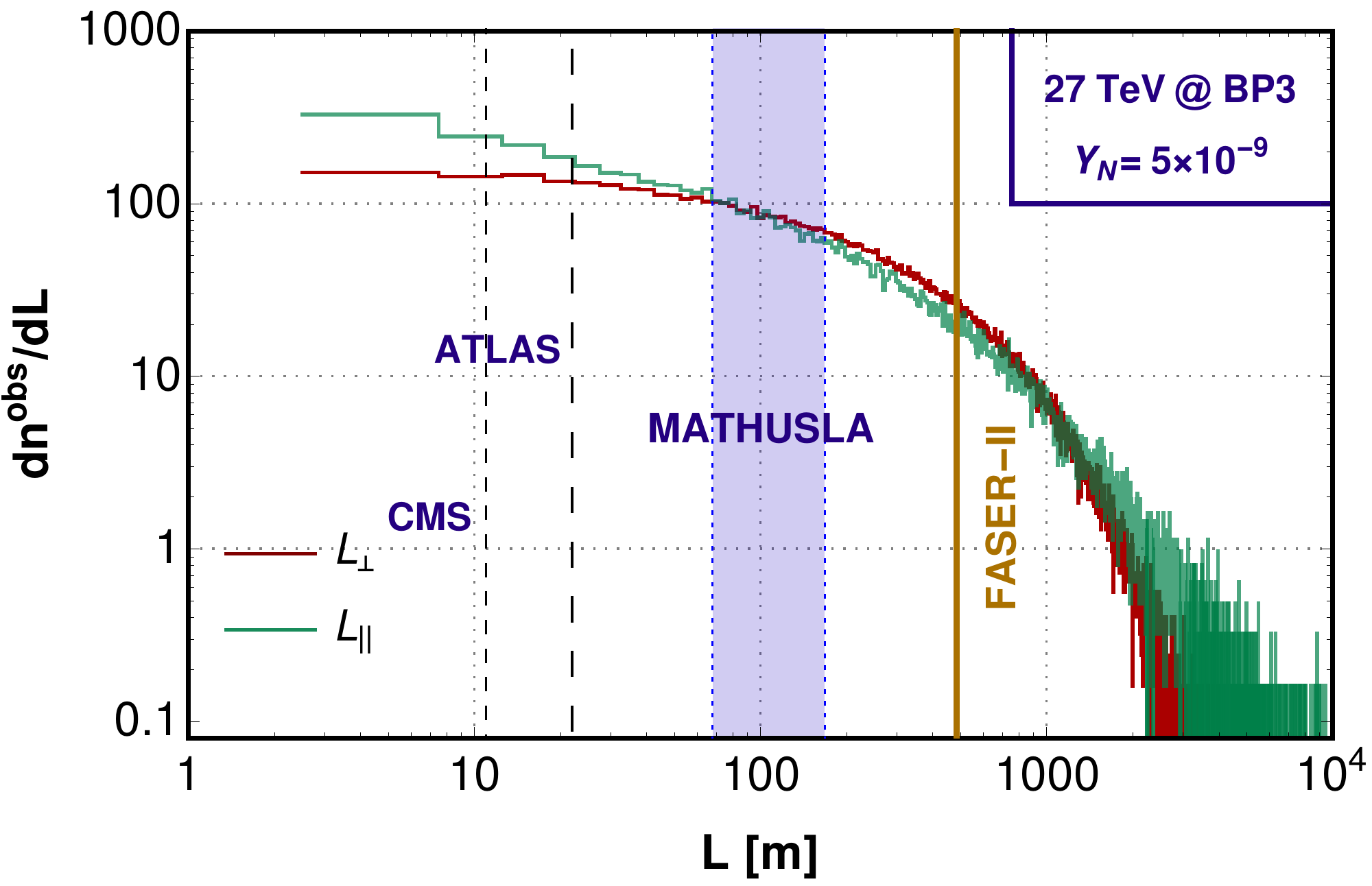}\label{}}\quad
			\subfigure[]{\includegraphics[width=0.35\linewidth,angle=-0]{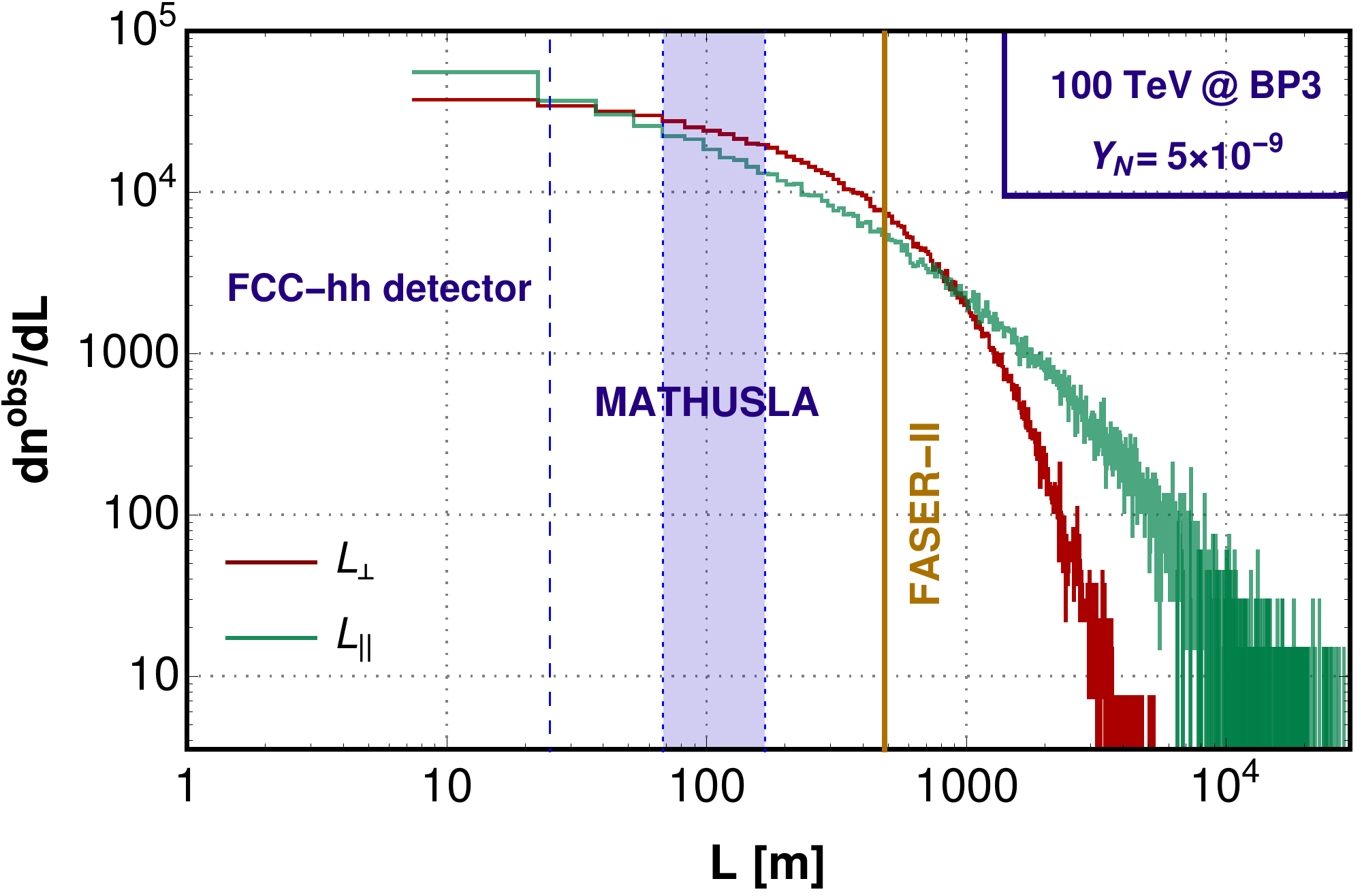}\label{}}}
		\caption{Displaced transverse ($L_\perp$ in red) and longitudinal ($L_{||}$ in green) decay length distributions for $N$, coming from the pair productions at the LHC with the centre of mass energies of 14\,TeV (a, d, g, j), 27\,TeV (b, e, h, k) and 100\,TeV (c, f, i, l) with the integrated luminosities of 3, 10, 30 ab$^{-1}$, respectively for BP2 (first, third rows) and BP3 (second, fourth rows). The choice of Yukawa couplings are $Y_N = 5 \times 10^{-8}$ (first, second rows) and $Y_N = 5 \times 10^{-9}$ (third, fourth rows).  The small-dashed and medium-dashed lines indicate the upper limits of CMS and ATLAS, respectively, whereas, the blue dashed line specify the upper limit of the proposed FCC-hh detector. The light blue band (68\,m$-$168\,m) denotes the MATHUSLA region and the golden yellow strip denotes the FASER-II region.}\label{DcyLT1gen}
	\end{center}
\end{figure*}
Repeating the previous analyses for this scenario, we obtain \autoref{DcyLT1gen}, which describes the differential distribution of  displaced decay lengths for the benchmark points at 14, 27 and 100 TeV centre of mass energies with the integrated luminosities of 3, 10, 30 ab$^{-1}$, for $Y_N= 5 \times 10^{-8}$ and $5\times 10^{-9}$.
\autoref{DcyLT1gen}(a, b, c) depict the distributions for BP2 with $Y_N= 5 \times 10^{-8}$ and it can be noticed that the displaced decay  lengths  can reach  to 10 km  for both transverse ($L_\perp$) and  longitudinal ($L_{||}$) ones.  The corresponding inclusive event numbers, that fall inside the MATHUSLA detector, are given in \autoref{MATHUSLATab1gen} (first row), at the centre of mass energies of 14, 27 and 100 TeV with the integrated luminosities of 3\,ab$^{-1}$, 10\,ab$^{-1}$ and 30\,ab$^{-1}$, respectively.
\autoref{DcyLT1gen}(d, e, f) present the  corresponding distributions for BP3 and since  the displaced decay lengths are mostly less than 100 m, at least for $L_\perp$, numbers are negligible. However, due to large longitudinal boost at 100 TeV, the corresponding distribution reaches MATHUSLA region and the number is also encouraging (\autoref{MATHUSLATab1gen} second row).

\begin{table*}[hbt]	
	\begin{center}
		\renewcommand{\arraystretch}{1.2}
		\begin{tabular}{|c|c|c|c|c|}
			\cline{2-5}
			\multicolumn{1}{c|}{} &  \multicolumn{4}{c|}{Events inside MATHUSLA}\\
			\cline{1-5}
			Yukawa & Benchmark &	\multicolumn{3}{c|}{Centre of mass energy} \\
			\cline{3-5}
			coupling & points & \multicolumn{1}{c|}{14\,TeV}&\multicolumn{1}{c|}{27\,TeV}&\multicolumn{1}{c|}{100\,TeV}\\
			\hline
			\multirow{2.0}{*}{$5\times 10^{-8}$} & BP2 & 1.4 & 81.2  & 6013.4  \\
			\cline{2-5}
			& BP3 & 0.0 & 1.7 & 534.2 \\
			\cline{1-5}
			\multirow{2.0}{*}{$5\times 10^{-9}$} & BP2 & 0.5  & 3.8 & 53.3  \\
			\cline{2-5}
			& BP3 & 1.6 & 113.1 & 8645.8 \\			
			\hline
		\end{tabular}
		\caption{The inclusive number of events inside the proposed detector MATHUSLA, for BP2 and BP3, with the Yukawa couplings of $5 \times 10^{-8}$, $5 \times 10^{-9}$, at the centre of mass energies of 14, 27, 100 TeV with the integrated luminosities of 3\,ab$^{-1}$, 10\,ab$^{-1}$ and 30\,ab$^{-1}$, respectively.} \label{MATHUSLATab1gen}
	\end{center}	
\end{table*}


\autoref{DcyLT1gen}(g, h, i) and \autoref{DcyLT1gen}(j, k, l) show the distributions for BP2 and BP3, respectively with $Y_N=5\times 10^{-9}$.  This  enhances the displacement, and thus we have more events for BP3 as can be read from  \autoref{MATHUSLATab1gen} (fourth row). For BP2, the displacements reaches up to $\sim 10^5$ m leaving very little events inside MATHUSLA range.  We also depict the FASER-II regions, but FASER-II prospect is not encouraging  due to very small range in this model.



\subsection{Number of displaced leptonic events}\label{result_SC2}
Let us now present the event numbers for the final states of displaced $4\ell,\, 3\ell +1j$ and $2\ell+2j$ at the center of mass energies of 14\,TeV, 27\,TeV and 100\,TeV with the integrated luminosities of $3\,\text{ab}^{-1}$, $10\,\text{ab}^{-1}$ and $30\,\text{ab}^{-1}$, respectively.  \autoref{FS1-1gen} shows the numbers of $4\ell$ events.  It is evident that  CMS, ATLAS and the proposed detector for FCC-hh can have healthy event numbers only for BP3 at higher centre of mass energies of 27 and 100 TeV. MATHUSLA registers some good number of events for BP3 with $Y_N=5\times 10^{-9}$ at 100 TeV. For BP2 the event numbers remain low at all three detectors as compared to BP3.
\begin{table}[h]	
	\begin{center}
		\hspace*{-1.0cm}
		\renewcommand{\arraystretch}{1.2}
		\begin{tabular}{ |c||c|c|c|c|c| }
			\hline
			{\multirow{2}{*}{\diagbox[width=4.5cm]{$4\ell$}{Displaced decay}}}&
			Yukawa & Benchmark &\multicolumn{3}{c|}{Centre of mass energy}\\
			\cline{4-6}
			& couplings ($Y_N$) & points & 14\,TeV & 27\,TeV & 100\,TeV  \\
			\hline
			\multirow{4}{*}{CMS} & \multirow{2.0}{*}{$5\times 10^{-8}$} & BP2 & 0.0  &  2.7  & $-$ \\
			\cline{3-6}
		    & & BP3 & 1.8  & 129.4  & $-$ \\
			\cline{2-6}
			& \multirow{2.0}{*}{$5\times 10^{-9}$} & BP2 & 0.0  & 0.0  & $-$  \\
			\cline{3-6}	
			\multirow{0.25}{*}{}& & BP3 & 0.1  & 7.3  & $-$ \\	
			\hline 
			\multirow{2.5}{*}{ATLAS} & \multirow{2.0}{*}{$5\times 10^{-8}$} & BP2 & 0.1  & 3.6   & 292.32  \\
			\cline{3-6}
			\multirow{2.5}{*}{\&}& & BP3 & 1.9  & 133.7  & 12566.8  \\
			\cline{2-6}
			\multirow{2.5}{*}{FCC-hh reference detector}& \multirow{2.0}{*}{$5\times 10^{-9}$} & BP2 & 0.0  & 0.0 & 4.1 \\
			\cline{3-6}
			\multirow{.5}{*}{}& & BP3 & 0.2  & 12.9  & 1082.2  \\	
			\hline 
			\multirow{4}{*}{MATHUSLA} & \multirow{2.0}{*}{$5\times 10^{-8}$} & BP2 &  0.0  & 1.5  & 90.5  \\
			\cline{3-6}
			& & BP3 & 0.0  & 0.3  & 12.6  \\
			\cline{2-6}
			& \multirow{2.0}{*}{$5\times 10^{-9}$} & BP2 & 0.0  & 0.1  & 3.2 \\	
			\cline{3-6}
			\multirow{0.25}{*}{}& & BP3 & 0.0  & 3.0  &  209.1 \\
			\hline
			
		\end{tabular}
		\caption{Number of events in  $4\ell $ final state for the benchmark points, corresponds to the Yukawa couplings ($Y_N=$) $5\times 10^{-8}$ and $5\times 10^{-9}$, with the center of mass energies of 14\,TeV, 27\,TeV and 100\,TeV at the integrated luminosities of $3\,\text{ab}^{-1}$, $10\,\text{ab}^{-1}$ and $30\,\text{ab}^{-1}$, respectively. The numbers are given separately for CMS, ATLAS, FCC-hh reference detector (for 100 TeV) and MATHUSLA.}  \label{FS1-1gen}
	\end{center}	
\end{table}

Next, we indulge in the final state of $3\ell + (\geq 1j)$ shown in \autoref{FS2-1gen}.  Similarly to the $4\ell$ final state, the events numbers are healthy only for BP3 at 27 and 100\,TeV centre of mass energies. BP2 looks promising only for $Y_N=5\times 10^{-8}$ at 100 TeV.

\begin{table}[h]	
	\begin{center}
		\hspace*{-1.0cm}
		\renewcommand{\arraystretch}{1.15}
		\begin{tabular}{ |c||c|c|c|c|c| }
			\hline
			{\multirow{2}{*}{\diagbox[width=4.5cm]{$3\ell + (\geq 1j)$}{Displaced decay}}}&
			Yukawa & Benchmark &\multicolumn{3}{c|}{Centre of mass energy}\\
			\cline{4-6}
			& couplings ($Y_N$) & points & 14\,TeV & 27\,TeV & 100\,TeV  \\
			\hline
			\multirow{4.0}{*}{CMS} & \multirow{2.0}{*}{$5\times 10^{-8}$} & BP2 & 0.2  & 5.7  & $-$  \\
			\cline{3-6}
			& & BP3 & 3.2 & 195.3  & $-$  \\
			\cline{2-6}
			& \multirow{2.0}{*}{$5\times 10^{-9}$} & BP2 & 0.0  & 0.1  & $-$ \\
			\cline{3-6}	
			\multirow{0.25}{*}{}& & BP3 & 0.4  & 14.2  & $-$  \\	
			\hline 
			\multirow{2.5}{*}{ATLAS} & \multirow{2.0}{*}{$5\times 10^{-8}$} & BP2 & 0.3  &  8.6  &  620.1  \\
			\cline{3-6}
			\multirow{2.5}{*}{\&} & & BP3 & 3.3  & 199.3  & 17121.7  \\
			\cline{2-6}
			\multirow{2.5}{*}{FCC-hh reference detector} & \multirow{2.0}{*}{$5\times 10^{-9}$} & BP2 & 0.0  & 0.2  & 25.2 \\
			\cline{3-6}
			\multirow{2.0}{*}{} & & BP3 & 0.6  & 23.1  & 1691.6  \\	
			\hline
			\multirow{4}{*}{MATHUSLA} & \multirow{2.0}{*}{$5\times 10^{-8}$} & BP2 & 0.0   & 1.7  &  134.1 \\
			\cline{3-6}
			& & BP3 &  0.0  & 0.5  & 33.6  \\
			\cline{2-6}
			& \multirow{2.0}{*}{$5\times 10^{-9}$} & BP2 & 0.0 & 0.3 & 16.8  \\	
			\cline{3-6}
			\multirow{0.25}{*}{}& & BP3 & 0.1  & 4.4 & 306.6 \\
			\hline
			
		\end{tabular}
		\caption{Number of events in  $3\ell + (\geq 1j)$ final state for the benchmark points, corresponds to the Yukawa couplings $Y_N=5\times 10^{-8}$ and $5\times 10^{-9}$, with the center of mass energies of 14\,TeV, 27\,TeV and 100\,TeV at the integrated luminosities of $3\,\text{ab}^{-1}$, $10\,\text{ab}^{-1}$ and $30\,\text{ab}^{-1}$, respectively. The numbers are given separately for CMS, ATLAS, FCC-hh reference detector (for 100 TeV) and MATHUSLA.}  \label{FS2-1gen}
	\end{center}	
\end{table}
\begin{table}[h]	
	\begin{center}
		\hspace*{-1.0cm}
		\renewcommand{\arraystretch}{1.15}
		\begin{tabular}{ |c||c|c|c|c|c| }
			\hline
			{\multirow{2}{*}{\diagbox[width=4.5cm]{$2\ell + (\geq 2j)$}{Displaced decay}}}&
			Yukawa & Benchmark &\multicolumn{3}{c|}{Centre of mass energy}\\
			\cline{4-6}
			& couplings ($Y_N$) & points & 14\,TeV & 27\,TeV & 100\,TeV  \\
			\hline
			\multirow{4.0}{*}{CMS} & \multirow{2.0}{*}{$5\times 10^{-8}$} & BP2 & 0.6  &  20.7 & $-$ \\
			\cline{3-6}
			& & BP3 & 5.1 & 303.0 & $-$ \\
			\cline{2-6}
			& \multirow{2.0}{*}{$5\times 10^{-9}$} & BP2 & 0.0 & 0.8 & $-$ \\
			\cline{3-6}	
			\multirow{0.25}{*}{}& & BP3 & 0.7 & 25.3 & $-$ \\	
			\hline 
			\multirow{2.5}{*}{ATLAS} & \multirow{2.0}{*}{$5\times 10^{-8}$} & BP2 & 0.9  & 33.8  & 2546.2  \\
			\cline{3-6}
			\multirow{2.5}{*}{\&} & & BP3 & 5.2 & 309.3 & 28186.1 \\
			\cline{2-6}
			\multirow{2.5}{*}{FCC-hh reference detector} & \multirow{2.0}{*}{$5\times 10^{-9}$} & BP2 & 0.0 & 1.8 & 232.2 \\
			\cline{3-6}
			\multirow{2.0}{*}{} & & BP3 & 1.0 & 40.4 & 2901.5 \\	
			\hline 
			\multirow{4}{*}{MATHUSLA} & \multirow{2.0}{*}{$5\times 10^{-8}$} & BP2 & 0.2  & 9.2 & 795.6 \\
			\cline{3-6}
			& & BP3 & 0.0 & 0.8 & 49.5 \\
			\cline{2-6}
			& \multirow{2.0}{*}{$5\times 10^{-9}$} & BP2 & 0.0 & 1.2 & 62.7 \\	
			\cline{3-6}
			\multirow{0.25}{*}{}& & BP3 & 0.1 & 6.7 & 553.3 \\
			\hline
			
		\end{tabular}
		\caption{Number of events in  $2\ell + (\geq 2j)$ final state for the benchmark points, corresponds to the Yukawa couplings ($Y_N=$) $5\times 10^{-8}$ and $5\times 10^{-9}$, with the center of mass energies of 14\,TeV, 27\,TeV and 100\,TeV at the integrated luminosities of $3\,\text{ab}^{-1}$, $10\,\text{ab}^{-1}$ and $30\,\text{ab}^{-1}$, respectively. The numbers are given separately for CMS, ATLAS, FCC-hh reference detector (for 100 TeV) and MATHUSLA.}  \label{FS3-1gen}
	\end{center}	
\end{table}

Finally the results for the final state of $2\ell + (\geq 2j)$ are presented in  \autoref{FS3-1gen}. Overall event numbers are comparatively healthy for both the benchmark points.  However, MATHUSLA is favorable to BP2 with $Y_N=5\times 10^{-8}$ and BP3 with $Y_N=5\times 10^{-9}$ for 100 TeV centre of mass energy.

\subsection{Sensitivity regions at different energies}\label{region_SC2}
\begin{figure}[hbt]
	\begin{center}
		\hspace*{-0.5cm}
		\mbox{\subfigure[]{\includegraphics[width=0.35\linewidth,angle=0]{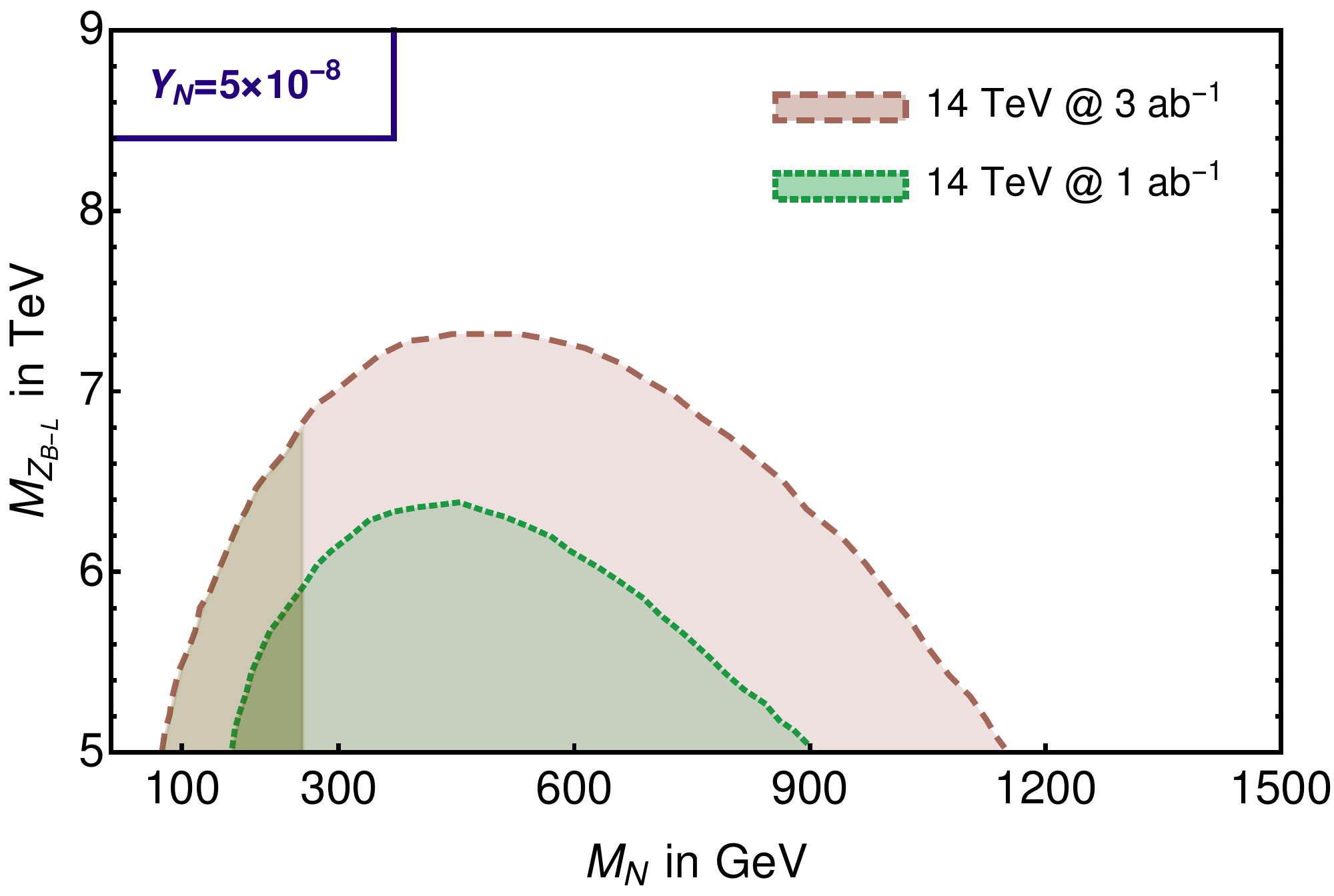}\label{14dcyl}}\quad
		\subfigure[]{\includegraphics[width=0.35\linewidth,angle=0]{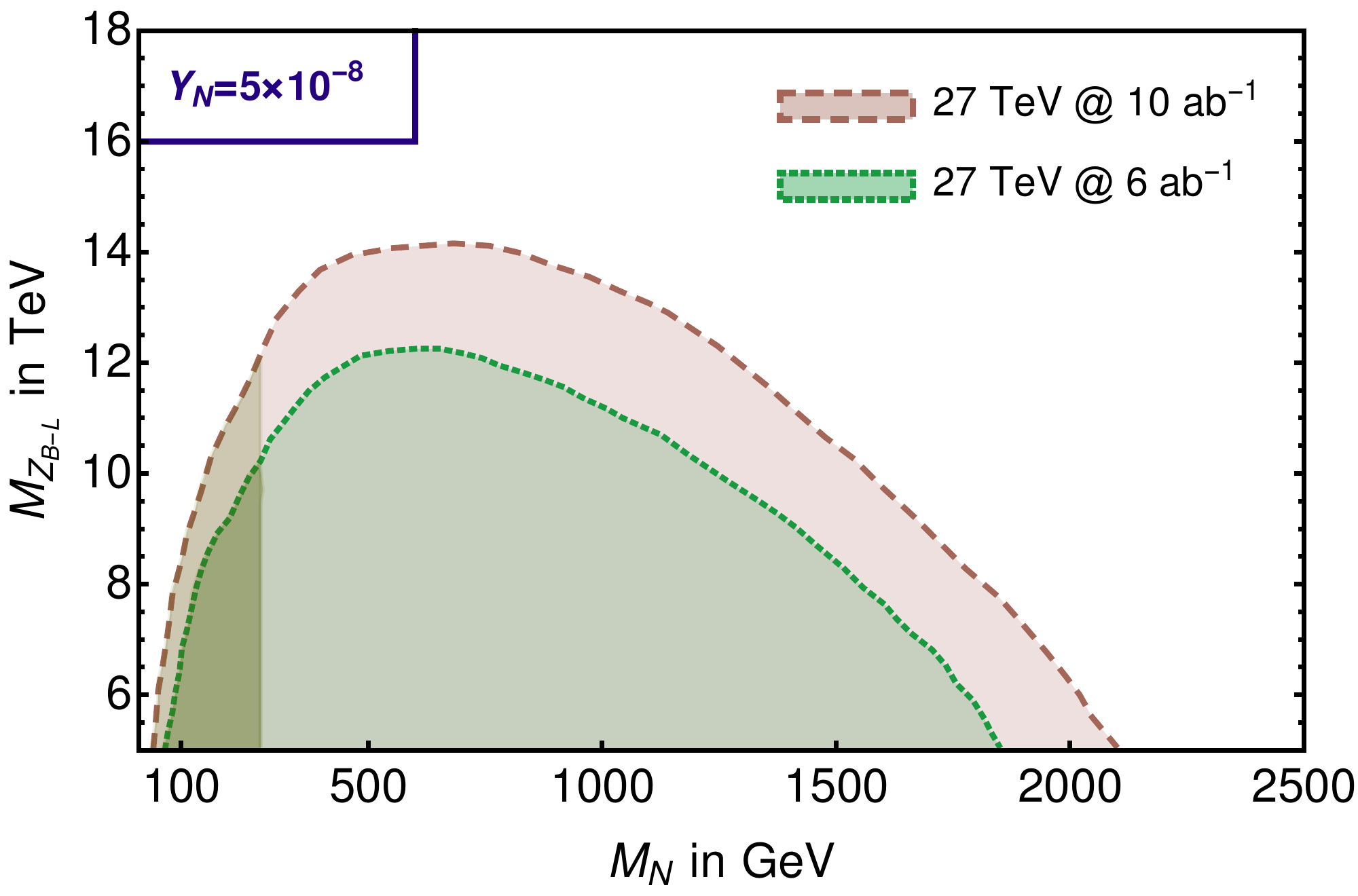}\label{14dcyl}}\quad
		\subfigure[]{\includegraphics[width=0.335\linewidth,angle=0]{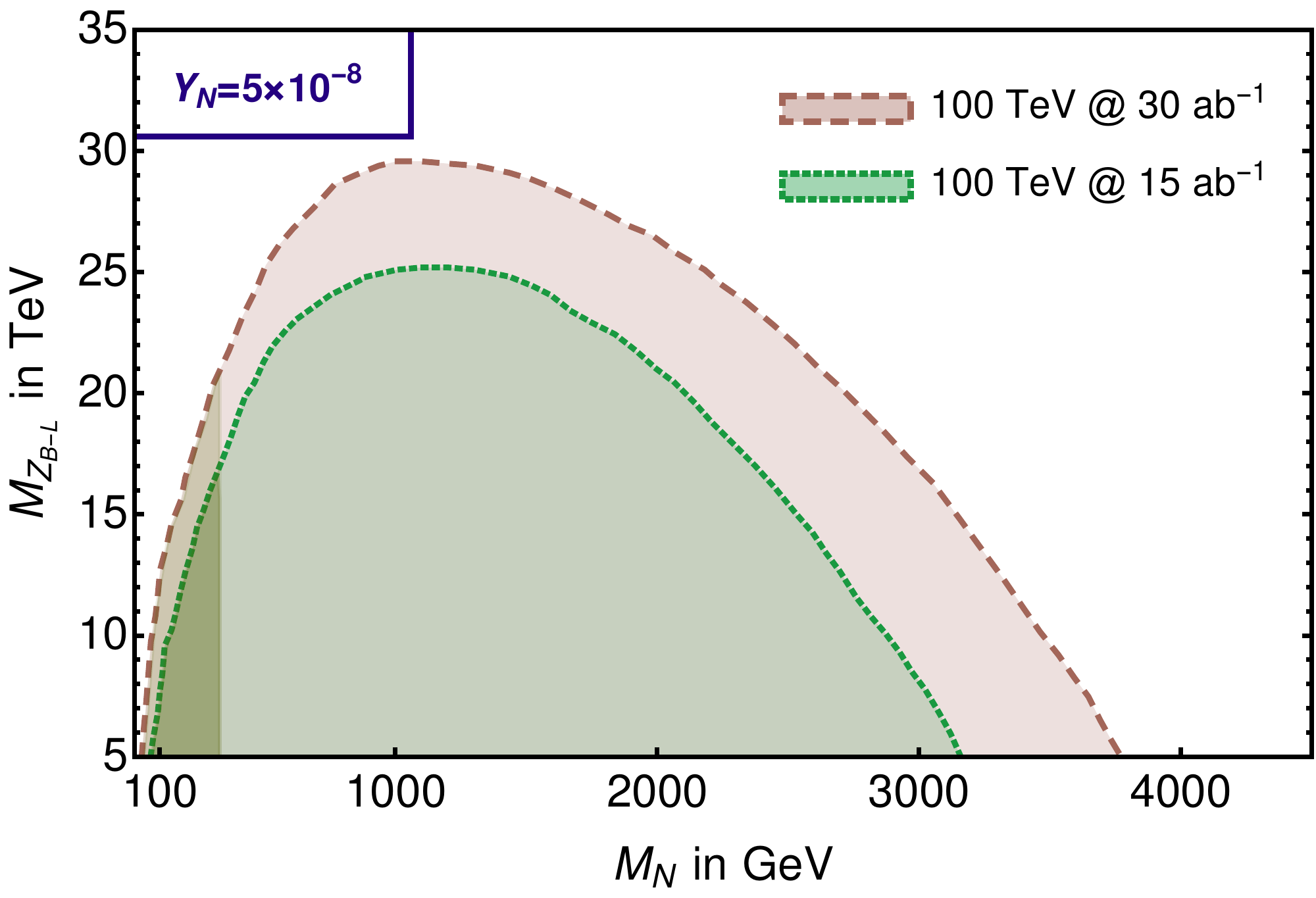}\label{14dcyl}}}
		\hspace*{-0.5cm}
		\mbox{\subfigure[]{\includegraphics[width=0.35\linewidth,angle=0]{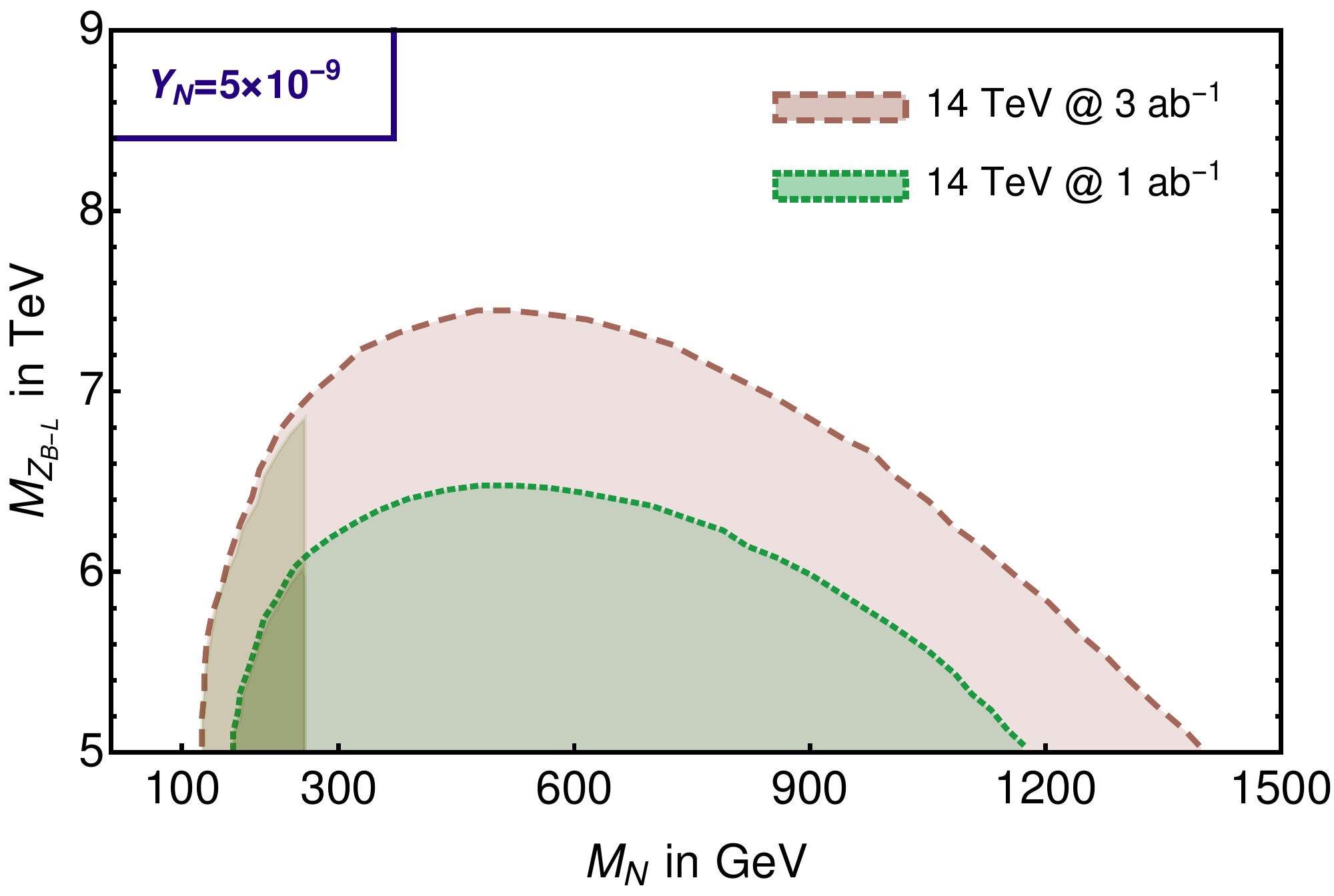}\label{14dcyl}}\quad
		\subfigure[]{\includegraphics[width=0.34\linewidth,angle=0]{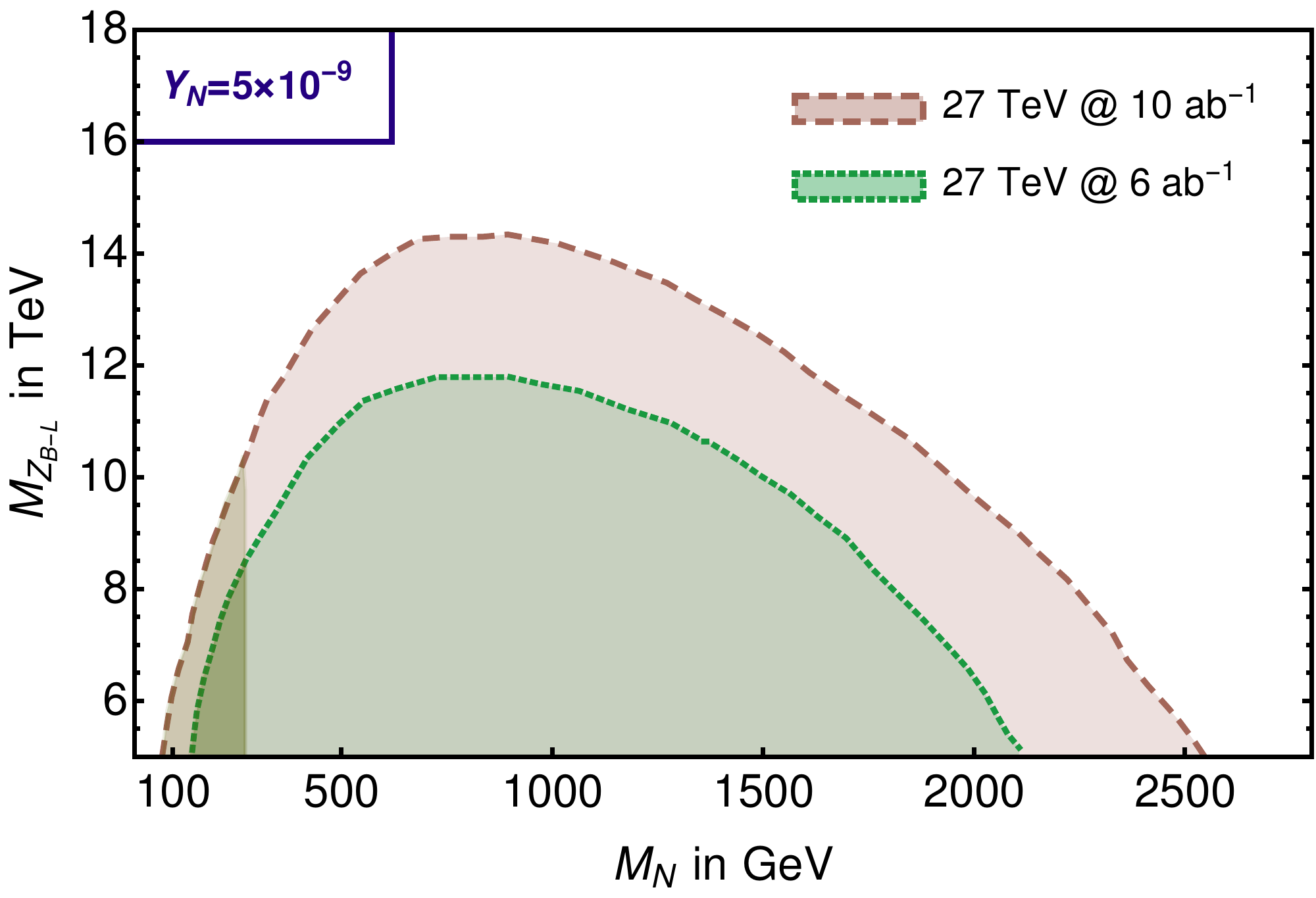}\label{14dcyl}}\quad
		\subfigure[]{\includegraphics[width=0.35\linewidth,angle=0]{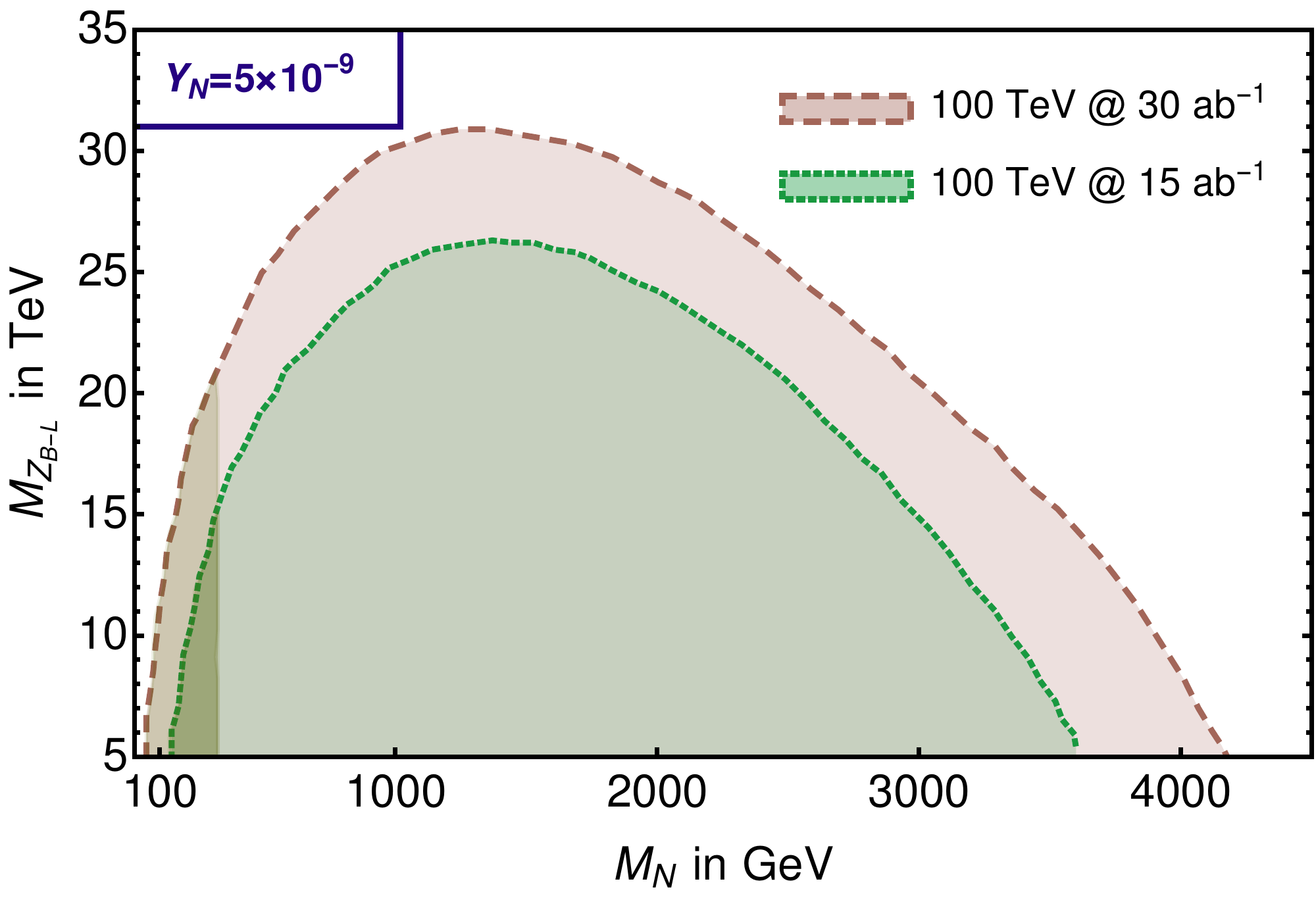}\label{14dcyl}}}
		\caption{Limits obtained via the displaced decays of RHNs to $2\ell + 2j$ final state (dark shaded region), $2\ell + 4j$ final state (light shaded region) and it is presented in $M_{Z_{B-L}}$ versus $M_N$ plane at $95\%$ CL. The first and second panel present the probable regions  for the Yukawa couplings $5\times 10^{-8}$ and $5\times 10^{-9}$, respectively. The shaded regions can be probed at any of the detectors CMS, ATLAS and MATHUSLA for  14\,TeV\,(a, d), 27\,TeV (b, e) centre of mass energies, and at either of the FCC-hh reference detector and MATHUSLA for 100\,TeV (c, f) centre of mass energy.}\label{ReachMzMn}
	\end{center}
\end{figure}

In the scenario-2, only one RHN having a very small Yukawa coupling is responsible for
the displaced decay phenomenology.  The other two can have relatively larger Yukawa couplings while reproducing the observed neutrino masses and mixing \cite{Sen:2021fha}. \autoref{ReachMzMn} describes the reaches in the $M_N-M_{Z_{B-L}}$ plane for three different centre of mass energies 14, 27 and 100 TeV. The colour conventions  are same  as in \autoref{ReachUPMNS}, however  we have chosen $Y_N=5\times 10^{-8}$ for the first row and  $Y_N=5\times 10^{-9}$  for the second row, respectively.  It can be seen that the reach  for  $M_N$ has increased substantially as compared to \autoref{ReachUPMNS} due to the smaller Yukawa couplings.  

For  $Y_N=5 \times 10^{-8}$ (in \autoref{ReachMzMn} first row) $M_N$ can be probed up to 1.15 TeV,  while $Z_{B-L}$  reach can  be  up to  7.3 TeV at 14 TeV centre of  mass energy. At 100 TeV, these are enhanced to 3.75 TeV, and 30 TeV, respectively. The reach on the lower end of $M_N$ is 10\,GeV, which is a little high as compare to \autoref{ReachUPMNS}.
The results for $Y_N=5 \times 10^{-9}$ are shown in the second row of  \autoref{ReachMzMn}. The choice of lower Yukawa pushed the probable regions towards the right side, i.e. for higher $M_N$. The maximum values that can be explored  are $M_N=1.42\, (4.18)$ TeV and $M_{Z_{B-L}}=7.4\, (30.5)$ TeV  at  14 (100) TeV centre of mass energy.

\begin{figure}[hbt]
	\begin{center}
		\hspace*{-1.0cm}
		\mbox{\subfigure[]{\includegraphics[width=0.35\linewidth,angle=0]{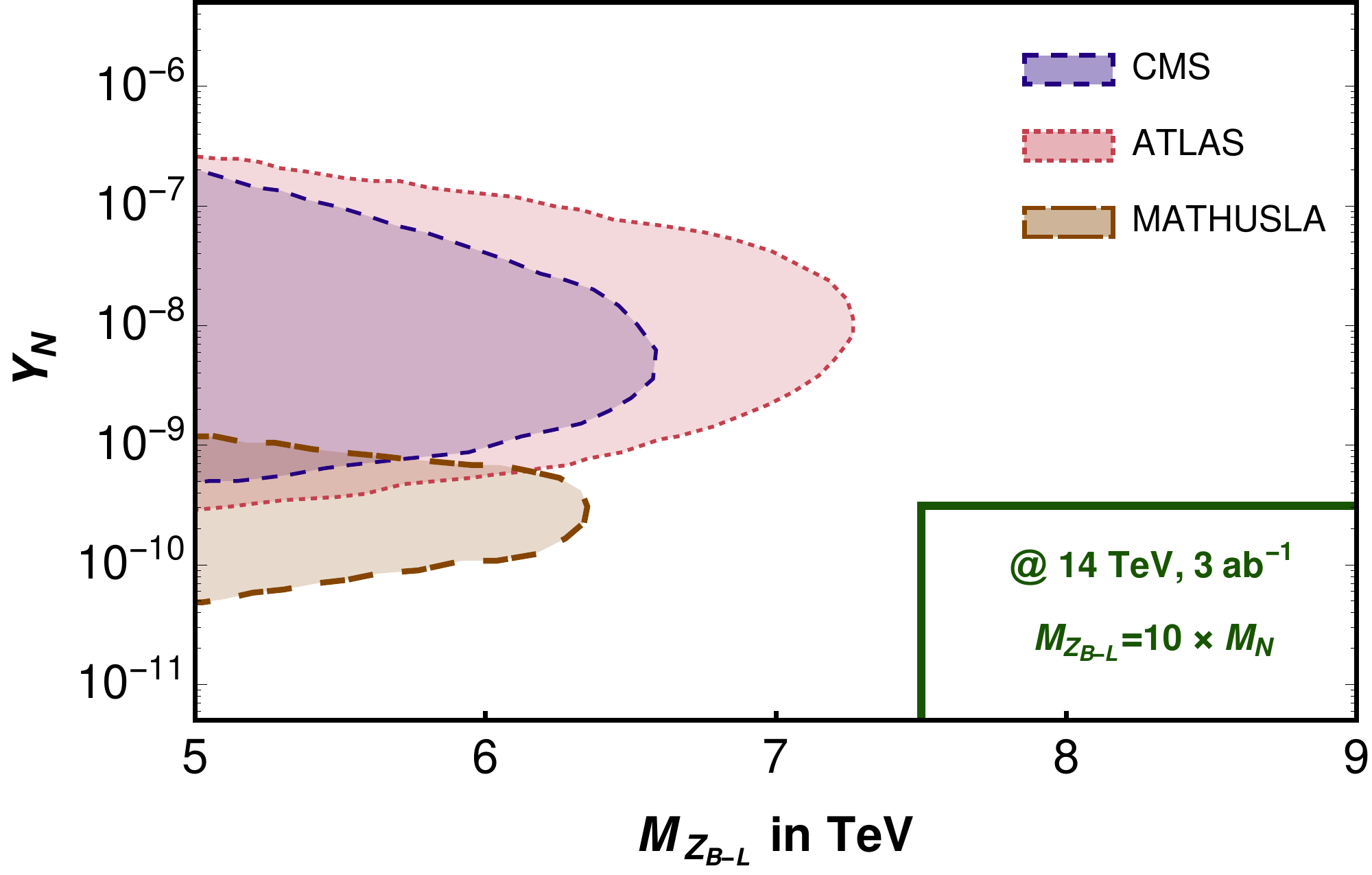}\label{14dcyl}}\quad
		\subfigure[]{\includegraphics[width=0.35\linewidth,angle=0]{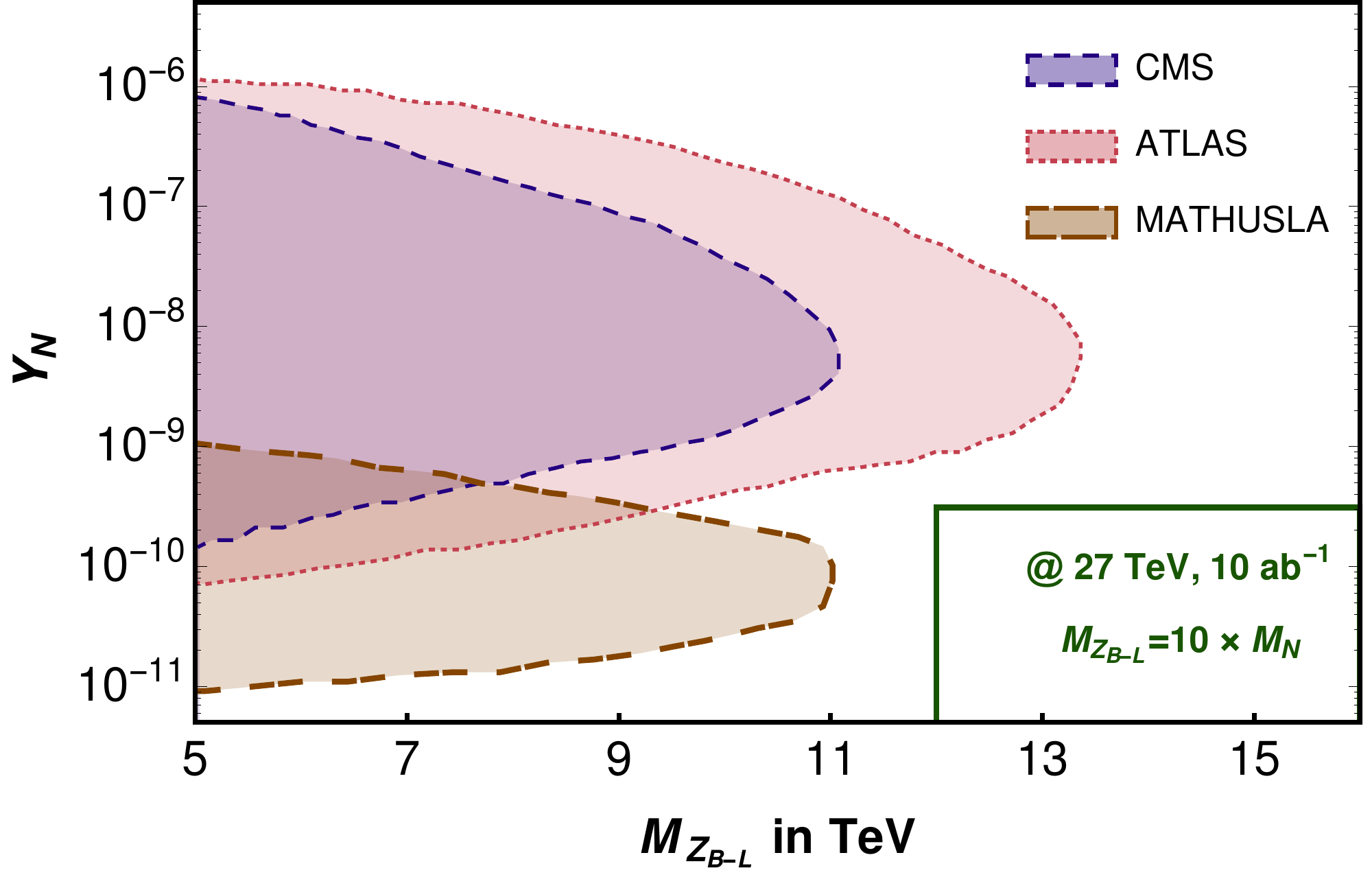}\label{30dcyl}}\quad
		\subfigure[]{\includegraphics[width=0.35\linewidth,angle=0]{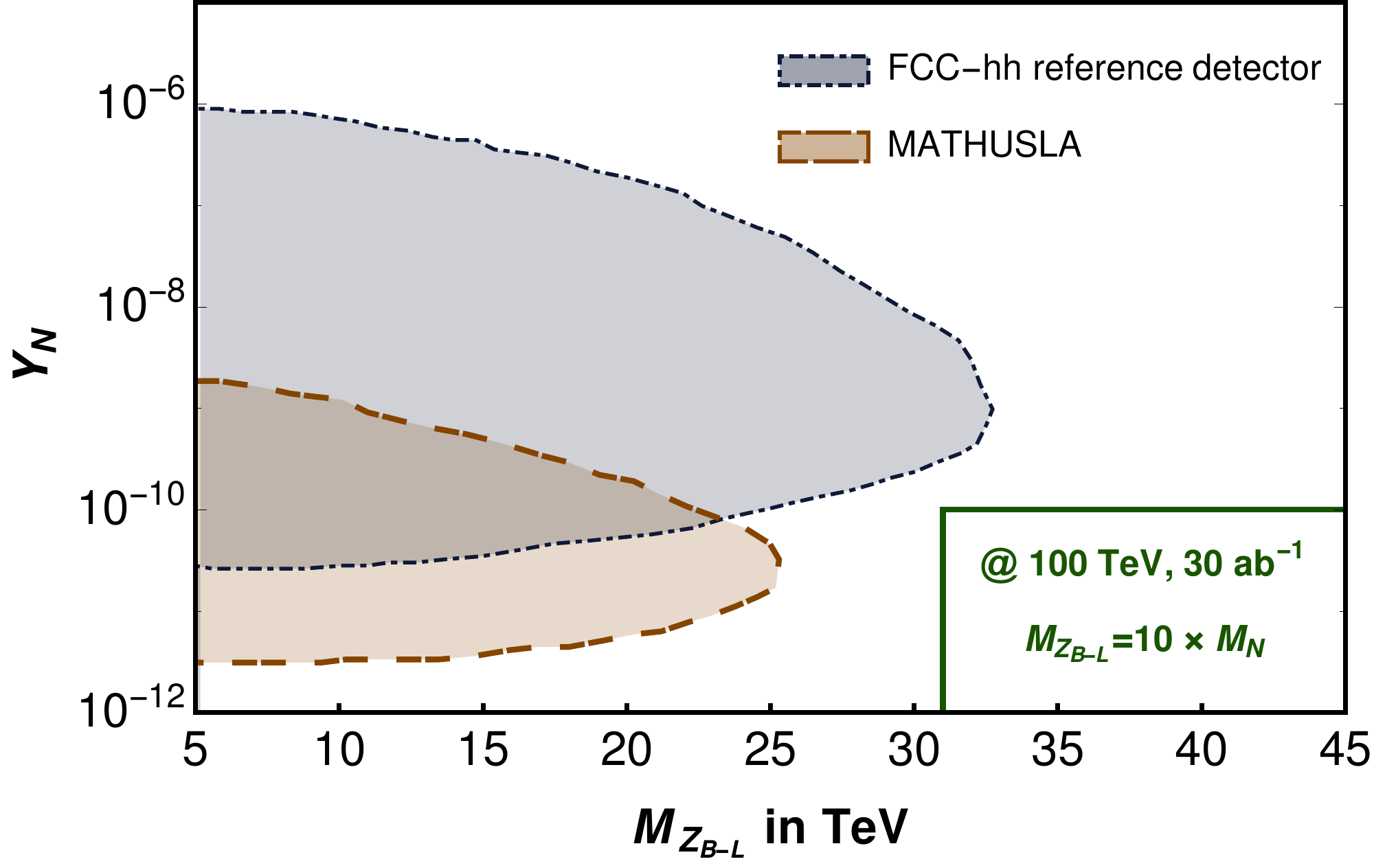}\label{30dcyl}}}        		
		\caption{Limits obtained via the displaced decays of RHNs to $2\ell + 4j$ final state and is presented in $Y_N$ versus $M_{Z_{B-L}}$  plane at $95\%$ CL, where $M_{Z_{B-L}}=10 M_N$. The probable regions are shown for different centre of mass energies, i.e. 14\,TeV, 27\,TeV and 100\,TeV with the integrated luminosities of 3, 10 and 30 ab$^{-1}$, respectively. The purple, pink, grey and light brown colours depict the reaches for CMS, ATLAS, FCC-hh reference detector and MATHUSLA, respectively.}\label{ReachYnMz}
	\end{center}
\end{figure}
As the Yukawa coupling $Y_N$ can be arbitrarily small in the scenario-2, it is also interesting to see how small $Y_N$ can be probed by the displaced events. In \autoref{ReachYnMz} we show the regions in $Y_N -M_{Z_{B-L}}$ plane, that can be probed by different detectors at the LHC/FCC for centre of mass energies of 14, 27 and 100 TeV with 3, 10 and 30 ab$^{-1}$ luminosities, respectively. For this we choose the mass ratio: $M_{Z_{B-L}}=10 M_N$, and consider the  $2\ell +4j$ final state. The purple, pink and brown colours depict  the regions that can be probed by CMS, ATLAS and MATHUSLA via the displaced decays, whereas, the grey regions represent the same for the proposed FCC-hh detector. Such detectability via different detectors have overlapping regions also as can be seen from \autoref{ReachYnMz}. MATHUSLA is $\mathcal{O}$(100\,m) from the interaction point and more sensitive for lower $Y_N$, as decay length varies inversely with $Y_N$.
In the figure, one can see that $Y_N \sim 3 \times 10^{-12}-  10^{-6}$ can be probed  combining the proposed FCC-hh detector and MATHUSLA for 100 TeV centre of mass energy via displaced decay. The corresponding reach for $M_{Z_{B-L}}$ is around 7.2\,(33) TeV for 14\,(100) TeV centre of mass energies considering the chosen mass hierarchy and the mentioned final state.

\begin{figure}[hbt]
	\begin{center}
		\hspace*{-1.0cm}
		\mbox{\subfigure[]{\includegraphics[width=0.35\linewidth,angle=0]{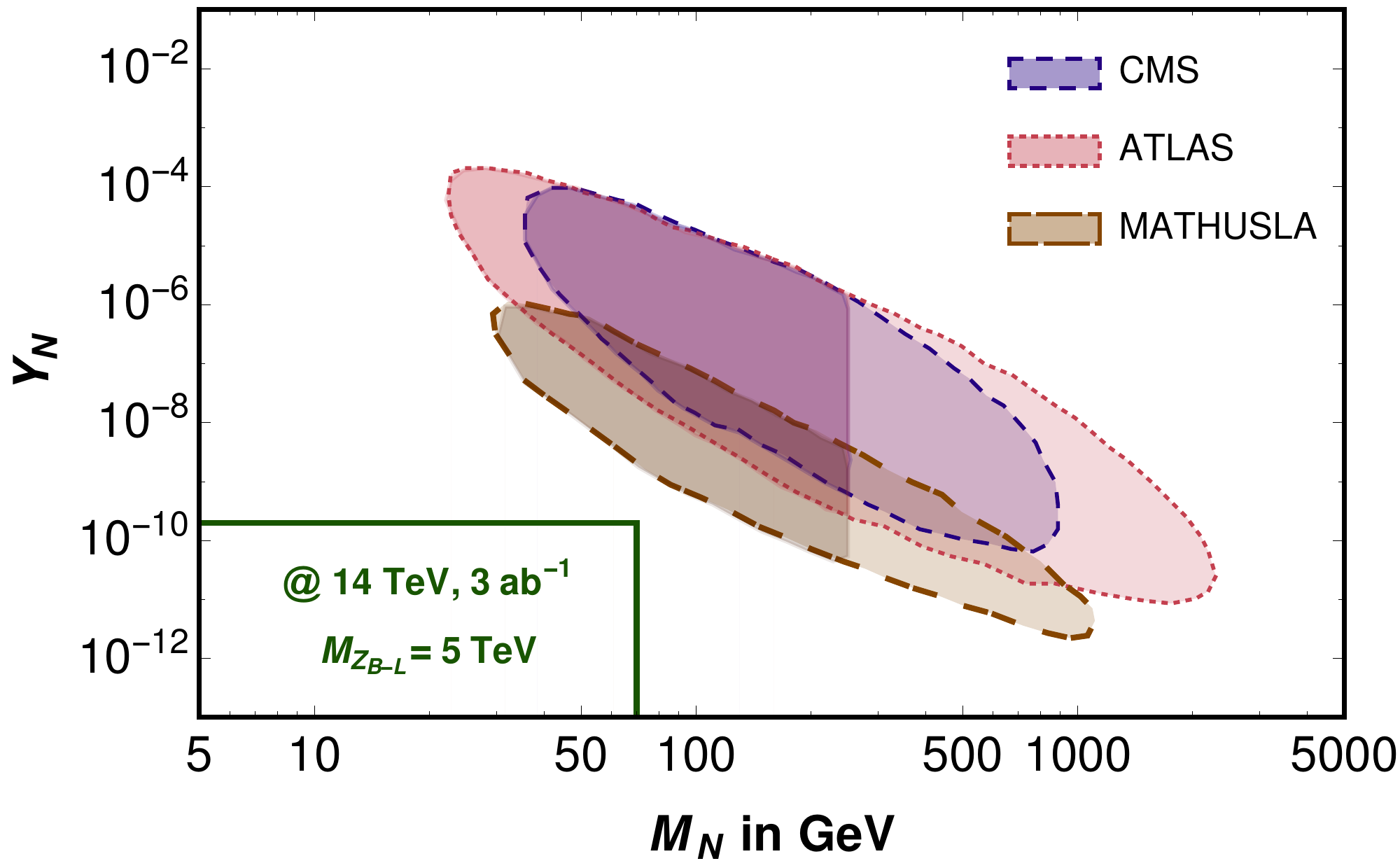}\label{14dcyl}}\quad
			\subfigure[]{\includegraphics[width=0.35\linewidth,angle=0]{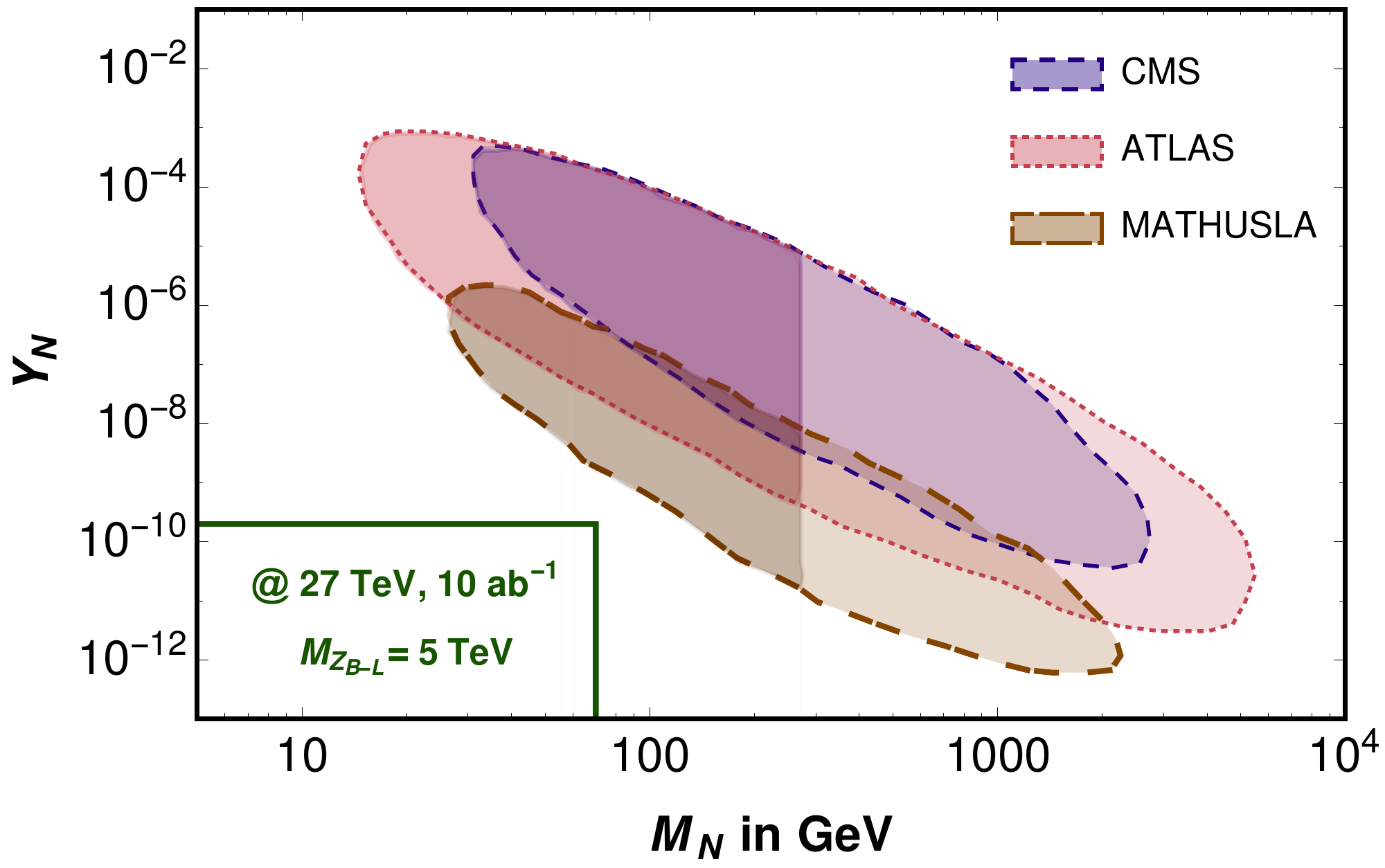}\label{30dcyl}}\quad
			\subfigure[]{\includegraphics[width=0.35\linewidth,angle=0]{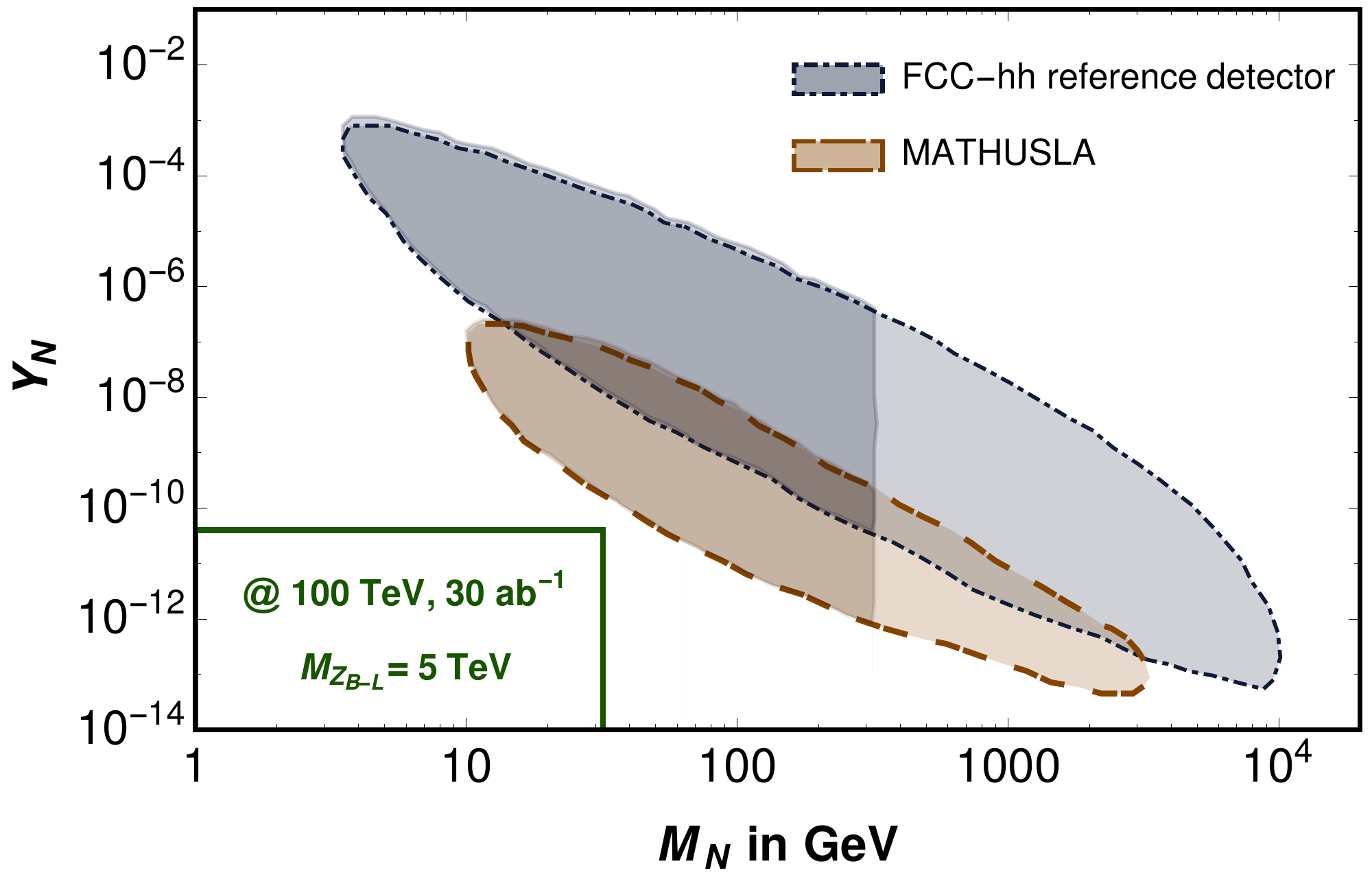}\label{30dcyl}}}        		
			\caption{Limits obtained via the displaced decays of RHNs to $2\ell + 2j$ final state (dark shaded region),  $2\ell + 4j$ final state (light shaded region) and it is presented in $Y_N$ versus $M_{N}$  plane at $95\%$ CL, where $M_{Z_{B-L}}$ is 5\,TeV. The probable regions are for different centre of mass energies, i.e. 14\,TeV, 27\,TeV and 100\,TeV at the integrated luminosities of 3, 10 and 30\,ab$^{-1}$, respectively. The purple, pink, grey and light brown colours depict the reaches for CMS, ATLAS, FCC-hh reference detector and MATHUSLA, respectively.}\label{ReachYnMn}
	\end{center}
\end{figure}

In the similar manner,  the parameter space can be explored in a different angle.  In \autoref{ReachYnMn}, the reaches in the $Y_N - M_N$ plane are shown for a fixed value of $M_{Z_{B-L}}=5$ TeV.  We present our result in a log-log scale for $ 2\ell +2j$ final state with darker shades, and  $2\ell +4j$ with lighter shades with the centre of mass energies of 14, 27, 100 TeV  at the luminosities of 3, 10, 30 ab$^{-1}$, respectively. The pink, blue and brown regions can be explored by ATLAS, CMS and MATHUSLA, respectively, whereas, the grey region can be probed via FCC-hh reference detector described in \autoref{DisVartex_SC1}. As the cross-section decreases with the increase of the RHN mass, the reach stays up to $M_N=2.5\,(10)\,\rm{TeV}$ for 14\,(100)\,TeV centre of mass energies at the LHC\,(FCC-hh).  Displaced decay can be observed for the RHN mass of $\sim 3$ GeV with substantially high Yukawa coupling up to $\sim 10^{-3}$.

\section{Boost effect on di-lepton final state}\label{boostfs}

In this section we investigate the boost effect to our most desired final states of two leptons, which arise as $2\ell+2 j$ and $2\ell +4j$. $2\ell$ final state can come from the leptonic decays of the gauge bosons. In case of  leptons coming from $Z$ boson, it gives rise to final states with opposite sign lepton and maximally two parton level jets. However, if we focus on the scenario where both the RHNs decay via $\ell^\pm W^\mp$, it gives rise to final state of $2\ell +4j$ (if the $W^{\mp}$ decays hadronically), where the leptons can be either same sign or opposite sign and ideally with 1:1 ratio.  The departure from the ideal case happens when the mass of the RHN is smaller as compared to the centre of mass energy of the collider,  producing boosted RHNs. In that case  two of the potential jets coming from the $W^\pm$, are collimated  and form a Fatjet  and we obtain $2\ell+2 j$ final state. 

\begin{figure}[h!]
	\centering
	\includegraphics[width=0.6\linewidth,angle=0]{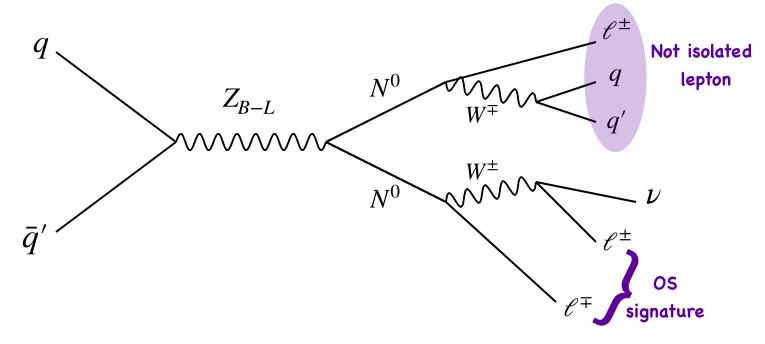}
	\caption{A schematic diagram of getting more OS over SS in boosted scenario.}\label{Fig:OSD}
\end{figure}


\begin{table}[h]	
	\begin{center}
		\hspace*{-1.2cm}
		\renewcommand{\arraystretch}{1.2}
		\begin{tabular}{ |c|c|c|c|c|c|c| }
			\cline{2-7}
			\multicolumn{1}{c|}{}&\multirow{3}{*}{Final states} &
			\multirow{2.0}{*}{$M_N$ in} &\multicolumn{4}{c|}{Centre of mass energy}\\
			\cline{4-7}
			\multicolumn{1}{c|}{} & & \multirow{2.5}{*}{GeV}  & \multicolumn{2}{c|}{27\,TeV} & \multicolumn{2}{c|}{100\,TeV}  \\
			\cline{4-7} 
			 \multicolumn{1}{c|}{} & &  & SS & OS & SS & OS \\
			\hline
			\multirow{4}{*}{CMS} & \multirow{3}{*}{$2\ell + 2j$} & 60 & 15.6\,(5.3\%) & 273.9\,(94.7\%) & $-$ & $-$ \\
			\cline{3-7}
			& & 100 & 87.7\,(10.2\%)  & 768.3\,(89.7\%) & $-$ & $-$ \\
			\cline{3-7}
			& & 250 & 231.4\,(17.7\%) & 1085.3\,(82.4\%) &  $-$ & $-$ \\	
			\cline{2-7}
			& \multirow{1}{*}{$2\ell + 4j$}& 500 & 54.7\,(20.5\%) & 211.4\,(79.4\%) &  $-$ & $-$ \\	
			
			\hline \hline
			\multirow{1.5}{*}{ATLAS} & \multirow{3}{*}{$2\ell + 2j$} & 60 & 29.5\,(6.1\%) & 453.1\,(93.9\%)  & 2057.9\,(4.3\%) & 45510.6\,(95.7\%) \\
			\cline{3-7}
			\multirow{1.1}{*}{\&} & & 100 & 87.7\,(10.2\%)  & 768.3\,(89.7\%) & 7582.7\,(9.3\%) & 73820.1\,(90.7\%) \\
			\cline{3-7}
			\multirow{0.5}{*}{FCC-hh reference} & & 250 & 231.4\,(17.7\%) & 1085.3\,(82.4\%) & 8011.4\,(22.4\%) & 27726.6\,(77.6\%) \\
			\cline{2-7}
			\multirow{0.5}{*}{detector} & \multirow{1}{*}{$2\ell + 4j$} & 500 & 54.7\,(20.5\%) & 211.4\,(79.4\%) & 566.8\,(23.0\%) & 1893.3\,(76.9\%) \\	
			\hline \hline 
			\multirow{4}{*}{MATHUSLA} & \multirow{3}{*}{$2\ell + 2j$} & 60 & 1.8\,(7.3\%) & 22.9\,(92.7\%) & 258.3\,(9.4\%) & 2461.7\,(90.6\%) \\
			\cline{3-7}
			& & 100 & $-$ & $-$ & $-$ & $-$ \\
			\cline{3-7}
			& & 250 & $-$  & $-$ & $-$ & $-$ \\	
			\cline{2-7}
			& \multirow{1}{*}{$2\ell + 4j$}& 500 & $-$  & $-$ & $-$ & $-$ \\	
			\hline
			
		\end{tabular}
		\caption{Number of events containing SS or OS and their percentage numbers (in bracket) for different RHN masses ($M_N$ in GeV) with the centre of mass energies of 27, 100\,TeV, at the integrated luminosities of $ 10, \,\, 30\,\text{ab}^{-1}$, respectively, considering the root sum square values of the Yukawa couplings discussed in \autoref{BPs_SC1}.} \label{ss_os_UPMNS}
	\end{center}	
\end{table}

The leptons coming from two RHNs decay can give rise to same sign di-lepton and opposite sign di-lepton and $N$ being Majorana in nature the ratio of SS:OS ideally should be 1:1 \cite{Chen:2011hc,Sirunyan:2018xiv,Das:2017hmg}. Such signature can be achieved if we demand both the $W^\pm$s coming from two RHNs decays hadronically and thus having a parton level final state of $2\ell +4j$. However, the lepton coming from RHN decays often are co-linear to the hadronic jet coming from the $W^\pm$ decays and thus cannot satisfy the isolation criteria, whereas, we still get di-lepton when the other $W^\pm$ from the other RHN decays leptonically as can be seen from \autoref{Fig:OSD}. In that case the second RHN leg fully contribute to di-lepton, which only contributes to OS and other ISR/FSR jets add up to the $4j$ criteria. In the absence of the hard ISR/FSR jets satisfying the cuts (\autoref{cuts}), such events forms a final state of $2\ell +2j$ and more likely to happen for lighter RHNs ($M_N\lesssim 300$ GeV). However, for higher $M_N$ ($\gsim 300$ GeV), $2\ell+ 4j$ final state is more likely. These two phenomena are explicitly mentioned in \autoref{ss_os_UPMNS} and \autoref{ss_os_1gen}. In both the cases, such collinearity leads to skewed ratio of SS:OS, and as we go for higher mass values, the ratio of SS:OS gets better.
	
Scenario-1 (\autoref{SC1}) deals with $U_{\rm PMNS}$ with three generations of RHNs in Casas-Ibarra parameterization, where $Y_N \sim \sqrt{M_N}$, that leads to relatively higher $Y_N$ for higher mass values. In \autoref{ss_os_UPMNS} we present the number of events for $2\ell+2j$ final state for $M_N=60,\, 100,\, 250, \,500$ GeV. The events for $M_N=10$ GeV are mostly outside the detectors i.e. CMS, ATLAS, FCC-hh reference detector and MATHUSLA. Whereas, for $M_N= 1000$ GeV, we get prompt leptons, which is not our interest for this article. $M_N=60$ GeV, has some events for $2\ell+2j$ final state inside MATHUSLA detectors, unlike $100, 250, 500$ GeV, which lie inside the CMS, ATLAS or proposed FCC-hh reference detector range.  

\autoref{ss_os_1gen} describes the similar situation for the scenario-2 (in \autoref{SC2}), where we consider only one generations of RHN with $Y_N= 5 \times 10^{-9}$. In this case even for $M_N= 1000$ GeV, we obtain the displaced signature for $2\ell +4j$ final state. However, for $M_N=1000$ GeV point, the displaced decay does not reach to MATHUSLA.

\begin{table}[h]	
	\begin{center}
		\hspace*{-1.2cm}
		\renewcommand{\arraystretch}{1.2}
		\begin{tabular}{|c|c|c|c|c|c|c| }
			\cline{2-7}
			\multicolumn{1}{c|}{}& \multirow{3.0}{*}{Final states} &
			\multirow{2.0}{*}{$M_N$ in} &\multicolumn{4}{c|}{Centre of mass energy}\\
			\cline{4-7}
			\multicolumn{1}{c|}{}& & \multirow{2.5}{*}{GeV}  & \multicolumn{2}{c|}{27\,TeV} & \multicolumn{2}{c|}{100\,TeV}  \\
			\cline{4-7} 
			\multicolumn{1}{c|}{}& &  & SS & OS & SS & OS \\
			\hline
			\multirow{4}{*}{CMS} & \multirow{2}{*}{$2\ell + 2j$} & 100 & 3.2\,(15.4\%) & 17.6\,(84.6\%) & $-$ & $-$ \\
			\cline{3-7}
			& & 250 & 10.3\,(26.3\%)  & 28.8\,(73.6\%) & $-$ & $-$ \\
			\cline{2-7}
			& \multirow{2}{*}{$2\ell + 4j$} & 500 & 16.1\,(32.6\%) & 33.3\,(67.4\%) &  $-$ & $-$ \\
			\cline{3-7}
			& & 1000 & 13.5\,(34.4\%) & 25.8\,(65.6\%) &  $-$ & $-$ \\	
			\hline \hline
			\multirow{1.5}{*}{ATLAS} & \multirow{2}{*}{$2\ell + 2j$} & 100 &  5.6\,(16.0\%) & 29.3\,(83.9\%)   & 397.8\,(18.2\%) & 1781.6\,(81.8\%) \\
			\cline{3-7}
			\multirow{1.1}{*}{\&} & & 250 &   13.1\,(25.2\%) & 38.8\,(74.7\%)  &  623.2\,(27.4\%) & 1653.8\,(72.6\%)  \\
			\cline{2-7}
			\multirow{0.5}{*}{FCC-hh reference} & \multirow{2}{*}{$2\ell + 4j$} & 500 & 18.4\,(32.7\%) & 37.7\,(67.2\%) &  702.0\,(36.7\%) & 1208.4\,(63.2\%)  \\
			\cline{3-7}
			\multirow{0.5}{*}{detector}& & 1000 &  14.4\,(34.6\%) & 27.2\,(65.4\%)  &  547.7\,(37.8\%) & 899.9\,(62.1\%)  \\	
			\hline \hline 
			\multirow{4}{*}{MATHUSLA} & \multirow{2}{*}{$2\ell + 2j$} & 100 &  0.7\,(15.2\%) & 3.9\,(84.7\%) &  73.9\,(14.8\%) & 422.1\,(85.1\%)  \\
			\cline{3-7}
			& & 250 &  1.6\,(23.5\%) & 5.2\,(76.3\%) &  84.2\,(22.9\%) & 282.5\,(77.1\%) \\
			\cline{2-7}
			& \multirow{2}{*}{$2\ell + 4j$} & 500 &   1.5\,(34.8\%) & 2.8\,(65.1\%) & 13.1\,(25.2\%) & 38.8\,(74.7\%) \\
			\cline{3-7}
			& & 1000 & $-$  & $-$ & $-$ & $-$ \\	
			\hline
			
		\end{tabular}
		\caption{Number of events containing SS or OS and their percentage numbers (in bracket) for different RHN masses ($M_N$ in GeV) with the centre of mass energies of 27, 100\,TeV, at the integrated luminosities of $10, \,\,30\,\text{ab}^{-1}$, respectively, and for $Y_N= 5 \times 10^{-9}$.} \label{ss_os_1gen}
	\end{center}	
\end{table}

It has been noted that without tagging the legs, i.e. the RHNs, event for higher mass values, it is not possible to get OS:SS as 1:1. Achieving Majorana nature thus relies on perfectly tagging of both the RHNs in a pair production which will be discussed in the next section.

\section{Majorana nature: OSD vs SSD}\label{ssd_osd}

To extract the information for the Majorana fermion, it is essential to reconstruct the the RHNs, i.e. the legs from which the charged leptons are coming. Charged leptons coming from such RHNs, without the contamination of the leptons from the other gauge bosons in the decay chain can easily satisfy the  same sign di-lepton (SSD): opposite sign di-lepton (OSD) equals to 1:1 \cite{Chen:2011hc,Sirunyan:2018xiv,Das:2017hmg}. It is worth mentioning here, since MATHUSLA will be situated in one of the hemispheres, it can only tag one RHN leg for a given event. Thus, measurement of OSD:SSD for a given event is not possible inside the MATHUSLA detector and we have to rely on CMS, ATLAS and the proposed FCC-hh detector.  
Ideally, $pp \to NN \to \ell^\pm \ell^\mp +2 W^\pm$ gives rise to $2\ell +4j$ final state when $W^\pm$ decays hadronically. However, due to the boost effect, the jets coming from the $W^\pm$ can be co-linear making the final state as $2\ell +2j$ as discussed in the previous section. This occurs mostly when $M_N \lesssim 300$ GeV.

\begin{table}[h]	
	\begin{center}
		\hspace*{-1.0cm}
		\renewcommand{\arraystretch}{1.2}
		\begin{tabular}{ |c|c|c|c|c|c|c| }
			\cline{2-7}
			\multicolumn{1}{c|}{}&\multirow{3}{*}{Final states} &
			\multirow{2.0}{*}{$M_N$ in} &\multicolumn{4}{c|}{Centre of mass energy}\\
			\cline{4-7}
			\multicolumn{1}{c|}{} & & \multirow{2.5}{*}{GeV}  & \multicolumn{2}{c|}{27\,TeV} & \multicolumn{2}{c|}{100\,TeV}  \\
			\cline{4-7} 
			\multicolumn{1}{c|}{}& & & SS & OS & SS & OS \\
			\hline
			\multirow{3}{*}{CMS} & \multirow{2}{*}{$2\ell + 2j$}  & 100 & 23.8\,(48.8\%)   & 25.0\,(51.2\%) & $-$ & $-$ \\
			\cline{3-7}
			& & 250 & 48.4\,(49.0\%) & 50.3\,(51.0\%) &  $-$ & $-$ \\
			\cline{2-7}
			& \multirow{1}{*}{$2\ell + 4j$} & 500 & 16.6\,(49.4\%) & 17.0\,(50.6\%) &  $-$ & $-$ \\	
			\hline \hline
			\multirow{1.5}{*}{ATLAS \&} & \multirow{2}{*}{$2\ell + 2j$}  & 100 & 23.8\,(48.8\%)  & 25.0\,(51.2\%) & 434.7\,(49.2\%) & 448.4\,(50.7\%)  \\
			\cline{3-7}
			\multirow{1.4}{*}{FCC-hh reference} & & 250 & 48.4\,(49.0\%) & 50.3\,(51.0\%) &  549.5\,(48.8\%) & 576.1\,(51.2\%) \\	
			\cline{2-7}
			\multirow{1.0}{*}{detector} & \multirow{1}{*}{$2\ell + 4j$} & 500 & 16.6\,(49.4\%) & 17.0\,(50.6\%) &  202.0\,(49.6\%) & 205.6\,(50.4\%) \\	
			\hline 			
		\end{tabular}
		\caption{Number of events containing SSD or OSD and their percentage numbers (in bracket) for different RHN masses ($M_N$ in GeV) with the centre of mass energies of 27, 100\,TeV, at the integrated luminosities of $ 10, \,\, 30\,\text{ab}^{-1}$, respectively, considering the root sum square values of the Yukawa couplings discussed in \autoref{BPs_SC1}.} \label{ssd_osd_UPMNS}
	\end{center}	
\end{table}

\begin{table}[h]	
	\begin{center}
		\hspace*{-1.2cm}
		\renewcommand{\arraystretch}{1.2}
		\begin{tabular}{|c|c|c|c|c|c|c| }
			\cline{2-7}
			\multicolumn{1}{c|}{} & \multirow{3.0}{*}{Final states} &
			\multirow{2.0}{*}{$M_N$ in} &\multicolumn{4}{c|}{Centre of mass energy}\\
			\cline{4-7}
			\multicolumn{1}{c|}{}& & \multirow{2.5}{*}{GeV}  & \multicolumn{2}{c|}{27\,TeV} & \multicolumn{2}{c|}{100\,TeV}  \\
			\cline{4-7} 
			\multicolumn{1}{c|}{}& &  & SSD & OSD & SSD & OSD \\
			\hline
			\multirow{4}{*}{CMS} & \multirow{2}{*}{$2\ell + 2j$} & 100 & 2.6\,(48.1\%) & 2.8\,(51.8\%) & $-$ & $-$ \\
			\cline{3-7}
			& & 250 & 7.9\,(49.0\%)  & 8.2\,(51.0\%) & $-$ & $-$ \\
			\cline{2-7}
			& \multirow{2}{*}{$2\ell + 4j$} & 500 & 12.9\,(48.7\%) & 13.6\,(51.3\%) &  $-$ & $-$ \\
			\cline{3-7}
			& & 1000 & 10.1\,(48.8\%) & 10.6\,(51.2\%) &  $-$ & $-$ \\	
			\hline \hline
			\multirow{1.5}{*}{ATLAS} & \multirow{2}{*}{$2\ell + 2j$} & 100 &  3.4\,(48.6\%) & 3.6\,(51.4\%)  & 223.6\,(50.0\%) & 223.6\,(50.0\%) \\
			\cline{3-7}
			\multirow{1.1}{*}{\&} & & 250 &   9.2\,(50.0\%) & 9.2\,(50.0\%)  &  347.4\,(49.7\%) & 353.0\,(50.3\%)  \\
			\cline{2-7}
			\multirow{0.5}{*}{FCC-hh reference} & \multirow{2}{*}{$2\ell + 4j$} & 500 & 15.2\,(48.7\%) & 16.0\,(51.3\%) &  391.6\,(49.3\%) & 402.5\,(50.6\%)  \\
			\cline{3-7}
			\multirow{0.5}{*}{detector}& & 1000 &  12.5\,(49.8\%) & 12.6\,(50.1\%)  &  247.9\,(48.7\%) & 260.6\,(51.2\%)  \\	
			\hline 			
		\end{tabular}
		\caption{Number of events containing SSD or OSD and their percentage numbers (in bracket) for different RHN masses ($M_N$ in GeV) with the centre of mass energies of 27, 100\,TeV, at the integrated luminosities of $10,\, 30\,\text{ab}^{-1}$ and for $Y_N= 5 \times 10^{-9}$.} \label{ssd_osd_1gen}
	\end{center}	
\end{table}

It is evident from the previous section, isolation cuts and the boost effect together can alter the ratio of SS:OS from 1:1. This prompt us to reconstruct the RHN legs via the stepwise reconstructions of the $W^\pm$ boson as well as RHN mass $M_N$. We incorporated advance cuts to ensure that the leptons are coming from opposite legs, i.e. from two different RHNs. This is achieved for $2\ell+4j$ final state by demanding that we observe two $W^\pm$ peaks in a event and then reconstruct two different RHNs by  reconstructing the invariant mass $m_{j_i\,j_j, \,\ell_k}$, where  $j_i,\, j_j$ corresponds to the  two jets coming from the 10  GeV window of the  $W^\pm$ peak in the invariant mass  distribution of the  di-jet,  i.e. $m_{j_i j_j}$, and $\ell_k$ correspond to the hard leptons for two different  $m_{j_i\,j_j, \,\ell_k}$ distributions. Once the $m_{j_i\,j_j, \,\ell_k}$ distributions are formed we demand two peaks around $M_N$ in a given event and demand the leptons to be inside the 10 GeV mass window around $m_{j_i\,j_j, \,\ell_k}=M_N$. For $2\ell+2j$, the Fatjet is formed with the jet mass at $M_{W^\pm}$ and $m_{j_i, \ell_j}$ peaks around $M_N$. Charges  of  these two leptons are then investigated to calculate SSD: OSD as tabulated in  \autoref{ssd_osd_UPMNS} for scenario 1 and \autoref{ssd_osd_1gen} for scenario 2 for both the final states ($2\ell + 2j$ and $2\ell + 4j$). We can remind ourselves that for scenarios 1, as we have considered $U_{\rm PMNS}$, for higher $M_N$ (for $2\ell+4j$), we do not get any displaced leptons and thus they are not considered here. From  \autoref{ssd_osd_UPMNS} and \autoref{ssd_osd_1gen} we realise that though the Majorana nature (SSD:OSD=1:1) can be restored but the requirements of the additional cuts result in very low number of events and only possible at the HE-LHC and FCC-hh with 27, 100 TeV centre of mass energies at the integrated luminosity of 10, 30 ab$^{-1}$, respectively. For lower mass values $M_N <M_{W^\pm}$, the $W^\pm$ remains off-shell which makes such reconstruction of RHNs difficult, along with that the lower statistics make it hard to reproduce SSD:OSD=1:1.

\section{Discussion and conclusion}\label{discussion}
In this article, we consider $Z_{B-L}$ model to produce the RHN pairs, which can have displaced decay depending on the Yukawa coupling $Y_N$. In the scenario-1, we consider Casas-Ibarra parameterization to incorporate the light neutrino mass and mixing angle considering three generations of RHNs. The longitudinal and transverse boost effects are investigated separately and their effects on the displaced decay lengths are extensively studied.  In this context the event numbers in CMS, ATLAS and MATHUSLA detectors are shown for HL-LHC and HE-LHC. The event number for the collision with a 100 TeV centre of mass energy is provided for the proposed FCC-hh detector and MATHUSLA as well.  Specific final states of $4\ell,\, 3\ell+\geq 1j,\, 2\ell +\geq 2j$ are also studied and we see that BP2, i.e $M_N=60$ GeV satisfying $U_{\rm PMNS}$ has some promise at the MATHUSLA. The prospect of FASER-II, though not much, but is shown in the displaced decay distributions.

One of the most interesting feature of Majorana fermions, i.e same sign di-lepton (SSD) and opposite sign di-leptons (OSD) come as the same numbers when the leptons directly come from the RHN decay. However, boosted decay products from the RHNs decays are often collinear and fail the jet-lepton isolation criteria and the leptons coming from $W^\pm$ bosons can be misidentified as the leptons coming from RHNs. This results into skewed ratio of SSD:OSD, giving rise to more OSD. A thorough investigation of such effects on RHNs of different masses and for 27, 100\,TeV centre mass energies are studied. It has been found that lesser the boost, less skewed is the ratio.  A remedy to get back the Majorana signature via successfully tagging the legs are also prescribed. However, such reconstructions of the different RHN legs suffer in the final state event numbers.

For scenario-1 we also have  drawn the regions which can be probed at the LHC/FCC with centre of mass energies of 14, 27, 100 TeV at integrated luminosities of  3, 10 and 30 ab$^{-1}$ in the $M_N - M_{Z_{B-L}}$ plane. We see that a lighter mass of RHN  i.e. $M_N=5$   can be probed with a maximum of 900 GeV, whereas $Z_{B-L}$ can be probed up to 15.7 TeV. 

In the later part of the article, we consider scenario-2, motivated by the collider searches, where we consider one generation of RHN with small Yukawa coupling, which is a free parameter, while the others can explain the light neutrino masses and mixing. We benchmark the scenario with $M_N=60, 100$ GeV with $Y_N= 5\times 10^{-8}$ and $5\times 10^{-9}$. Final states of $4\ell,\, 3\ell+\geq 1j,\, 2\ell +\geq 2j$ are also studied as we found that MATHUSLA regions are preferred by BP2, BP3 for $Y_N=5\times 10^{-8},\, 5\times 10^{-9}$, respectively. To complete the study we drew  the regions plots in $M_N-M_{Z_{B-L}}$ (in \autoref{ReachMzMn}),  $M_{Z_{B-L}}-Y_N$ (in \autoref{ReachYnMz}) and $M_N-Y_N$  (in \autoref{ReachYnMn}) planes, where in \autoref{ReachYnMz} we assume a hierarchy of $M_{Z_{B-L}}=10 M_N$. $M_N=4200$ GeV along with $M_{Z_{B-L}}=30.5$ TeV can be explored with $Y_N=5\times 10^{-9}$ at the FCC-hh with centre of mass energy  of  100 TeV. Similarly, $Y_N$ can be probed as low as $\mathcal{O}(10^{-13})$.  While considering lighter mass of RHN,  $M_N\sim 3$ GeV can also be explored with  substantially  high values of $Y_N\sim 10^{-3}$.

The study performed in this article can be easily attributed to other $U(1)^\prime$ models with right-handed neutrino \cite{Bandyopadhyay:2011qm,Bandyopadhyay:2015iij,Bandyopadhyay:2014sma, Bandyopadhyay:2017bgh,Accomando:2016rpc,Accomando:2016sge} and other displaced neutral decays \cite{SabanciKeceli:2018fsd, Bandyopadhyay:2010cu,Bhattacherjee:2021rml}. The segregation of transverse and longitudinal displaced decays manifests the longitudinal boost effect at higher centre of mass energies. Finally general SS:OS signature can be skewed due to the boost effect. However, Majorana nature can be explored via the tagging of the RHNs legs at the CMS, ATLAS and the proposed FCC-hh detector. Due to presence of MATHUSLA detector in one of the hemispheres, it cannot be resolved there and we have to rely on CMS, ATLAS or FCC-hh detector.

\section*{Acknowledgements}
PB acknowledges SERB CORE Grant CRG/2018/004971 and MATRICS Grant MTR/2020/000668 for the financial and computational support towards the work. PB also acknowledges Anomalies 2019 for the initiation of the project. CS would like to thank MOE Government of India for the SRF. PB and CS would like to express their gratitude
to Korea Institute For Advanced Study for arranging a collaborative visit during the last
phase of the project.

\bibliography{Reference}

\end{document}